\documentclass[twocolumn,epjc3]{svjour3}

\usepackage[english]{babel}
\usepackage{latexsym}
\usepackage{amsmath}
\usepackage{amssymb}
\usepackage{amsfonts}
\usepackage{graphicx}
\usepackage{epstopdf}
\usepackage{xcolor}

\usepackage{cite}           

\usepackage{cuted}  
\usepackage{orcidlink}

\allowdisplaybreaks

\newcommand{\cc}[1]{\mathnormal{#1}}

\renewcommand{\vec}[1]{{\mathbf{#1}}}
\newcommand{\nuc}[2]{$^{#1}${#2}}

\newcommand{\iunit}{\mathrm{i}}

\newcommand{\vnabla}{\boldsymbol{\mathbf\nabla}}
\newcommand{\vsigma}{\boldsymbol{\mathbf\sigma}}

\newcommand{\rmd}{\text{d}}

\newcommand{\fourmat}[4]{{\begin{pmatrix}
                         {#1} & {#2} \\ {#3} & {#4}
                         \end{pmatrix}}}
\newcommand{\twospinor}[2]{{\begin{pmatrix}
                           {#1} \\ {#2}
                           \end{pmatrix}}}

\newcommand{\nn}{\nonumber}

\newcommand{\LOPP}[2]{LO$_{\mathrm{#1}}^{\mathrm{#2}}$}

\begin{document}

\title{Parameter adjustment of nuclear leading-order local pairing energy density functionals}

\author{Michael Bender   \orcidlink{0000-0001-8707-3410} 
        \and 
        Karim Bennaceur  \orcidlink{0000-0002-6722-491X}
        \and 
        Valentin Guillon \orcidlink{0009-0006-1476-402X}
       }

\institute{Universit{\'e} Lyon 1, 
         CNRS / IN2P3,
         IP2I Lyon, 
         UMR 5822, Villeurbanne, France
}

\date{11 October 2026}

\maketitle
%

\begin{abstract} 
Phenomenological local pairing energy density 
functionals (EDFs) are widely used in the literature to complement 
Skyrme and other forms of effective EDFs that model the in-medi\-um
particle-hole interaction in nuclei.
Because the amount of pair correlations induced by such pairing
EDF scales with the level density of the underlying single-particle 
spectrum and the size of the pairing-active space, its 
parameters have to be specifically adjusted for each underlying 
choice of mean field and pairing regulator.
The adjustment of pairing parameters that yield consistent results across
many different Skyrme parameter sets is not a trivial task, mainly
because of the large scatter in predictions for single-particle
spectra. 

This study reports on the benchmarking of a protocol for the 
adjustment of the parameters of a local leading-order (LO) $T=1$
(like-particle) pairing EDF that consists in adjusting the density-dependence 
of the ${}^{1}S_{0}$ pairing gap at the chemical potential in 
infinite nuclear matter (INM). When using a suitably chosen 
reference calculation, this protocol leads to consistent results 
for the odd-even staggering of masses of spherical and heavy deformed nuclei 
and also for the rotational moments of inertia calculated 
in a time-reversal-breaking cranked HFB approach.

The implementation of the HFB solver for infinite matter at
arbitrary isospin asymmetry used for this study is sketched
in appendices. Additional points that are discussed concern 
(i) the illustration that the gaps at the chemical potential are not 
necessarily sufficient to completely characterise the pairing 
interaction in infinite matter, and that adjusting LO pairing EDF
to reproduce gaps obtained from finite-range pairing 
interactions in HFB or from more microscopic calculations can 
lead to unrealistic predictions for finite nuclei;
(ii) the finding that some regions of the parameter space for
the density dependence of the $T=1$ LO pairing EDF lead to a 
spurious transition to a Bose-Einstein condensate of di-nucleons in 
spite of producing realistic pairing correlations for well-bound
nuclei;
(iii) the correlation between effective mass and the parameters of the LO 
pairing EDF that control the form factor of the density dependence when 
reproducing pairing gaps in infinite matter,
(iv) the significant impact on the odd-even staggering of masses
made by either including or not the spin-gradient terms 
in the particle-hole part of the Skyrme EDF;
(v) the sizeable impact of keeping or not the contribution from the 
density-dependent LO pairing EDF to the mean fields on the odd-even 
staggering of radii.

\end{abstract}
\maketitle
%
%

\section{Introduction}
\label{intro}

Pairing correlations affect virtually all aspects of phenomena in nuclear
matter and finite nuclei \cite{Ring80a,Brink05a,Broglia13a,Dean03a,Chamel08a}, 
from the phases of matter to be found in neutron stars, over the odd-even staggering 
of nuclear masses, the energy difference between various minima and/or maxima
of deformation energy surfaces, the generic difference between the spectra of
low-lying states of even-even, odd-mass, and odd-odd nuclei, 
the moments of inertia of collective rotation and other modes 
of large-amplitude collective motion, 
to nuclear reactions, most prominently pair transfer.

Methods based on a phenomenological energy density functional 
(EDF)  that models the effective in-medi\-um interaction provide a widely-used 
framework to describe the aforementioned phenomena \cite{Bender03a,Schunck19a}. 
Many different forms of EDFs are employed in the literature. For some of these,
the particle-hole (ph) and particle-particle (pp, also often called pairing
or anomalous)
parts of the EDF are generated from the same underlying pseudo-potential with 
common coupling constants consistently derived from the same parameters.
For others, and actually in the majority of cases, the ph and pairing parts of the EDF are constructed 
independently from one another with unconnected coupling constants.
Gogny interactions are an example of the former, whereas the vast majority
of parametrisations of the Skyrme EDF are examples of the latter. 
But even when using separate phenomenological functionals for the 
ph and pairing parts of the EDF, in order to describe data, the 
parameters of the pp EDF cannot be chosen independently from those 
of the ph part of the EDF. The main reason is that the magnitude of 
pair correlations found in a nuclear system is correlated with 
the density of single-particle levels \cite{Brink05a} that in turn is
mainly determined by the ph part of the EDF. 
The global trend of the density of single-particle levels around the 
chemical potential is determined by the effective mass 
\cite{Jensen86a}, and it is 
well-known that parameter sets of the ph EDF that yield different
effective masses require different strength of the
pairing interaction \cite{Bender03a,Jensen86a,DaCosta24a}. 

The actual shell structure of finite nuclei, however, always
deviates from these global trends, and no available ph EDF is capable
of describing all details of the phenomenological level sequences
\cite{Bonneau07a,Lesinski07a,Tarpanov14a,Dobaczewski15a,Bender09t}. 
When adjusting the 
parameters of the pairing EDF to observables from a few given nuclei 
calculated for different parametrisations of the ph EDF that have the
same effective mass, one may still end up with rather different pairing 
strength for each because the underlying mean fields differ in their 
(imperfect) reproduction of the single-particle spectra.

This can pose consistency problems when constructing series of 
parametrisations that, in a controlled way, differ in some property
that affects single-particle spectra. We think it is therefore opportune 
to investigate if the parameters of the pairing EDF can be adjusted in 
a more robust way that is only sensitive to the global trends of 
single-particle spectra, but not their details.

There have been attempts to construct non-empiri\-cal parametrisations of the 
pairing EDFs through the direct mapping of two-body matrix elements obtained
within a microscopic method on those defining the EDF without passing 
through many-body observables \cite{Duguet08a,Hebeler09a,Lesinski09a}. 
While such studies greatly help to understand the origin and 
global features of the effective in-medi\-um pairing interaction, for the 
time being the results do not yet have the predictive power needed 
for detailed nuclear structure studies.

Another possibility that has been used in the past is to reproduce
phenomenological data for the nucleon-nucleon scattering length
in the vacuum \cite{Bertsch91a,Esbensen97a,Matsuo06a}. Such procedure,
however, has also its limits as it does not control the in-medium aspects 
of the effective pairing interaction.

An alternative possibility is offered by the pairing properties 
of homogeneous infinite nuclear matter (INM). 
The continuous single-particle spectrum of INM only depends on 
the (in general momentum-dependent) effective mass of nucleons in 
the medium that varies smooth\-ly 
with density, proton-neutron asymmetry and spin polarisation. 
With this, homogeneous INM provides a robust framework for the study of the
features of pairing interactions that is free of the complications of
finite nuclei with their discrete single-particle spectra whose bunching 
around the chemical potential that strongly varies with proton and neutron
number and also deformation.

Calculations of paired INM also offer the possibility to adjust the
parameters of the EDF to some reference pairing gaps obtained with 
more microscopic models that construct a correlated wave function 
of paired INM \cite{Garrido99a,Duguet04a,Chamel10a}.
When aiming at the predictive EDF description of pairing correlations
in nuclei, however, it has to be recalled that the effective pairing 
interaction, and with that the pairing gap, is 
affected by the coupling of nucleon pairs to excitation modes 
that have a different nature in infinite matter 
\cite{Shen05a,Ramanan18a,Urban20a,Ramanan21a}
and systems with a surface such as finite nuclei
\cite{Brink05a,Baldo04a,Pastore08a,Baldo10a,Idini12a,Idini16a,Litvinova20a},
see also \cite{ESNT25a}.
For the purpose of this study, we will not touch upon this issue
but address the question of how the simplicity of HFB calculations of paired 
INM can be used to transfer the predictive power of a given reference
pairing EDF constructed for a specific ph EDF to calculations of finite
nuclei with a different ph EDF.
The main goals of this study are
\begin{enumerate}
\item
the set-up of the tools for the calculation of isospin-asymmetric paired INM 
under conditions that match the schemes used to calculate paired finite nuclei 
with the Skyrme ph EDF, used in both its standard NLO and extended N2LO forms;

\item
the analysis of the effect of each of the parameters of the widely used density-dependent
local LO $T=1$ pairing EDF on pairing correlations in INM;

\item
the analysis of the connection between pairing correlations in INM and finite nuclei,
and how this link can be used to adjust the parameters of LO pairing EDFs that 
have comparable performance irrespective of the underlying ph mean fields;

\item
the analysis of results for three observables that are known to be 
sensitive to pairing correlations for isotopic chains of spherical and 
deformed nuclei obtained with symmetry-breaking 3d HFB calculations, 
which are the odd-even staggering of masses, the odd-even 
staggering of charge radii, and the rotational moment of inertia in 
the yrast band of even-even nuclei. 

\end{enumerate}
Along the way, we will also investigate a number of consequences
of some choices frequently made for the \textit{particle-hole} EDF 
on the performance of the \textit{pairing} EDF, in particular 
concerning spin-gradient terms in the ph Skyrme EDF and the contribution
from density-depen\-dent pairing EDFs to the ph mean fields. 

We also confirm the earlier observation that for some regions of its
parameter space the $T=1$ density-depen\-dent LO pairing EDF 
produces a nonphysical instability towards Bose-Einstein condensate 
of di-nucleons that jeopardises HFB calculations for weakly-bound nuclei, 
and this in spite of producing realistic pairing correlations 
for well-bound nuclei. 

%
\section{The energy density functional}
\label{sec:edf}

%
\subsection{The mean-field EDF}
\label{sec:edf:ph}

Calculations presented in what follows are performed using parametrisations of 
the standard Skyrme EDF at next-to-leading order (NLO) and of an extended Skyrme 
EDF at next-to-next-to-leading order (N2LO). The generator of the NLO Skyrme EDF 
contains two-body terms with up to two gradients, to which the generator of the 
N2LO Skyrme EDF adds terms with four gradients.

Throughout this paper, we use the notation and conventions of Ref.~\cite{Ryssens21a}.
The complete expression for the Skyrme EDF, which for the calculations of odd-mass 
nuclei and rotational bands comprises all time-even and time-odd terms, the local 
densities as well as potentials can be found in that reference.

For our study we consider the SLy4~\cite{Chabanat98a}, 
1T2T(0.70), 1T2T(0.80) (which is also called SLy7*), 
and 1T2T(0.85) parametrisations~\cite{DaCosta24a} of the EDF at NLO
as well as the SN2LO1 parameter set \cite{Becker17a} of the EDF at N2LO.
All of these were adjusted with similar fitting protocols. 
Compared to the widely-used SLy4, the recent 1T2T(X) parameter sets
have a more realistic surface energy \cite{DaCosta24a}. They 
differ in their isoscalar effective $k$-mass that has been set to the 
values $m^*_0/m = 0.7$, 0.8, and 0.85 as indicated between the parentheses. 
SN2LO1 is the first N2LO parametrisation that has been adjusted to 
describe properties of finite nuclei \cite{Ryssens21a,Becker17a}.

SLy4 and SN2LO1 follow the widely-used practice of considering only 
the one-body part of the centre-of-mass (CM) correction, whereas 
the 1T2T(X) employ the full one-body and two-body contributions \cite{DaCosta24a}. 

For the 1T2T(X) and SN2LO1 parameter sets the complete set of particle-hole
terms obtained from the respective Skyrme generator is kept in the ph part
of the EDF, including the spin-tensor 
terms and all time-odd terms, with coupling constants as obtained from a Skyrme
generator. By contrast, for SLy4 the spin-tensor terms are omitted, as is 
done for many parametrisations of the NLO Skyrme EDF. For reasons of Galilean 
invariance and internal consistency, the time-odd terms that contain two 
gradients and two Pauli matrices are then also usually neglected as done in
Refs.~\cite{Dobaczewski15a,Rigollet99a,Duguet01a,Duguet01n}. 
As explained in~\ref{sec:sly4+ulb}, this choice made for SLy4 has some 
unexpected consequences for properties of finite nuclei that are
relevant for our study.

When keeping the time-odd terms of the Skyrme EDF with their native 
coupling constants, many standard parametrisations 
exhibit finite-size instabilities in the spin channels at densities probed by finite 
nuclei \cite{Hellemans12a,Hellemans12b,Pastore15a}. The presence of these instabilities 
jeopardises the possibility to converge blocked configurations and self-consistently cranked
calculations for collective rotational bands. Such nonphysical behaviour can 
be avoid\-ed when adjusting the parameter sets with a constraint on the response 
properties of infinite matter \cite{Pastore13a}, which has been done
during the adjustment of the 1T2T(X) and SN2LO1. 
For SLy4, the absence of spin-gradient terms in the EDF sidesteps this problem~\cite{Hellemans12a}. 

Running the calculations for the present study, however, we noticed that 
some of the calculations of high-spin states of Yb isotopes presented in 
Sec.~\ref{sec:paired:nuclei} diverge when being run with SN2LO1. Further 
analysis revealed that SN2LO1 is on the edge of the onset of a finite-size
spin instability in the $T=0$ channel.
In the notation of Ref.~\cite{Ryssens21a}, a small reduction of the 
coupling constant of the time-odd isoscalar spin-gradient term 
$A^{(2,1)}_{0,\mathrm{o}} \, \vec{D}^{1,\sigma}_{0} (\vec{r}) \cdot \Delta \vec{D}^{1,\sigma}_{0} (\vec{r})$
from its Skyrme-force generated value 
$A^{(2,1)}_{0,\mathrm{o}} = 32.925 \, \text{MeV} \, \text{fm}^{5}$ to 
$28.0 \, \text{MeV} \, \text{fm}^{5}$ removes the problem, at least 
for the resolution of the coordinate-space mesh used for our calculations. 
The coupling constant of this term can be changed without compromising 
the Galilean invariance of the Skyrme EDF in the same spirit as what 
is often done for the NLO spin gradient terms when facing an 
instability~\cite{Hellemans12a}.
This modification, however, introduces a small violation 
of the exchange symmetry of the EDF (in addition to those already present because of 
density-dependent terms and the use of different EDFs for the particle-hole
and particle-particle channels) \cite{Stringari78a,Lacroix09a,Bender09p}.
All calculations of time-reversal-in\-vari\-ance breaking configurations reported
here were done with this modification of the SN2LO1 parameter set that leaves
time-reversal-conserving states unaffected.

For all parameter sets, in calculations of finite nuclei the Coulomb exchange 
energy \cite{Bender19a} is calculated in Slater approximation \cite{Ryssens15a}, 
while the Coulomb pairing energy \cite{Bender19a} is neglected. The strategy 
for the adjustment of the parameters of the pairing EDF is the main subject of 
our study.

%
%
\subsection{The pairing EDF}
\label{sec:edf:pair}

Using the notation and definitions of Ref.~\cite{Ryssens21a} for the local 
isoscalar density $D^{1,1}_{0}(\vec{r})$ and the local pair densities 
$\tilde{D}^{1,1}_q(\vec{r})$ and $\tilde{C}^{1,1}_q(\vec{r})$ of protons 
and neutrons, the local $T=1$ LO pairing EDF of Ref.~\cite{Rigollet99a} reads
\begin{align}
\label{eq:EDF:pair:ULB}
E_{\text{pair}}
& = \int \! \rmd^3 r \, \frac{V_0}{4} \,
    \Bigg[ 1 
           - \eta \bigg( \frac{D^{1,1}_{0}(\vec{r})}{\rho_{\text{ref}}} \bigg)^{\sigma}
    \Bigg]
    \nn \\
& \quad \times
  \sum_{q=n,p} 
  \Big[ \tilde{D}^{1,1 *}_q(\vec{r}) \,  \tilde{D}^{1,1}_q(\vec{r})
       +\tilde{C}^{1,1 *}_q(\vec{r}) \,  \tilde{C}^{1,1}_q(\vec{r})
  \Big] \, ,
\end{align}
where $D^{1,1}_q(\vec{r})$ and $\tilde{D}^{1,1}_q(\vec{r})$ 
are even under time reversal, whereas $\tilde{C}^{1,1}_q(\vec{r})$ 
is time-odd. The pair densities $\tilde{D}^{1,1}_q(\vec{r})$ and 
$\tilde{C}^{1,1}_q(\vec{r})$ are in general complex objects, as are other 
ingredients of HFB theory \cite{Ring80a}. However, with the symmetries 
imposed in our HFB solvers for finite nuclei and INM, relative phases 
can be chosen such that 
the normal and anomalous density matrices $\rho_{ij}$ and $\kappa_{ij}$,
as well as the matrix elements of the the single-particle Hamiltonian 
$h_{ij}$ and the pairing field $\Delta_{ij}$, are all real in the basis 
that diagonalises $h_{ij}$ and also in the canonical basis that 
diagonalises $\rho_{ij}$. These two bases are the ones that we will use to illustrate
the inner workings of our calculation. The occupation amplitudes $u_k$ 
and $v_k$ in the canonical basis are then real as well.
By construction, the pair densities $\tilde{D}^{1,1}_q(\vec{r})$ and 
$\tilde{C}^{1,1}_q(\vec{r})$ also become real in this case~\cite{Ryssens21a}.
We recall that they then represent the real and imaginary parts of the 
pair density $\tilde{\rho}_q(\vec{r}) = \tilde{D}^{1,1}_q(\vec{r}) 
+ \iunit \tilde{C}^{1,1}_q(\vec{r})$ in traditional notation that
becomes complex in time-reversal-breaking calculations. We nevertheless
keep the complex conjugations of objects entering the HFB equations 
throughout the paper to better separate the independent variational 
degrees of freedom.

In Eq.~\eqref{eq:EDF:pair:ULB}, the
term bilinear in $\tilde{D}^{1,1}_q(\vec{r})$ represents the 
time-even part of the EDF, whereas the term bilinear in 
$\tilde{C}^{1,1}_q(\vec{r})$ is the time-odd part, a distinction 
that cannot be made in traditional notation~\cite{Ryssens21a}. 
This pairing EDF represents the anomalous part of the HFB expectation value of
the density-dependent $S=0$, $T=1$ two-body contact pairing interaction
\begin{align}
\label{eq:force:pair:ULB}
& v_{\text{pair}} (\vec{r}_1 \sigma_1 \tau_1,     \vec{r}_2  \sigma_2 \tau_2, 
                   \vec{r}_1' \sigma_1' \tau_1' , \vec{r}_2' \sigma_2' \tau_2')
    \nn \\
& \quad 
  = V_0 \hat{\Pi}_{S=0}
  \Bigg[ 1 - \eta \bigg( \frac{D^{1,1}_{0} ( \vec{R}_{12})}{\rho_{\text{ref}}} \bigg)^{\sigma} \Bigg]
  \delta_{\vec{r}_1,\vec{r}_2} \,
  \delta_{\vec{r}_1,\vec{r}_1'} \,
  \delta_{\vec{r}_2,\vec{r}_2'} \, ,
\end{align}
where $\vec{R}_{12} = (\vec{r}_1 + \vec{r}_2)/2$ is the position of the 
centre-of-mass of the two nucleons at $\vec{r}_1$ and $\vec{r}_2$. The operator 
$\hat{\Pi}_{S=0} = \tfrac{1}{2} \, \big( 1 - \hat{P}^{\sigma}_{12} \big)$, 
where $\hat{P}^{\sigma}_{12}$ is the spin-exchange operator, projects on $S=0$ 
two-body states \cite{Ragnarsson95a}, which for a gradientless two-body contact 
force implies simultaneous projection on $T=1$. 

Besides the overall strength $V_0$, this pairing EDF has two 
additional parameters that affect the spatial dependence of the 
pairing potential in finite nuclei. 
For historical reasons, there is a redundant factor 
$\rho_{\text{ref}} = 0.16 \, \text{fm}^{-3}$ in the 
density dependence of Eq.~\eqref{eq:EDF:pair:ULB} that 
represents the saturation density of INM, such 
that $\eta$ parametrises the mixture between pure volume 
($\eta = 0$) and pure surface ($\eta = 1$) character of 
the pairing interaction, with the intermediate case 
$\eta = 1/2$ usually being called ``mixed pairing''.
By contrast, for $\eta > 0$, the parameter $\sigma$ controls the 
width of the peak of the pairing potential 
at the nuclear surface, see the illustrations in Sec.~\ref{sec:fit:fit}, 
where the effect of these two parameters on the density-dependence of the 
pairing gap in INM will be analysed.

Many authors use a generalisation of the EDF of Eq.~\eqref{eq:EDF:pair:ULB} 
for which different strength parameters $V_p$ and $V_n$ are used for 
protons and neutrons 
\cite{Dobaczewski15a,Duguet01n,Bender00a,Bertsch09a,Klupfel09a,Kortelainen14a}
that slightly improves the description of empirical data.
Although this choice is widely used in studies of finite nuclei,
it will not be considered here because using such an EDF for the study
of pairing correlations in INM leads to different pairing
gaps of protons and neutrons in matter with $\rho_p = \rho_n$
that therefore would not be symmetric. We therefore stick to a form
of the pairing EDF with a single strength $V_0$.

Some authors have instead added a dependence on the isovector density 
\cite{Chamel10a,Margueron08a,Bertulani09a,Yamagami12a,Zhang19a}
to the pairing EDF of Eq.~\eqref{eq:EDF:pair:ULB}. While this is 
clearly needed to simultaneously reproduce microscopic results for 
pairing gaps in symmetric and neutron matter 
\cite{Chamel10a,Zhang10a,Goriely16a,Grams23a,Grams24x,Grams26x}, 
it turns out that the available experimental information on pairing 
gaps across the chart of nuclei is not sufficient for the reliable
determination of a possible isovector density dependence of the pairing EDF
\cite{Yamagami12a}. For reasons that will become clear in the 
further discussion, for the present study we will stick to 
the simpler form of Eq.~\eqref{eq:EDF:pair:ULB}.

Before analysing the effect of the various parameters of the pairing
EDF \eqref{eq:EDF:pair:ULB} on the pairing gaps in INM, we recall
that for the observed odd-even staggering of masses of heavy nuclei 
there is no clear-cut difference in global performance between volume and 
surface character of this pairing EDF when the strength $V_0$ is adjusted 
within the same protocol \cite{Bertsch09a,Changizi15a}. 
By contrast, the evolution of moments of inertia in rotational bands 
slightly prefers a surface character \cite{Rigollet99a,Duguet01n}.
A phenomenological difference between the surface and volume character 
of the pairing EDF becomes also evident in pair transfer reactions 
\cite{Khan09a,Pllumbi11a}, but the comparison with data requires 
beyond-mean-field techniques, as discussed for example in 
Refs.~\cite{Gambacurta12a,Shimoyama13a}.

Unless explicitly stated otherwise, for all calculations discussed in what 
follows, the contribution of the pairing EDF to the mean-field potentials 
is kept and added to the terms obtained from the variation of the Skyrme EDF
\begin{align}
\label{eq:F11:paircontribution}
& F^{1,1}_{q}(\vec{r})
    \nn \\
& = \frac{\delta \mathcal{E}_{Sk}}{\delta D^{1,1}_{q}(\vec{r})}
    \nn \\
& \quad    
   - \frac{V_0}{4} \,
   \sum_{q} \Big[ \tilde{D}^{1,1 *}_q(\vec{r}) \, \tilde{D}^{1,1}_q(\vec{r}) 
                 +\tilde{C}^{1,1 *}_q(\vec{r}) \, \tilde{C}^{1,1}_q(\vec{r})
            \Big]  
    \nn \\
& \qquad    
   \times
     \frac{\eta \sigma}{\rho_{\text{ref}}^{\sigma}} \,     
     \big[ D^{1,1}_{0}(\vec{r}) \big]^{\sigma-1}
\, .
\end{align}
The term containing pair densities brings a small positive contribution 
to the single-particle potentials $F^{1,1}_{q}(\vec{r})$ 
that depends on the magnitude of pairing correlations as measured 
by the size of $\tilde{D}^{1,1}_q(\vec{r})$ and
$\tilde{C}^{1,1}_q(\vec{r})$. With pairing
correlations being in general stronger in
systems with an even proton or neutron number than an odd one, this introduces 
a small odd-even effect on the potentials' depth that through
self-consistency leads to a small odd-even effect of nuclear radii 
\cite{Fayans94a,Fayans96a,Fayans00a,Fayans01a,Saperstein11a,Borzov22a,Reinhard17a,Reinhard24a} 
that will be analysed in Sec.~\ref{sec:nuclei:stagger:r}. 
For homogeneous INM, the pairing contribution to $F^{1,1}_{q}$ only
yields a constant shift of the single-particle spectra $\varepsilon_k$ and
chemical potentials $\lambda_q$ without affecting any of the observables 
discussed in this paper, see \ref{sec:EDF:pp}.

The contact interaction \eqref{eq:force:pair:ULB} has an infinite
range and a non-decreasing strength
in momentum space and therefore couples bound nucleons to states
at arbitrarily large momentum. This leads to a divergence of the 
HFB equations with increasing number of active single-particle levels,
which can be avoided by either renormalising the pairing strength 
for a given model space, or by introducing a cutoff in energy
\cite{Bulgac02a,Duguet05a,Borycki06a}.

There are several possible strategies for introducing a cutoff. Their 
numerical efficiency depends on the numerical representation
and the algorithm chosen to solve the HFB equations.
For reasons of its simplicity when using the two-basis method in 
coordinate space as done here, we use a cutoff that acts 
in the basis that diagonalises the single-particle Hamiltonian $\hat{h}$
\cite{Rigollet99a,Ryssens15a,Bender00a,Bonche85a,Krieger90a}.
In this basis, the pairing energy and the matrix elements of
the pairing field are
\begin{align}
\label{eq:Epair:1}
E_{\text{pair}}
&=\tfrac{1}{2} \sum_{ijkl} v_{ijkl} \, \kappa^*_{ij}\, \kappa_{kl} \,,   \\
\label{eq:Delta:1}
\Delta_{ij}
&=\sum_{kl} v_{ijkl} \, \kappa_{kl} \,.
\end{align}
Using the cutoff factors $f_i$, these quantities can be regularized and
expressed in the form~\cite{Ryssens19a}
\begin{align}
\label{eq:Epair:cutoff:1}
E_{\text{pair}}
&=\tfrac{1}{2} \sum_{ijkl} v_{ijkl} \, f_i \, \kappa^*_{ij} f_j \, f_k \, \kappa_{kl} \, f_l \, ,
   \\
\label{eq:Delta:cutoff:1}
\Delta_{ij}
&= \tfrac{1}{2} \, f_i f_j \sum_{kl} v_{ijkl} \, f_k \, \kappa_{kl} \, f_l  \, .
\end{align}
The cutoff factors $f_i$ can be absorbed into the definition of the local pair densities
\cite{Ryssens21a}, see \ref{sec:EDF:pairdensities} for the special case
of INM calculations.

One possible functional form for the cutoff is a root of a Fermi function 
parametrised by a range $E_{\text{cut}}$ and the width of the interval
over which the cutoff falls to zero~\cite{Bonche85a,Krieger90a} that cuts 
above and below the chemical potential $\lambda_q$
\begin{align}
\label{eq:cutoff}
f_i(\varepsilon_i)
& = \big[1 + e^{( \varepsilon_i - \lambda_q - E_{\text{cut}})/\mu} \big]^{-1/4}
    \nn \\
& \quad \times 
    \big[1 + e^{(-\varepsilon_i + \lambda_q - E_{\text{cut}})/\mu} \big]^{-1/4}
  \, .
\end{align}
The same cutoff is used for calculations of INM and finite nuclei. 

A widely-used reference parametrisation of the pairing 
EDF \eqref{eq:EDF:pair:ULB} with the cutoff~\eqref{eq:cutoff}
and originally adjusted in Ref.~\cite{Rigollet99a} to describe the 
evolution of the moment of inertia of the yrast superdeformed 
band of \nuc{194}{Hg} in calculations with the SLy4
parametrisation of the NLO Skyrme EDF, is given by 
$V_0 = -1250~\text{MeV} \, \text{fm}^{3}$,
$\eta = \sigma = 1$, 
$\rho_{\text{ref}} = 0.16~\text{fm}^{-3}$,
$E_{\text{cut}} = 5~\text{MeV}$,
$\mu = 0.5~\text{MeV}$. This choice of parameters has also
been very successfully used together with other parametrisations
of the Skyrme EDF with same effective mass that were adjusted with 
slight variations of the same fit protocol \cite{DaCosta24a,Jodon16a}.
For the rest of this paper, we will call this specific
parametrisation of the EDF~\eqref{eq:EDF:pair:ULB}
``ULB pairing'' as already done in Ref.~\cite{Duguet05a}.

To avoid the occasional breakdown of pairing correlations in finite nuclei, 
the original ULB pairing was adjusted within a Lipkin-Nogami 
(LN) scheme~\cite{Rigollet99a}. The LN scheme has, however,
the disadvantage that it is non-variational and therefore might 
not converge to the energetically lowest configuration. For
the calculations of finite nuclei reported here we use instead
the stabilisation of the pairing EDF as proposed in Ref.~\cite{Erler08a} 
\begin{align}
\label{eq:Epair:stab:1}
E_{\text{pair}}^{\text{stab}}
& = E_{\text{pair}} \, \Bigg( 1 + \frac{E_{\text{cut},b}^2}{E_{\text{pair}}^2} \Bigg) 
\end{align}
with $E_{\text{pair}}$ given by Eq.~\eqref{eq:Epair:cutoff:1} and
$E_{\text{cut},b} = 0.3~\text{MeV}$ for the cutoff parameter.
Neither the LN scheme nor the stabilisation of Ref.~\cite{Erler08a} 
can however be meaningfully used for INM calculations.

There is no observable that directly allows for quantifying the 
amount of pairing correlations present in a given system. Some 
quantities that can nonetheless be used to quantify pairing from the change of some observable
between two different states are discussed in Sec.~\ref{sec:nuclei:stagger}.
The amount of pairing correlations predicted by the HFB calculation
can be quantified with weighted averages of the matrix elements $\Delta_{\bar{\jmath} i}$.
One possibility is using $\kappa_{i\bar{\jmath}}$ as the weight factor
\cite{Bender00a,Erler08a,Sauvage81a}, which in the HF basis of a fully
paired time-reversal invariant quasiparticle vacuum yields
\begin{equation}
\label{eq:uvgap:gen}
\langle uv \Delta \rangle_q
\equiv \frac{\sum_{i,j > 0} \, \kappa_{i\bar{\jmath}} \, f_i f_{j} \, \Delta_{\bar{\jmath} i} }
            {\sum_{i,j > 0} \, \kappa_{i\bar{\jmath}} \, f_i f_{j} }
\, .
\end{equation}
In the canonical basis 
$\kappa_{i\bar{\jmath}} = v_i u_j \, \delta_{\bar{\jmath} i}$,
which explains the symbol used. When the cutoff is defined in the HF basis 
as done here, however, there is no advantage of summing in the canonical basis.
Note that only one matrix element contributes per pair, otherwise the 
denominator would be zero since
$\kappa_{i\bar{\jmath}} = - \kappa_{\bar\jmath i}$.
Both the numerator and the denominator 
should include the cutoff factor as done in \cite{Erler08a} in order to limit 
the sum to levels that can contribute to pairing correlations, which is not 
always done. For the pairing EDF of Eq.~\eqref{eq:EDF:pair:ULB} with cutoffs
\eqref{eq:Epair:cutoff:1} this gap can also be expressed with pair densities
that absorb the cutoff factors \eqref{eq:Dp11cut}
\begin{equation}
\label{eq:uvgap:int}
\langle uv \Delta \rangle_q
= \frac{\int \! \rmd^3r \, \tilde{D}_q(\vec{r}) \, \tilde{F}^{1,1}_q(\vec{r})} 
       {\int \! \rmd^3r \, \tilde{D}_q(\vec{r})}
\, ,
\end{equation}
where 
$\tilde{F}^{1,1}_q(\vec{r}) \equiv 
\delta E_{\text{pair}} / 
\delta \tilde{D}^{1,1 *}_q(\vec{r})$ 
is the local pair potential from 
which the pairing gaps $\Delta_{\bar{\jmath}i}$ are calculated, see 
Eq.~\eqref{eq:Fp11} for INM and Ref.~\cite{Ryssens21a} for finite nuclei.
In this expression, the additional relative phases of the single-particle states 
remove all ambiguities about the single-particle states that have to be summed over.

%
\subsection{HFB calculations of finite nuclei}
\label{sec:finite:nuclei}

Unless specified otherwise, the calculations for finite nuclei reported 
here are performed with the \textsf{MOCCa} code~\cite{RyssensThesis,Mocca} that 
solves the HFB equations with the two-basis method as described in Ref.~\cite{Ryssens19a} 
in a 3d coor\-di\-nate-space representation. This code uses Lagrange-mesh techniques that
provide a very high numerical accuracy for derivatives and integrals even when
working with a coarse discretisation \cite{Ryssens15b}. 
As often done, we impose a point-group symmetry on the single-particle states 
that are chosen to be eigenstates of the signature $\hat{R}_z$,
parity $\hat{P}$ and time-simplex $\check{S}_y^T$, which introduces three plane
symmetries for the densities, as discussed in Appendix~D of Ref.~\cite{Ryssens21a}.
Ground states of even-even nuclei are calculated imposing in addition
time-reversal invariance, whereas the calculations of blocked one-quasiparticle
(1qp) configurations of odd nuclei and rotational states of even-even nuclei
are calculated allowing for time-reversal-breaking configurations.

The 1qp configurations of odd nuclei are constructed through
self-consistently blocking the quasiparticle that has the largest overlap 
with a given eigenstate of the single-particle Hamiltonian. We limit 
ourselves here to nuclei for which a fully paired non-blocked calculation 
of the ``false vacuum'' with odd neutron number yields either a spherical or 
axially-symmetric configuration such that the blocked quasiparticles
have a half-integer expectation value of $\langle \hat{\jmath}_z \rangle = K$
in the false vacuum. This is the angular-momentum quantum number that we
attribute also to the blocked many-body configuration.

\begin{figure*}[t!]
\centerline{
\includegraphics[width=5.45cm]{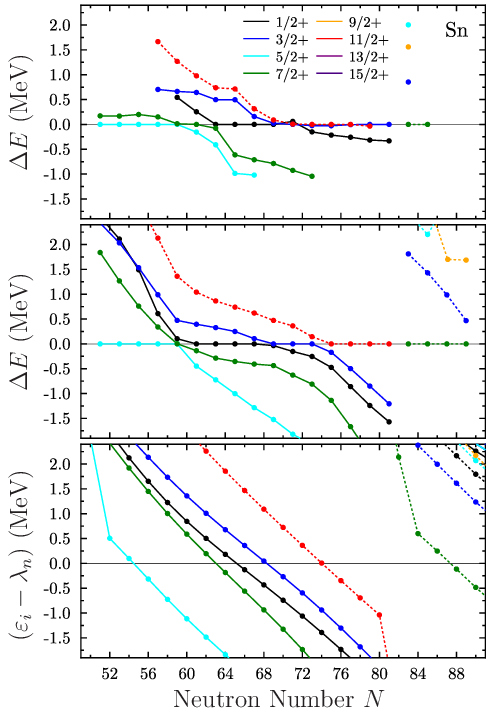}
\includegraphics[width=5.45cm]{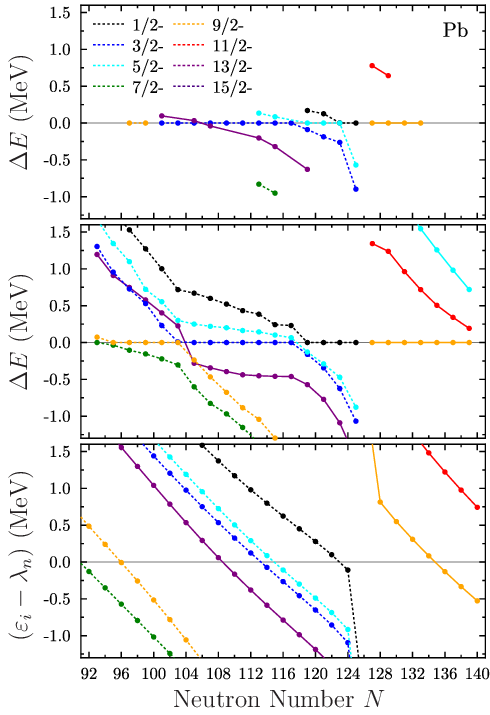}
\includegraphics[width=5.45cm]{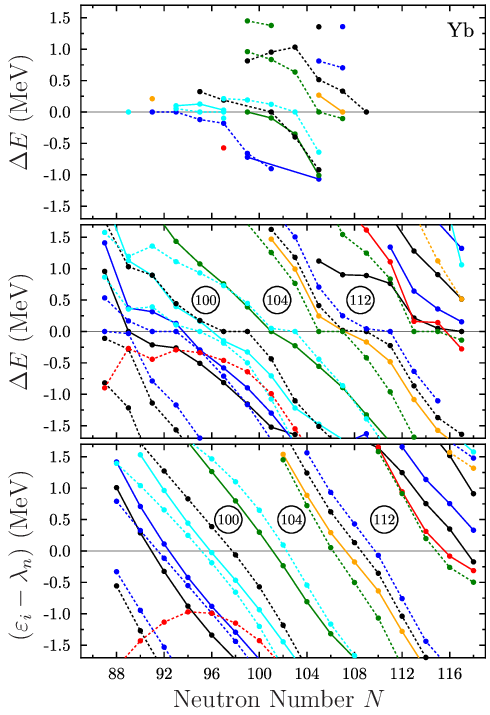}
}
\caption{\label{fig:gbandheads:1t2t80:ULB}
Total energies $\Delta E$ of 1qp configurations in odd-mass Sn, Pb, and Yb isotopes 
relative to the ground state extracted from \cite{NuDat,ESNDF} (upper row) compared
with calculations with 1T2T(0.80) (also known as SLy7*) (middle row) for states with 
$K = \langle \hat{J}_z \rangle$ and parity as indicated.
Following the convention of Ref.~\cite{Dobaczewski15a}, the sign indicates 
if the dominant single-particle component of the blocked qp is above (positive) 
or below (negative) the chemical potential (see text). Solid (dotted) lines 
indicate configurations of positive (negative) parity, whereas colour 
indicates angular momentum.
The lower row displays the difference between the eigenvalues of the 
single-particle Hamiltonian $\varepsilon_i$ and the chemical potential 
of neutrons $\lambda_n$ in the even-even isotopes in between. For 
Yb isotopes, several deformed neutron shell closures predicted by 
1T2T(0.80) are indicated.
}
\end{figure*}

Because of rearrangement and polarisation effects in self-consistent 
HFB calculations, 
however, at convergence of a blocked HFB calculation with a 3d code 
the many-body expectation value of total angular momentum 
$\langle \hat{J}_z \rangle$ will in general slightly differ from 
the single-particle expectation value of the angular momentum 
$\langle \hat{\jmath}_z \rangle$ of the blocked neutron in the canonical 
basis, and both will in general differ from the value of $K$ in 
the false vacuum used to initialise the calculation. This particularly 
becomes an issue when there are near-degenerate (or energetically 
lower) quasiparticles with same signature and parity that can mix 
with the targeted one in a 3d calculation. To stabilise the 
convergence of the blocked HFB calculations towards the targeted 
configuration and to facilitate comparison with experiment, we impose
the condition $\langle \hat{J}_z \rangle = K$ with a principal-axis 
cranking constraint that adjusts the rotational frequency $\omega_z$ to 
the targeted value of $\langle \hat{J}_z \rangle$.

For the well-deformed Yb isotopes that can be assumed to satisfy the 
strong-coupling limit where $K = J$ \cite{Ragnarsson95a,Rowe70a}, 
the nucleus is oriented in the numerical box such that its long axis 
is aligned with the $z$ axis. Only the $\langle \hat{\jmath}_z \rangle > 0$ 
level from each Kramers-degenerate pair with single-particle energy near 
the chemical potential of neutrons in the false vacuum is blocked;
blocking its partner state would converge to the time-reversed many-body 
configuration that has the same energy, but an angular momentum that points 
into the opposite direction.

For the spherical Sn and Pb isotopes, only the state with projection 
$\langle \hat{\jmath}_z \rangle = j$ out of a given $j$-shell is blocked.
It is to be noted that blocking a different sub-state of a given spherical $j$ shell
with half-integer $|K| < j$ in the false vacuum with the same procedure 
yields in general a slightly different total energy at convergence. This is 
because of the broken alispin-symmetry (see Ref.~\cite{Schunck10a} for
its definition) in a code that imposes point-group symmetries as done here. 
As a consequence, blocking different magnetic sub-states of a given $j$-shell 
does not generate configurations whose current and spin density distributions are 
simply related by a global rotation.

Some calculations of even-even Sn and Pb isotopes are performed 
with a spherical code \cite{pelops} that uses the same numerical representation 
as \textsf{MOCCa}, i.e.\ it also solves the HFB equations with the two-basis method 
in a Lagrange-mesh coordinate-space representation, but in this case for a 
radial mesh. For spherical nuclei this code and \textsf{MOCCa} yield almost 
identical results within a few keV for binding energies.

For the purpose of this study we also have set up a new tool to calculate 
paired isotropic homogeneous infinite matter that will be described in 
Sec.~\ref{sec:paired:inm} and Appendices A-D.

%
%
\subsection{The origin of ambiguities when adjusting the parameters of the pairing EDF to finite nuclei}
\label{sec:ambiguity}

Whatever the chosen form of the pairing EDF, to be predictive, its 
parameters have to be adjusted to some selected observables. 
For a given nucleus, pairing correlations depend very sensitively on details 
of the single-particle spectrum around the chemical potential.
It is well-known, however, that even the best-per\-form\-ing pa\-ra\-me\-tri\-sa\-tions
of any form of nuclear EDF have deficiencies with respect to predicting
all details of shell structure 
\cite{Bender03a,Bonneau07a,Lesinski07a,Tarpanov14a,Dobaczewski15a,Bender09t}.
This is illustrated by Fig.~\ref{fig:gbandheads:1t2t80:ULB} 
for the Skyrme NLO parameter set 1T2T(0.80) (also known as SLy7*). The calculation is
done with \LOPP{}{ULB} pairing parameters adjusted as explained in Sec.~\ref{sec:fit:fit}.

The upper two rows of panels of the figure show results obtained for low-lying 
states of odd-mass isotopes in the Sn, Pb, and Yb chains that
can be interpreted as being dominantly non-collective one-quasi-particle 
configurations. The upper row displays experimental data for which 
such configuration is indicated in the ESNDF database \cite{ESNDF}, 
whereas the middle row displays total energies from fully self-consistent 
calculations of blocked 1qp configurations as described in 
Sec.~\ref{sec:finite:nuclei}.

In both cases, the total energies of these 1qp configurations are plotted 
relative to the respective ground state of each nucleus, where for the 
calculated 1qp states the sign of the energy indicates if in the 
respective false vacuum the blocked single-particle state is above 
(positive energy) or below (negative energy) the chemical potential. 
For the experimental data the sign is decided from the assumption that 
the energies decrease with increasing $N$. This might not be the case for 
all levels, a notable exception being the $11/2^-$ in the light Yb isotopes
that is pushed up from below the chemical potential by a change in deformation.
Such convention~\cite{Dobaczewski15a} for plotting the energies of the 
1qp configurations has the advantage that it can be read in the same way 
as a spectrum of single-particle energies as a function of $N$.

The definition of single-particle energies is in general model dependent. 
In what follows, we use this notion for the eigenvalues $\varepsilon_i$ 
of the single-particle Hamiltonian $\hat{h}$.
The lower panels of Fig.~\ref{fig:gbandheads:1t2t80:ULB} show for each 
isotopic chain the spectrum of the neutron single-particle energies
$\varepsilon_i$ from which, for easier comparison with the upper panels, 
the chemical potential $\lambda_q$ is subtracted.

When being calculated with 1T2T(0.80), over the range of isotopes shown, 
the HFB ground states of even-even Sn and Pb isotopes take a spherical shape, 
$\beta_{20} = 0.0$, whereas the Yb take an axial prolate shape. In all cases,
this is consistent with the available spectroscopic information about these 
isotopic chains. The calculated ground-state quadrupole deformation $\beta_{20}$ of 
the even-even Yb isotopes slowly increases from 0.14 for $N=86$ 
to 0.32 for $N=94$ and then remains nearly constant up to $N=106$, beyond 
which $\beta_{20}$ slightly drops to values around 0.28. The available data
\cite{NuDat,ESNDF} for $88 \leq N \leq 106$ have a very similar trend. 
In the calculations, the light Yb isotopes up 
to about $N=92$ are quite $\gamma$-soft, however, such that a single 
1qp configuration might not capture well the structure of the band-heads
of these nuclei. Because of the blocked quasiparticle and the 
polarisation it induces, none of odd-mass Sn and Pb isotopes can remain
strictly spherical, but in all cases discussed here the deformation of the
self-consistent 1qp configurations remains close to the one of the adjacent 
even-even nuclei.

The blocked quasiparticle configuration of lowest total energy does not 
necessarily correspond to the single-particle level $\varepsilon_i$ 
closest to the chemical potential $\lambda_q$. When there are 
near-degenerate configurations, the lowest one is determined 
by the balance between dis\-tance of the dominant single-particle 
level from $\lambda_q$, the size of the dominant matrix 
elements of $\Delta_{i j}$, and the size of polarisation 
effects induced by the blocked quasiparticle. Also, because
of these polarisation and self-con\-sis\-ten\-cy effects when 
minimising the total energy, each 1qp configuration in a given nucleus
adopts a slight\-ly different deformation. This effect tends to compress 
the spectrum obtained from total energies that in the end differs
significantly from the spectrum of eigenvalues $E_k$ of the HFB 
Hamiltonian $\mathcal{H}$ in the false vacuum, that in turn also differs
from the spectrum of single-par\-ti\-cle energies $\varepsilon_i$.

When comparing the calculated spectra of total energy differences between 
different states of odd nuclei with experimental data, it is evident that
there are significant deviations concerning the relative distance of levels.
These deviations between the total energies can be attributed to systematic 
deficiencies of the single-parti\-cle spectra. For the Sn chain, it can be 
deduced that the $1h_{11/2^-}$ intruder level should be much closer to the
nor\-mal-parity levels, and also that the relative distances between the 
normal-parity levels should be slightly different, in particular the 
$2d_{5/2^+}$ should be higher up. Similar discrepancies, although somewhat 
less dramatic, are found found for Pb isotopes. What is common for the Sn
and Pb chains is that the calculated spectrum is much to spread out 
for isotopes just above the spherical shell closures, i.e.\ $N=50$ and 82
for Sn and $N=126$ for Pb. In fact, for Pb isotopes with $N > 126$, the 
single-particle energies $\varepsilon_i - \lambda_n$ of the $3d_{5/2^+}$ and 
$1j_{15/2^-}$ shells above the $N=126$ gap remain outside the energy interval 
shown, taking the values of 1.7 and 1.6 MeV, respectively for the Pb 
isotope with $N=140$. 

For the deformed Yb isotopes, 
for which the 1qp configurations constitute the band heads of 
low-lying rotational bands, only a small subset of the possible bands has been 
identified for most odd-mass isotopes, which complicates the comparison
with theory. The somewhat erratic behaviour of the relative distance between 
the calculated total energies of the lightest isotopes is a consequence 
of their rapidly changing deformation, from about $\beta_2 \simeq 0.17$ 
for $N=87$ to $\beta_2 \simeq 0.31$ for $N=93$.
The calculations with 1T2T(0.80) predict several quite
pronounced deformed neutron shell closures at $N=100$, 104 and 112 that are
indicated in Fig.~\ref{fig:gbandheads:1t2t80:ULB}.
Experimental data for two-neutron separation energies rather point towards
deformed shell effects at $N=98$ and $N=104$ instead. The sparse data for 
the heavier Yb isotopes also point towards some structural change at about 
$N \simeq 108$.

For the majority of odd-mass isotopes displayed on Fig.~\ref{fig:gbandheads:1t2t80:ULB}
neither the angular momentum of the ground-state nor the bunching of low-lying
1qp states is always well described, which is a common feature of all
parametrisations of all types of nuclear EDF models. Although these often give
a fair description of global features of shell structure and predict the
configuration with correct $J$ to be in general one of the low-lying 
states \cite{Bonneau07a}, the many deviations in detail can have consequences 
when the fit protocol of the pairing interaction relies on the
odd-even mass staggering of masses or average pairing matrix elements.
First, the matrix elements of $\Delta_{i j}$ depend on the spatial
wave function, so blocking the incorrect configuration implies that the 
matrix elements that make the largest difference have not the correct size.
In addition, levels with incorrect distance from the chemical potential 
will in general not have the correct occupation numbers.
The parameters of the pairing EDF will partially absorb these 
imperfections when adjusting them to properties of finite nuclei, which
then leads to problems when extrapolating the such adjusted pairing EDF to 
other nuclei or to other phenomena.

We found similar deviations from data for SLy4, 1T2T(0.70), 1T2T(0.85), and SN2LO1, 
and on general grounds this can be expected for virtually all existing
parametrisations of the nuclear EDF, but there can be quite large differences 
in performance for a given level in a given nucleus.

%
\section{HFB study of paired homogeneous infinite nuclear matter}
\label{sec:paired:inm}

As a robust alternative to calculations of finite nuclei, there also
is the possibility to adjust the parameters of the pairing EDF to
properties of the model system of homogeneous infinite nuclear matter.
%
%
\subsection{Historical remarks}

After first perturbative calculations of the pairing properties of INM
\cite{Henley64a,Kennedy64a}, to the best of our knowledge, the first full 
self-consistent solution of the HFB equation for INM taking into account 
the feedback of pairing correlations onto the mean fields was 
reported in Ref.~\cite{Kennedy68a}. Solving the simpler BCS pairing problem 
of INM for a fixed single-particle spectrum nevertheless remains common
practice. Over the years, there have been numerous BCS and HFB studies 
using bare NN interactions such as 
Refs.~\cite{Dean03a,Baldo90a,Chen93a,Khodel96a,Elgaroy98a,Yin23a,Hebeler10a},
which, however, in the context of a modelling from first principles
is only the first step towards more complete microscopic 
calculations of paired INM that also take into account other correlations
\cite{Cao06a,Rios17a,Sedrakian19a,Gandolfi08a,Gandolfi09a,Gandolfi22a}.

The pairing properties of INM have also been studied for the widely-used
effective non-relativistic EDFs of Gogny \cite{Kucharek89a,Davesne25a}, 
M3Y \cite{Davesne25a}, and Skyrme \cite{Takahara94a,Aguirre12a} type,
as well as for relativistic mean-field models \cite{Kucharek91a,Serra01a}.
There also are numerous studies of the pairing properties in INM obtained
with specifically tailored phenomenological pairing interactions
\cite{Duguet08a,Hebeler09a,Lesinski09a,Garrido99a,Duguet04a,Chamel10a}
that use one of these EDFs as the ph interaction to determine the 
underlying mean fields of the nucleons, which is also what will be done here.

%
%
\subsection{HFB calculations of INM}

Our study is limited to homogeneous non-polarised INM, which is completely 
characterised by two pa\-ra\-me\-ters. These are the total density $\rho$ and the 
isospin asymmetry $I$, which are related to proton and neutron densities through
\begin{align}
\label{eq:rho}
\rho
= \rho_n + \rho_p 
= D^{1,1}_n + D^{1,1}_p \, ,
  \\
\label{eq:I}
I
= \frac{\rho_n - \rho_p}
       {\rho_n + \rho_p}
= \frac{D^{1,1}_n - D^{1,1}_p}
       {D^{1,1}_n + D^{1,1}_p} \, ,
\end{align}
or
\begin{align}
\rho_n
& = D^{1,1}_n
  = \tfrac{1}{2} \big( 1 + I \big) \, \rho \, ,
    \\
\rho_p
& = D^{1,1}_p
  = \tfrac{1}{2} \big( 1 - I \big) \, \rho \, .
\end{align}
Throughout this paper, we make the distinction where the densities 
$\rho_p$,  $\rho_n$, and $\rho$ in traditional notation represent the targeted 
parameters of INM, whereas the local densities $D^{A,B}_q$ in the notation 
of Ref.~\cite{Ryssens21a} are those that enter the Skyrme EDF and that 
are numerically calculated.
With these symmetries, the HFB equations~\cite{Ring80a,Bender19a,Blaizot86a}
\begin{equation}
\label{eq:HFB:equation}
\left( \begin{array}{cc}
h - \lambda & \Delta \\
-\Delta^* & -h^*  + \lambda
\end{array} \right)
\left( \begin{array}{c}
U \\ 
V
\end{array} \right)
= E \, 
\left( \begin{array}{c}
U \\ 
V
\end{array} \right) 
\end{equation}
can be mapped onto a set of pairwise coupled equations for given mean fields
$h$ and $\Delta$ and momenta, which enormously simplifies the numerical solution
of the problem compared to the treatment of finite nuclei. 
Contrary to the case of finite nuclei, in INM calculations the chemical potential
$\lambda_q$ that enters the HFB equation~\eqref{eq:HFB:equation} as a Lagrange parameter 
is used to adjust the density $\rho_q$, and not an average particle number.
The necessary steps to solve the HFB equations for $^{1}S_{0}$ 
pairing correlations in the context of Skyrme EDFs are detailed in the Appendices.
We outline here just the specificities of some of the objects that are relevant 
for the further discussion.

In a plane-wave basis the single-particle Hamiltonian $h$ of INM that enters
Eq.~\eqref{eq:HFB:equation} is diagonal, and for a N2LO Skyrme EDF its 
eigenvalues $\varepsilon_{\vec{k} \sigma q}$ that we call single-particle 
energies are given by
\begin{align}
\label{eq:eps}
\varepsilon_{\vec{k} \sigma q}
& = \bigg( \frac{\hbar^2}{2m} + F^{(\nabla,\nabla)}_q \bigg) \vec{k}^2
    + F^{1,1}_q 
     \nn \\
& \quad 
    + \sum_{\mu} F^{\nabla,\nabla}_{q,\mu\mu} \, k_{\mu}^2
    + F^{\Delta,\Delta}_q \, \vec{k}^4 \, ,
\end{align}
see \ref{sec:sph} for details. In the presence of N2LO terms, this expression cannot
be rewritten as the sum of a kinetic term with $k$-dependent effective mass and 
a $k$-independent potential, see \ref{sec:mstar:N2LO}.

As an alternative to $\rho_p$ and $\rho_n$, INM can also be characterised in terms 
of the Fermi momenta\footnote{Strictly speaking, $k_{\text{F},q}$
is a wave vector and not a momentum that would be $p_{\text{F},q} = \hbar k_{\text{F},q}$, 
but we will follow here the usual practice in the 
low-energy nuclear physics literature.} 
$k_{\text{F},q}$ of protons and neutrons in a non-paired Fermi gas that has the 
same density $\rho$ and asymmetry $I$ as the paired INM. Because of the well-known 
relation 
\begin{equation}
\label{eq:kF}
k_{\text{F},q}
= \big( 3 \pi^2 \, \rho_q \big)^{1/3} \, ,
\end{equation}
for a non-interacting Fermi gas \cite{Gross91a,Maruhn10a} that also 
applies to the HF solution of INM, 
$k_{\text{F},q}$ can be used as a proxy for the density $\rho_q$. At the 
corresponding Fermi energy $\varepsilon_{\text{F},q}$, the occupation 
probabilities of an interacting Fermi gas sharply fall from one to zero.

By contrast, in the paired phase obtained from a HFB calculation, the occupation 
numbers $v_{k,q}^2$ have a smooth distribution that falls off from one to zero 
over a range of single-particle energies characterised by the gap $\Delta_q$, 
with the occupation number taking the value 1/2 at the chemical potential 
$\lambda_q$. For the discussion of our results, we found it useful to define 
the momentum $k_{\lambda,q}$ for which the single-particle energy defined 
through Eq.~\eqref{eq:eps} equals the chemical potential
\begin{align}
\label{eq:klambda}
\varepsilon_{k_{\lambda,q} \sigma q}
& = \lambda_q \, ,
\end{align}
such that $v_{k_{\lambda,q},q}^2 = u_{k_{\lambda,q},q}^2 = 1/2$.
Not all authors seem to make such distinction, though. To avoid any confusion, 
we propose to use different symbols for these two quantities.

We note that the definition of $k_{\lambda,q}$ through
Eq.~\eqref{eq:klambda} also applies to the case 
of solutions with $\lambda_q < F^{1,1}_q$ that for some parametrisations of 
the effective $T=1$ pairing EDF can be found at very low density 
$\rho$~\cite{Guillon24m}.
An example will be discussed in Sec.~\ref{sec:INM:SLy4+ULB}.

Other characteristic properties of paired INM are the values of the
matrix elements $\Delta_k$ defined in Eq.~\eqref{eq:Deltamat} for the nucleon species $q$ at either
$k_{\text{F},q}$ or $k_{\lambda,q}$ that will be denoted as
$\Delta_q(k_{\text{F},q})$ and $\Delta_q(k_{\lambda,q})$, respectively.

We recall that the reason why INM is usually discussed in terms of characteristic 
momenta instead of characteristic energies (such as the chemical potential) is 
that, for INM, momentum is a quantum number of the plane wave states 
that is independent on the specific solution of the HFB calculation. This has 
to be contrasted with $\varepsilon_{\text{F},q}$ and $\lambda_q$ that are 
multivariate functions of $\rho$ and $I$ whose precise form is determined by 
the parametrisation of the EDF.

It is to be noted that in the literature on paired condensed matter 
$\varepsilon_{\text{F},q}$ usually represents the Fermi energy of a 
\textit{non-inter\-acting} Fermi gas,\footnote{The same convention is 
also sometimes used in nuclear physics papers, examples being
Refs.~\cite{Fayans94a,Fayans96a,Fayans00a,Fayans01a,Saperstein11a,Bulgac02a}.
} 
$\varepsilon_{\text{F},q} 
= \hbar^2 k_{\text{F},q}^2/(2m) 
= \hbar^2 (3 \pi \rho_q)^{2/3}/(2m)$ \cite{Strinati18a}.
For the discussion of nuclear matter calculated in the HFB approach, this is 
not very practical as nucleons then adopt an effective mass that is usually 
significantly different from the bare mass, and also are subject to the mean fields. 
In the nuclear physics literature, one sometimes finds a hybrid convention where the effective 
mass at the Fermi momentum is used to calculate the single-particle spectrum and 
therewith $\varepsilon_{\text{F},q}$, but all other terms in the single-particle 
Hamiltonian (including a possible $k$ dependence of the effective mass and 
single-particle potential) are ne\-glect\-ed. For Skyrme NLO EDFs doing so just 
leads to a global constant shift of the single-particle energies in INM. But for 
higher-order Skyrme EDFs and EDFs with finite-range interactions for which the
effective mass and the effective potential are $k$ dependent, the 
single-particle spectrum in such reference Fermi gas is not the same as the 
one obtained in a full HF calculation. 
First, the change of the effective mass with $k$ is missing, and second, the 
effective mass does in general not absorb the entire $k$-dependence of the
single-particle Hamiltonian. This is sketched for N2LO Skyrme EDFs in~\ref{sec:mstar:N2LO}.

%
\subsection{Pairing correlations from SLy4 + ULB}
\label{sec:INM:SLy4+ULB}

\begin{figure}[t!]
\centerline{
 \includegraphics[width=7cm]{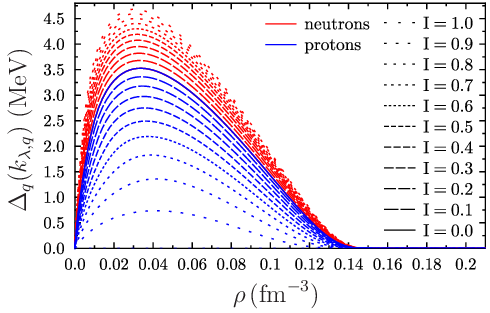}
 }
\caption{\label{fig:Delta:rho:SLy4+ULB} 
Gap $\Delta_q(k_{\lambda,q})$ of protons and neutrons at the respective
$k_{\lambda,q}$ in paired INM calculated with SLy4+ULB at asymmetries 
between $I = 0$ (symmetric matter) and $I=1$ (neutron matter) 
in steps of 0.1 as a function of the total density $\rho$.
}
\end{figure}

\begin{figure}[t!]
\centerline{
 \includegraphics[width=7cm]{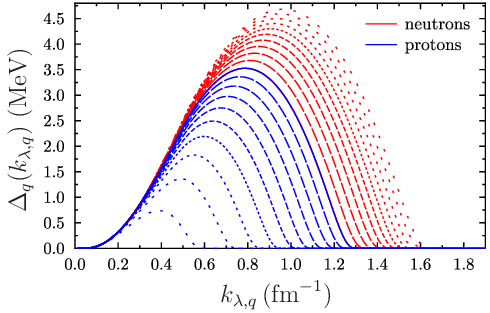}
 }
\caption{\label{fig:Delta:kl:SLy4+ULB} 
Same as Fig.~\ref{fig:Delta:rho:SLy4+ULB}, but as a function 
of the $k_{\lambda,q}$ of the respective nucleon species.
}
\end{figure}

In this subsection, we illustrate some properties of pairing correlations 
in INM obtained with the SLy4+ULB parametrisation as defined in 
Sec.~\ref{sec:edf:pair}. Figures~\ref{fig:Delta:rho:SLy4+ULB}  
and~\ref{fig:Delta:kl:SLy4+ULB} display the pairing gap $\Delta_q(k_{\lambda,q})$ 
of protons and neutrons at the respective $k_{\lambda,q}$ 
as a function of total density for asymmetries 
ranging from $I = 0$ (symmetric matter) to $I=1$ (neutron matter), either as a 
function of $\rho$ or as a function of $k_{\lambda,q}$. 
Plotting these as either a function of $\rho$ or of $k_{\lambda,q}$ 
highlights different aspects of the curves.

For this parametrisation of the EDF, pairing correlations are only found at
sub-saturation density, which is expected for a parametrisation of the 
pairing EDF that by construction is of pure surface character and therefore 
necessarily vanishes at $\rho_{\text{ref}} = 0.16 \, \text{fm}^{-3}$. 
That the pairing becomes weak around saturation density is however a common 
feature of all more microscopic calculations of the pairing gap in INM
\cite{Brink05a,Dean03a,Baldo90a,Chen93a,Khodel96a,Elgaroy98a,Yin23a,Hebeler10a,Cao06a,Rios17a,Sedrakian19a,Gandolfi08a,Gandolfi09a,Gandolfi22a}
and was the original motivation to construct effective pairing interactions 
of surface character \cite{Baldo04a,Fayans94a,Fayans96a,Fayans00a,Fayans01a}.

%
For symmetric matter ($I=0$), the gaps of protons and neutrons are equal, 
whereas the proton gap is trivially identical to zero in neutron matter ($I=1$).
For this parametrisation, the neutron pairing becomes stronger in neutron-rich
matter, whereas proton pairing becomes weaker.

There is no direct empirical information on the pairing gap in INM
at any density. BCS and HFB calculations with bare two-nucleon 
interactions give all very similar predictions for the density dependence of 
the $^{1}S_{0}$ pairing gap 
\cite{Dean03a,Baldo90a,Chen93a,Khodel96a,Elgaroy98a,Yin23a} 
with a maximum value of about 3 MeV for both symmetric and 
pure neutron matter. There are only small differences in detail concerning
the absolute and relative size. Adding N2LO three-body forces to
N3LO two-body potentials from chiral effective field theory 
reduces the BCS gap in neutron matter by a few 100 keV \cite{Hebeler10a}.
In more refined microscopic calculations based on bare interactions,
the pairing gap in INM is however significantly modified by correlations. 
But unfortunately, the different methods for the many-body treatment -- that 
often are also used with a different Hamiltonian -- each 
lead to significantly differing predictions 
\cite{Cao06a,Rios17a,Sedrakian19a,Gandolfi08a,Gandolfi09a,Gandolfi22a}.
Almost all of these calculations only address neutron matter, and 
find that the position of the peak of the gap is shifted
to values between $k = 0.6$ and 0.9 fm$^{-1}$ with a 
maximum value between 1.0 and 2.5 MeV.

To have an effective description of the fully correlated system through 
a simple HFB calculation, the impact of correlations on the pairing gap 
has to be mimicked by the EDF. When aiming at the efficient EDF description of 
homogeneous INM, then one should of course compare with the gap obtained 
from the most complete microscopic calculation, which is what is done 
in Refs.~\cite{Chamel10a,Zhang10a,Goriely16a,Grams23a,Grams24x,Grams26x}.
However, if the aim is to use INM as a proxy for pairing correlations 
that reproduce experimental results for finite nuclei within an HFB calculation 
-- which is what we aim at in this study -- then the gap from a microscopic 
INM calculation might not be a useful reference. 

Indeed, as mentioned in the introduction, in finite nuclei there 
is an induced pairing interaction from the coupling of nucleon pairs
to surface modes, whereas in INM there is an induced pairing interaction 
from the coupling to bulk modes. In an effective pairing EDF for 
predictive HFB calculations of both, the effect of these correlations 
has to be modeled separately through specific volume ans surface terms,
where the latter have to contain derivatives of some density in order to
be inactive in INM. This, however, necessitates that the pairing EDF is 
at least of NLO in gradients, and also requires the adjustment of at least some 
of the coupling constants to information on finite nuclei as done in
Refs.~\cite{Goriely16a,Grams23a,Grams24x,Grams26x} -- which for time
being we want to avoid for the reasons outlined in
Sec.~\ref{sec:ambiguity}.

These complications can be avoided when accepting that the 
effective pairing EDF will not be realistic for INM. Indeed, as we 
illustrate in Sec.~\ref{sec:paired:nuclei}, when being combined with
a wide range of mean-field parametrisations, a simple gradientless 
LO pairing EDF adjusted to yield the same HFB gaps in symmetric INM 
as SLy4+ULB gives a very satisfying description of pairing correlations 
in heavy nuclei.

In spite of the difficulty to identify a reliable reference, 
the significant increase of the HFB gap from SLy4+ULB  when 
going from symmetric to neutron matter found in 
Figs.~\ref{fig:Delta:rho:SLy4+ULB} and~\ref{fig:Delta:kl:SLy4+ULB} 
is probably not very realistic. This possibly points to the need
to augment the pairing EDF of Eq.~\eqref{eq:EDF:pair:ULB} by a 
dependence on the isovector density as done in 
Refs.~\cite{Chamel10a,Margueron08a,Bertulani09a,Yamagami12a,Zhang19a}.
This question, however, cannot be decorrelated from the
evolution of the effective masses of protons and neutrons in
asymmetric matter, for which microscopic calculations indicate
that for SLy4 they split the wrong way round \cite{Lesinski06a}.

\begin{figure}[t!]
\centerline{
\includegraphics[width=7cm]{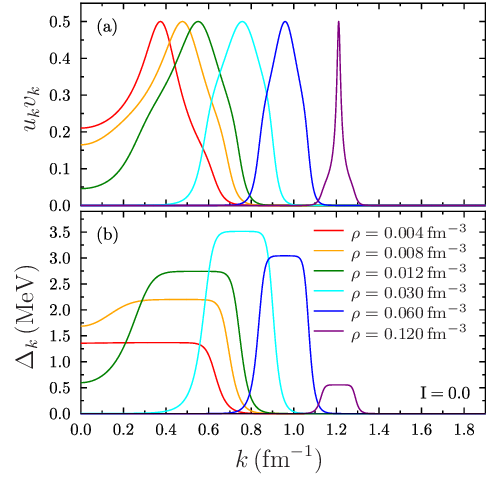}
}
\caption{\label{fig:gap+uv:k:SLy4+ULB} 
Product of occupation amplitudes $u_k v_k$ (panel a) and
matrix elements $\Delta_k$ (panel b) and as a function 
of $k$ in a calculation with SLy4+ULB 
in symmetric INM at the densities as indicated.
}
\end{figure}

\begin{figure}[t!]
\centerline{
 \includegraphics[width=7cm]{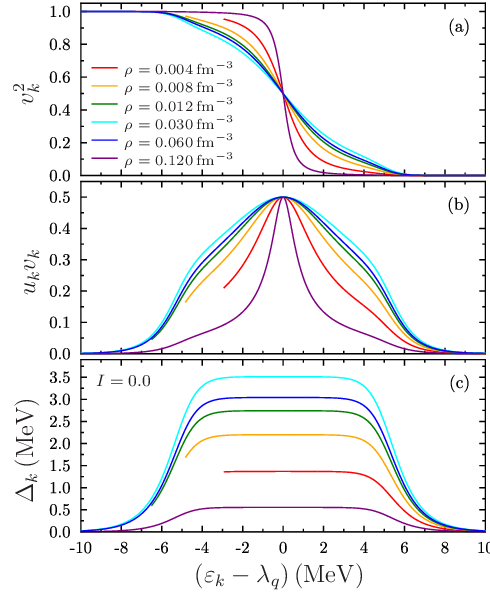}
 }
\caption{\label{fig:gap+uv:eps:SLy4+ULB} 
Occupation probabilities $v_k^2$ (panel a),
product of occupation amplitudes $u_k v_k$ (panel b), and
matrix elements $\Delta_k$ (panel c)
as a function of $\varepsilon_k-\lambda_q$ in a 
calculation with SLy4+ULB in symmetric INM at the densities as indicated.
}
\end{figure}

Figure~\ref{fig:gap+uv:k:SLy4+ULB} displays the matrix elements $\Delta_k$ 
entering the HFB equations and the product of the resulting occupation 
amplitudes $u_k v_k$ in symmetric matter as a function of $k$ at a few 
representative densities, while Fig.~\ref{fig:gap+uv:eps:SLy4+ULB} shows 
$\Delta_k$, $u_k v_k$, and the occupation probabilities $v_k^2$ as a function 
of the distance of the single-particle energy $\varepsilon_k$ to the chemical 
potential $\lambda_q$.

For a LO pairing EDF obtained from a gradientless contact generator 
\eqref{eq:force:pair:ULB} such as the ULB pairing EDF, 
without a cutoff the values of $\Delta_k$ of a given 
nucleon species in INM would take the same constant value for all 
$k$ that then would be equal to the gap $\Delta_q(k_{\lambda,q})$
plotted in Figs.~\ref{fig:Delta:rho:SLy4+ULB}  
and~\ref{fig:Delta:kl:SLy4+ULB}. The $k$-dependence of $\Delta_k$ that 
can be seen in Fig.~\ref{fig:gap+uv:k:SLy4+ULB} is the consequence of the 
cutoff~\eqref{eq:cutoff} that acts above and below the chemical potential 
$\lambda_q$. Because of the quadratic dependence of the single-particle 
energies $\varepsilon_k$ on $k$ and the very different potential depth 
at different densities, the curves for $\Delta_k$ appear to be distorted 
when being plotted as a function of $k$ in Fig.~\ref{fig:gap+uv:k:SLy4+ULB}. 
By contrast, when plotted as a function of $\varepsilon_k - \lambda_q$ as
in Fig.~\ref{fig:gap+uv:eps:SLy4+ULB}, these curves are all equally 
centred at zero.

The constancy of $\Delta_k$ inside the pairing window 
implies that when using a gradientless pairing EDF like the one
of Eq.~\eqref{eq:EDF:pair:ULB}, in the usual BCS pairing regime of INM  
the values of $\Delta_q(k_{\text{F},q})$ and $\Delta_q(k_{\lambda,q})$ 
are in general equal, in spite of $k_{\lambda,q} \neq k_{\text{F},q}$.
The situation will of course be different when using more elaborate pairing EDFs 
that yield a $k$-dependence of $\Delta_k$, such as the ones containing 
the kinetic pair density $\tilde{D}_q^{(\nabla,\nabla)}$ as generated by 
contact pairing interactions with gradients 
\cite{Dobaczewski84a,Sadoudi13a,Sadoudi13b,Hinohara18a},
or the ones obtained from a finite-range generator.

At $\rho = 0.004~\text{fm}^{-3}$, with 
$F^{1,1}_q = -4.447~\text{MeV}$ the potential is so shallow that
the entire range of the single-particle spectrum up to the 
chemical potential $\lambda_q = -1.520~\text{MeV}$ falls into the
pairing window, which explains why the corresponding curve in 
Fig.~\ref{fig:gap+uv:eps:SLy4+ULB} starts only in the middle of 
the plot. At $\rho = 0.008~\text{fm}^{-3}$, the distance 
between $F^{1,1}_q = -8.125~\text{MeV}$ and 
$\lambda_q = -3.301~\text{MeV}$ is just large enough so that the  
lowest single-particle levels are located where the cutoff starts to
fall off, whereas at $\rho = 0.012~\text{fm}^{-3}$, single-particle
levels at $k = 0$ are already in the tail of the cutoff. At higher 
density, the distance between the bottom of the potential $F^{1,1}_q$ 
and the chemical potential quickly grows to several tens of MeV such
that for low-lying single-particle levels the cutoff puts the values
of $\Delta_k$ to zero.

The occupation $v_{k,q}^2$ of a single-particle level at the chemical potential
$\varepsilon_{k,q} = \lambda_q$ is $1/2$, see Eq.~\eqref{eq:v2}.
If the pairing gap $\Delta_k$ remains constant over this interval, 
as is the case for the LO pairing EDF, then it follows from the same equation 
that the energy range over which the occupations fall from 0.9 to 0.1 
equals $8/3$ times the value of $\Delta_q(k_{\lambda,q})$, which
thereby directly indicates the magnitude of pairing correlations.

Because $\Delta_k = 0$ implies $v_k = 1$ and $u_k = 0$ for levels below
the chemical potential and $v_k = 0$ and $u_k = 1$ for levels above the
chemical potential, see Eqs.~\eqref{eq:v2} and~\eqref{eq:u2}, the distribution 
of $u_k v_k$ cannot exceed the range over which the $\Delta_k$ are non-zero. 
The cutoff used here has been tailored for calculations of finite nuclei,
for which the pairing gap rarely exceeds 1.5 MeV. The gap in INM, however, 
can become much larger, which then leads to the shoulder in the distribution 
of the $u_k v_k$ at about $\lambda_q \pm 5 \, \text{MeV}$, beyond which 
the $u_k v_k$ fall off very quickly.

\begin{figure}[t!]
\centerline{
 \includegraphics[width=7cm]{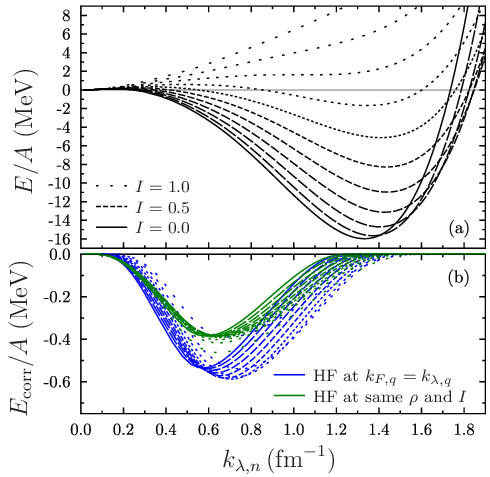}
 }
\caption{\label{fig:EA:kl:SLy4+ULB} 
Panel (a): Energy per particle $E/A$ in paired INM calculated with SLy4+ULB 
at asymmetries between $I = 0$ (symmetric matter) and $I=1$ (neutron matter) 
in steps of 0.1 plotted as a function of $k_{\lambda,n}$, where linestyles are 
the same as in Fig.~\ref{fig:Delta:rho:SLy4+ULB}.
Panel (b): corresponding pairing correlation energy per particle
$E_{\text{corr}}/A$, i.e.\ the difference in $E/A$ between a HFB and 
a HF calculation, (also known as pair condensation energy per particle)
calculated either at same $\rho$ and $I$ or at 
$k_{\lambda,q} = k_{\text{F},q}$.
}
\end{figure}

Panel~(b) of Fig.~\ref{fig:EA:kl:SLy4+ULB} displays the energy 
per particle $E/A$ as a function of total density for asymmetries 
ranging from $I = 0$ (symmetric matter) to $I=1$ (neutron matter) 
as a function  $k_{\lambda,n}$.
Panel (b) of the same figure displays the pairing correlation energy
per particle, i.e.\ the gain in $E/A$ when running a HFB calculation
instead of a HF calculation \cite{Ragnarsson95a}
\begin{align}
\frac{E_{\text{corr}}}{A}
& =  \frac{E_{\text{tot}}^{\text{HFB}}}{A}
   - \frac{E_{\text{tot}}^{\text{HF}}}{A}
\end{align}
at either the same $\rho$ and $I$, or at $k_{\lambda,q} = k_{\text{F},q}$. 
These two choices for the HF reference calculation are not equivalent 
because for paired states there is no relation between $k_{\lambda,q}$ 
and the nucleonic densities.
It is clearly the former of these two choices that should be made.
Note that $E_{\text{corr}} / A$ is also known as
pair condensation energy per particle in the literature \cite{Shelley20a}.

As for this parametrisation the pairing correlations disappear at saturation 
density in both symmetric and isospin-asymmetric matter, taking into account 
pairing correlations cannot affect the usual nuclear matter properties 
of SLy4 at saturation density.
But even when using a pairing EDF for which there still is a paired solution 
at saturation density, the effect of pairing on all widely-discussed INM properties 
at the saturation point remains small, see Ref.~\cite{Margueron14a}.
By contrast, pairing correlations can significantly affect the energy per 
particle of homogeneous INM at sub-saturation density, in particular in the 
region where INM becomes bound at small asymmetries, see also Ref.~\cite{Zhang19a}. 
Indeed, for the ULB pairing interaction the pairing correlation energy
per particle peaks at densities at which the total binding energy per particle
in paired INM is of comparable size. The detailed behaviour might of course 
sensitively depend on the choice of the EDF and its parametrisation, a few
examples for which just the parameters of the pairing EDF 
\eqref{eq:EDF:pair:ULB} are modified are discussed in 
Sec.~\ref{sec:role:of:params}.

We have to recall, however, that, because of the so-called ``spinodal'' 
finite-size instability \cite{Pastore15a,Ducoin07a}, 
there actually is no stable low-density phase of homogeneous INM 
symmetric matter or matter at small asymmetry.
HFB calculations of such matter at sub-saturation densities serve as a 
laboratory for the controlled analysis of pairing interactions, but 
have no direct application to a system that could be found in nature.

%
\subsection{Role of the parameters in the pairing EDF}
\label{sec:role:of:params}

As the next step, we will compare the role of the parameters $V_0$, $\eta$
and  $\sigma$ of the LO pairing EDF \eqref{fig:gap+uv:eps:SLy4+ULB}
for the pairing correlations found in finite nuclei and infinite matter.

\begin{table}
\caption{\label{tab:sly4:fit:mix}
Parameters of the pairing EDF \eqref{eq:EDF:pair:ULB}
with varied parameter $\eta$ at fixed $\sigma = 1.0$
that with SLy4 yield the same average neutron gap $\langle uv \Delta \rangle_n$ 
for \nuc{120}{Sn} 
as the original ULB parametrisation with $\eta = 1.0$. 
}
\begin{center}
\begin{tabular}{lccc}
\hline\noalign{\smallskip}
  & \multicolumn{1}{c}{$V_0$} 
  & \multicolumn{1}{c}{$\eta$} 
  & \multicolumn{1}{c}{$\sigma$} \\
  & \multicolumn{1}{c}{(MeV fm$^{3}$)}
  & & \\
\noalign{\smallskip}\hline\noalign{\smallskip}
surface     & $-1250.0$   & 1.00     & 1.0      \\
            & $-873.2$    & 0.75     & 1.0      \\  
mixed       & $-670.0$    & 0.50     & 1.0       \\
            & $-543.2$    & 0.25     & 1.0       \\
volume      & $-456.7$    & 0.00     & 1.0       \\ 
\noalign{\smallskip}\hline
\end{tabular}
\end{center}
\end{table}

We start with the analysis of the role of the parameter $\eta$ that regulates 
between the ``surface'' ($\eta = 1$) and ``volume'' ($\eta = 0$) character of 
the pair potential. In order to have comparable results, we constructed a series 
of parametrisations with decreasing $\eta$ that together with SLy4 give the 
same average neutron pairing gap as defined through Eq.~\eqref{eq:uvgap:gen} of
$\langle uv \Delta \rangle_n = 1.924$ MeV for \nuc{120}{Sn} as the original ULB 
parametrisation of the pairing EDF. The same parametrisation \eqref{eq:cutoff} of 
the cutoff is used in all cases, the contribution \eqref{eq:F11:paircontribution} 
of the pairing EDF to the normal potentials $F^{1,1}_q (\vec{r})$ is kept, and 
all calculations of \nuc{120}{Sn} are done with the pairing stabilisation 
\eqref{eq:Epair:stab:1} at fixed $\sigma =1$ using the spherical HFB code 
mentioned above. 
The resulting parameters are listed in Table~\ref{tab:sly4:fit:mix}. 
With decreasing $\eta$, the overall pairing strength becomes much smaller. 
We recall that for fixed $\eta$ and $\sigma$ the size of $V_0$ inversely 
scales with the pairing cutoff and can become much smaller when 
renormalising the pairing strength to a larger pairing window.

\begin{figure}[t!]
\centerline{
 \includegraphics[width=7cm]{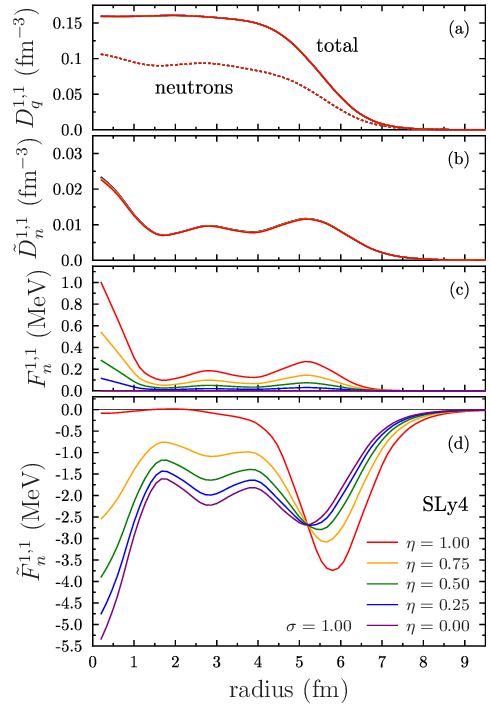}
 }
\caption{\label{fig:sn120:SLy4+ULB-mix} 
Normal density of neutrons and nucleons (panel a),
pair density of neutrons (panel b), 
contribution of the pairing EDF to the normal potential of neutrons (panel c),
and pair potential of neutrons (panel d)
for \nuc{120}{Sn} calculated with SLy4
and five variants of the pairing EDF \eqref{eq:EDF:pair:ULB} with different 
value of the parameter $\eta$ that regulates between surface and volume 
character of the EDF. The red curve represents the original ULB parametrisation.
}
\end{figure}

Figure~\ref{fig:sn120:SLy4+ULB-mix} displays the pairing potential of neutrons 
$\tilde{F}^{1,1}_n (\vec{r})$, the contribution of the pairing EDF to the normal
potential $F^{1,1}_n (\vec{r})$  of neutrons, the pair density 
$\tilde{D}^{1,1}_n (\vec{r})$ of neutrons as well as the neutron and total 
normal densities, $D^{1,1}_n (\vec{r})$ and $D^{1,1}_0 (\vec{r})$, 
for \nuc{120}{Sn} calculated with SLy4 and all parametrisations of the 
pairing EDF defined in Table~\ref{eq:EDF:pair:ULB}.

In the case of pure volume pairing ($\eta = 0$), $\tilde{F}^{1,1}_n(r)$ 
is simply proportional to the neutron pair density
$\tilde{D}^{1,1}_n(r)$ and the contribution of the pairing EDF to the
normal potential \eqref{eq:F11:paircontribution} vanishes. 
By contrast, in the case of surface pairing ($\eta = 1$), the contribution 
of the pairing EDF to the normal potential of neutrons is proportional 
to the square of the neutron pair density $\tilde{D}^{1,1}_n(r)$. 
Note that for an attractive pairing interaction of the form 
\eqref{eq:EDF:pair:ULB}, its contribution to the normal potentials 
\eqref{eq:F11:paircontribution} is strictly positive, i.e.\ repulsive.

The pairing potential $\tilde{F}^{1,1}_n(r)$ of neutrons illustrates the expected 
change in radial dependence when going from $\eta = 1$ to $\eta = 0$: 
For $\eta = 1$, the pair potential practically vanishes in the nuclear
interior and has a peak at the nuclear surface where the total density
falls to zero. We recall that because of the volume element $4\pi \, \rmd r \, r^2$, 
a pair potential peaked inside the nuclear volume has to be larger than 
a surface-peaked one in order to give the same matrix element for 
a single-particle state that is spread over the entire nucleus.

The single-particle orbits, however, are not equally spread over the
nucleus; therefore surface-peaked and volume-type pair potentials 
will in general lead to different individual matrix elements $\Delta_{km}$
\cite{Bender00a}. Within a given major shell, a surface-peaked pair 
potential typically gives larger $\Delta_{km}$ between high-$\ell$ 
single-particle states, whereas a volume-type pair potential gives
larger $\Delta_{km}$ between low-$\ell$ states. 

Still, as long as all pairing parameter sets yield the same average 
neutron gap $\langle uv \Delta \rangle_n$ for \nuc{120}{Sn}, the 
normal and also pair density distributions of this nucleus 
remain nearly identical for all values of $\eta$,
although there are small differences that cannot be resolved on
Fig.~\ref{fig:sn120:SLy4+ULB-mix} originating mostly from 
slightly different distributions of the occupation amplitudes 
$u_k$ and $v_k$ in each case. Similar results have been obtained 
before in Refs.~\cite{Sandulescu05a,Tolokonnikov11a}. 

The main reasons for the weak sensitivity of $\tilde{D}^{1,1}_q(r)$ 
to the details of the pairing form factor entering the $\Delta_{km}$ in
the HFB equation are that the readjustment of the pairing strength keeps 
the energy interval of eigenstates of $\hat{h}$ for which $\kappa_{km}$ 
is large almost constant, as well as that the spatial wave functions 
of these states are mainly affected through rearrangement effects from the self-consistent changes of the densities that enter the particle-hole EDF.
The only direct impact of the pairing form factor on $\hat{h}$ through 
the pairing contribution~\eqref{eq:F11:paircontribution} to 
the normal potential $F^{1,1}_q$ is typically about two orders of magnitude 
smaller than the contribution of the particle-hole EDF and its change with
$\eta$ therefore only leads to a minor redistribution of the radial wave 
functions that can barely be seen when plotting the densities as in
Fig.~\ref{fig:sn120:SLy4+ULB-mix}. But despite the  smallness of this effect, it is probed by
the odd-even staggering of charge radii, see Sec.~\ref{sec:nuclei:stagger:r}.
For very weakly-bound nuclei with $- |\Delta| \simeq \lambda $, 
the impact of the form factor might be more pronounced than what
is found for \nuc{120}{Sn}.

\begin{figure}[t!]
\centerline{
 \includegraphics[width=7cm]{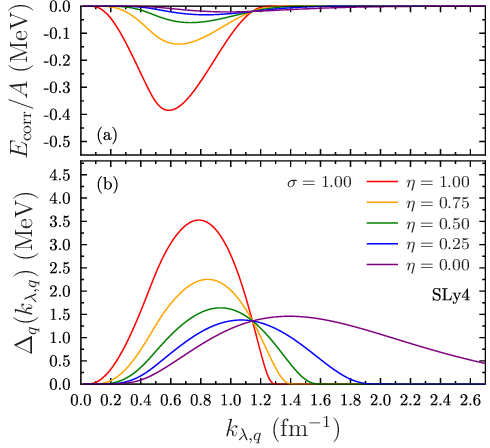}
 }
\caption{\label{fig:gap:kl:SLy4+ULB-mix} 
Gap $\Delta_q(k_{\lambda,q})$ (panel b) and
pairing correlation energy per particle $E_{\text{corr}}/A$ 
calculated as the difference in $E/A$ 
between a HFB and a HF calculation at same $\rho$ and $I$ (panel a),
as a function of $k_{\lambda,q}$ for symmetric INM calculated with SLy4 
and the same five variants of the pairing EDF \eqref{eq:EDF:pair:ULB} 
with different value of the parameter $\eta$  as in 
Fig.~\ref{fig:sn120:SLy4+ULB-mix}, using the same colour code.
}
\end{figure}

While the parametrisations of Table~\ref{tab:sly4:fit:mix} all give the
same average neutron pairing gap for \nuc{120}{Sn}, changing the form factor of 
the pairing EDF from surface to volume character dramatically changes the 
pairing gaps in INM, see Fig.~\ref{fig:gap:kl:SLy4+ULB-mix}. The width 
of the curve representing $\Delta_q(k_{\lambda,q})$, its value at the 
maximum, and also the position of the maximum all change.

With decreasing $\eta$, the peak becomes wider, and its maximum is shifted 
towards larger $k_{\lambda,q}$, which simply follows from the effective 
pairing strength changing from being zero at saturation density for pure
surface pairing ($\eta = 1$) to being independent on the density for pure
volume pairing ($\eta = 0$). The highest $k_{\lambda,q}$ on 
Fig.~\ref{fig:gap:kl:SLy4+ULB-mix} corresponds to 
$\rho \simeq 0.63 \, \text{fm}^{-3}$, i.e.\ about four times saturation 
density. For pure volume pairing, there are still pairing correlations 
found even at $\rho \simeq 4 \, \text{fm}^{-3}$, which is about 25 times
saturation density.
Curiously, for an unidentified reason all curves cross at about 
$k_{\lambda,q} \simeq 1.145 \, \text{fm}^{-1}$, which corresponds to 
$\rho \simeq 0.1015 \, \text{fm}^{-3} \simeq 0.63 \, \rho_{\text{sat}}$ 
with a value of $\Delta_q(k_{\lambda,q}) \simeq 1.35 \, \text{MeV}$. 
In any event, Fig.~\ref{fig:gap:kl:SLy4+ULB-mix} confirms that producing 
the same density dependence of $\Delta_q(k_{\lambda,q})$ as found in 
microscopic calculations of paired INM with an LO contact pairing EDF 
requires surface-dominated pairing with a value of $\eta$ near one.

The pairing correlation energy per particle $E_{\text{corr}}/A$ also
dramatically changes when going from surface to volume pairing, see
panel (a) of Fig.~\ref{fig:gap:kl:SLy4+ULB-mix}. As already 
observed on Fig.~\ref{fig:EA:kl:SLy4+ULB}, for surface pairing, $\eta = 1$, 
it might be as large as $0.4~\text{MeV} / A$. Decreasing $\eta$,
however, the pairing correlation energy also dramatically decreases.
For pure volume pairing, $\eta = 0$, it does not exceed a few tens 
of keV at any density.

\begin{table}
\caption{\label{tab:sly4:fit:exp}
Parameters of the pairing EDF \eqref{eq:EDF:pair:ULB}
with varied power of the density dependence $\sigma$ for 
surface pairing with fixed $\eta = 1$ that with SLy4 yield the same
average neutron gap $\langle uv \Delta \rangle_n$ for \nuc{120}{Sn} 
as the original ULB parametrisation with $\sigma = 1.0$. 
}
\begin{center}
\begin{tabular}{ccc}
\hline\noalign{\smallskip}
   \multicolumn{1}{c}{$V_0$} 
  & \multicolumn{1}{c}{$\eta$} 
  & \multicolumn{1}{c}{$\sigma$} \\
  \multicolumn{1}{c}{(MeV fm$^{3}$)}
  & & \\
\noalign{\smallskip}\hline\noalign{\smallskip}
$-1250.0$    & 1.0  &   1.0 \\
$-1333.9$    & 1.0  &   0.9 \\ 
$-1438.8$    & 1.0  &   0.8 \\
$-1573.7$    & 1.0  &   0.7 \\ 
$-1753.7$    & 1.0  &   0.6 \\ 
$-2005.7$    & 1.0  &   0.5 \\  
$-2383.9$    & 1.0  &   0.4 \\  
$-3014.4$    & 1.0  &   0.3 \\ 
$-4275.6$    & 1.0  &   0.2 \\
\noalign{\smallskip}\hline
\end{tabular}
\end{center}
\end{table}

\begin{figure}[b!]
\centerline{
 \includegraphics[width=7cm]{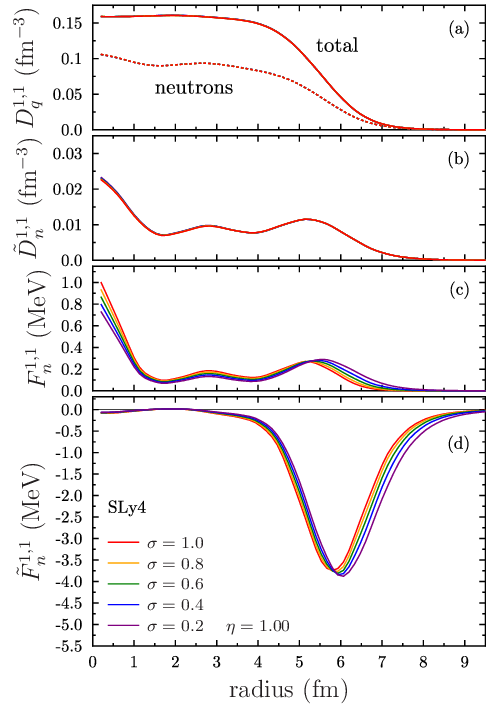}
 }
\caption{\label{fig:sn120:SLy4+ULB-exp} 
Same as Fig.~\ref{fig:sn120:SLy4+ULB-mix}, but for five variants of 
the LO pairing EDF with varied power $\sigma$ of the density dependence
from Table~\ref{tab:sly4:fit:exp}.
}
\end{figure}

\begin{figure}[t!]
\centerline{
 \includegraphics[width=7cm]{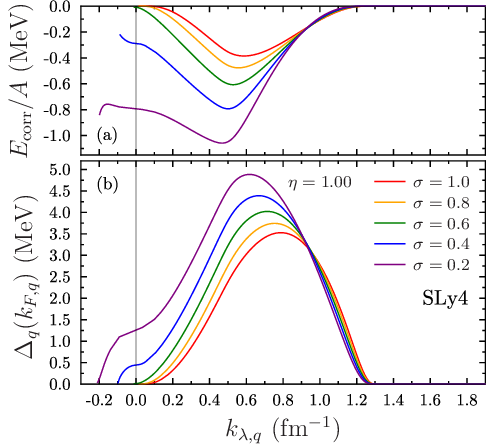}
 }
\caption{\label{fig:gap:kl:SLy4+ULB-exp} 
Gap $\Delta_q(k_{\text{F},q})$ (panel b) and pairing correlation 
energy per particle $E_{\text{corr}}/A$ (panel a) as a function of 
$k_{\lambda,q}$ for symmetric INM calculated with SLy4 and the same 
five variants of the the pairing EDF \eqref{eq:EDF:pair:ULB}  
with systematically varied power $\sigma$ of the density dependence 
used to prepare Fig.~\ref{fig:sn120:SLy4+ULB-exp}.
Negative values of $k_{\lambda,q}$ represent the imaginary solutions
of Eq.~\eqref{eq:klambda} that correspond to a chemical potential $\lambda_q$ 
below the physical single-particle spectrum that can be found in a 
Bose-Einstein condensate (see text).
}
\end{figure}

Instead of varying $\eta$ at fixed $\sigma = 1$, it is also illustrative
to vary $\sigma$ at fixed $\eta = 1$. Again, we generated a series of 
parametrisations of the LO pairing EDF \eqref{eq:EDF:pair:ULB} 
that together with SLy4 give the same average pairing gap of
$\langle uv \Delta \rangle_n = 1.924$ MeV for \nuc{120}{Sn} as the original ULB 
parametrisation of the LO pairing EDF. The resulting parameters are
listed in Table~\ref{tab:sly4:fit:exp}.
 
Figure~\ref{fig:sn120:SLy4+ULB-exp} displays the neutron and total
density, neutron pair density, neutron pair potential, and pairing 
contribution to the neutron potential $F^{1,1}_n(\vec{r})$ of \nuc{120}{Sn}
for these parameter sets in the same manner as in Fig.~\ref{fig:sn120:SLy4+ULB-mix}.
Contrary to what was found there for the parameter sets with varying $\eta$
but same $\sigma$, the parameter sets with same $\eta$ but different
$\sigma$ not only yield very similar results for the densities 
$D^{1,1}_n(\vec{r})$ and $\tilde{D}^{1,1}_n(\vec{r})$, but also 
for the potentials $\tilde{F}^{1,1}_n(\vec{r})$ and $F^{1,1}_n(\vec{r})$.
This indicates that results for well-bound nuclei like 
\nuc{120}{Sn} are rather insensitive to the power of the density dependence.
The only visible evolution is that with smaller $\sigma$ the peak of the
pair potential $\tilde{F}^{1,1}_n(\vec{r})$ becomes a little wider and is 
slightly shifted towards larger radii in the nuclear surface.

\begin{figure}[b!]
\centerline{
 \includegraphics[width=7cm]{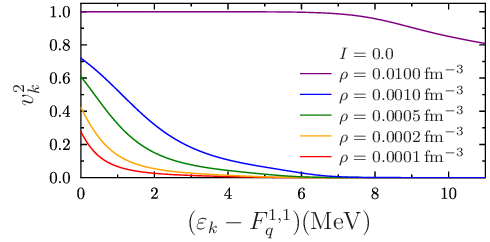}
 }
\caption{\label{fig:v2:eps:SLy4+ULB-exp} 
Occupation numbers $v_k^2$ for solutions of HFB calculations of symmetric INM 
with SLy4 and the pairing parameter set with $\sigma = 0.2$ from
Table~\ref{tab:sly4:fit:exp} at different small total densities as indicated as a function 
of single-particle energy relative to the bottom of the potential well 
$\varepsilon_k - F^{1,1}_q$.
}
\end{figure}

The gap and pairing correlation energy per particle found 
with these five parameter sets for symmetric INM 
are displayed in Fig.~\ref{fig:gap:kl:SLy4+ULB-exp}. 
With decreasing $\sigma$, at low density -- or equivalently 
small (positive) $k_{\lambda,q}$ -- both the gap at $k_{\lambda,q}$ and 
$E_{\text{corr}}/A$ increase to the point that one still finds a paired
solution at $k_{\lambda,q} = 0$. 

The structure of this kind of unusual solution at low density can be easily 
understood from the corresponding distribution of occupation numbers $v_k^2$.
Using that the canonical and HF bases are identical in INM, 
Fig.~\ref{fig:v2:eps:SLy4+ULB-exp} displays $v_k^2$ as a function of the 
single-particle energy relative to the bottom of the potential well obtained
in HFB calculations with SLy4 plus the pairing parameter set with $\sigma = 0.2$
for five different densities.
The single-particle state with $\varepsilon_k - F^{1,1}_q = 0$ is the lowest state 
in the single-particle spectrum with $k = 0$. All distributions of $v_k^2$
still follow~\eqref{eq:v2}, but what is different from the usual situation found for 
finite nuclei and paired INM is that, for $\rho \lesssim 0.007~\text{fm}^{-3}$, 
they start at values of $v_k^2$ that are much smaller than 1. 
For $\rho \lesssim 0.0003~\text{fm}^{-3}$, the occupation number at 
$k=0$ is even smaller than $0.5$, which implies that the chemical potential
is outside of the physical single-particle spectrum, $\lambda_q < F^{1,1}_q$. 
In such case, the formal solutions of Eq.~\eqref{eq:klambda} for $k_{\lambda,q}$ 
become imaginary and do not represent the momentum of a single-particle state
anymore. For an NLO EDF like SLy4, there are two solutions of opposite sign. 
We find that the negative modulus of the such obtained $k_{\lambda,q}$ 
is a useful quantity to plot the properties of these unusual solutions of the 
HFB equations, which is what is done in Fig.~\ref{fig:gap:kl:SLy4+ULB-exp} 
to plot the gap and $E_{\text{corr}}/A$ at small density.
As $k_{\lambda,q}$ does not correspond to a physical state, the gap $\Delta_k$
cannot be calculated at this momentum. For this reason,  
Fig.~\ref{fig:gap:kl:SLy4+ULB-exp} displays the gap at $k_{\text{F},q}$ instead.
We mention in passing that plotting $\Delta_q(k_{\text{F},q})$ and 
other properties of paired INM as a function of $\rho$ or $k_{\text{F},q}$ 
instead of $k_{\lambda,q}$ completely masks the emergence of these 
unusual solutions as they are found only for a tiny interval of either 
of these two coordinates.

\begin{figure}[t!]
\centerline{
 \includegraphics[width=8cm]{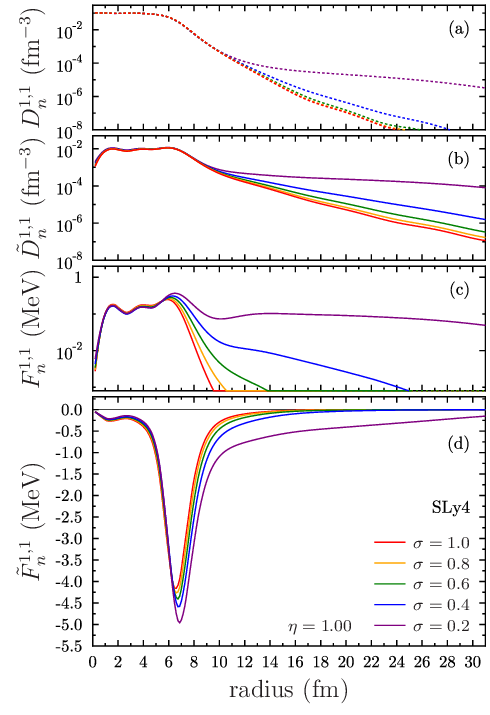}
 }
\caption{\label{fig:sn160:SLy4+ULB-exp} 
Same as Fig.~\ref{fig:sn120:SLy4+ULB-exp} but for the weakly-bound 
\nuc{160}{Sn}, plotting all densities and potentials but 
$\tilde{F}^{1,1}_n$ on a logarithmic scale to better resolve 
their behaviour at large $r$.
}
\end{figure}

The kind of solution of the HFB equations for which $v_{k=0}^2$ is much 
smaller than 1 can be associated with the formation of a Bose-Einstein 
condensate (BEC), see Ref.~\cite{Strinati18a} and references therein.
In the case discussed here the BEC is formed of bound di-nucleons of 
same isospin coupled to spin zero.

In similar calculations using the bare strong NN interaction that 
also consider $T=0$ pair correlations, a BEC of deuterons is  
found in the low-density limit of symmetric matter 
\cite{Rubtsova17a}. Di-neutrons, however, are not bound in 
nature and therefore should not form a condensate, and neither should 
hypothetical di-protons without electric charge. The appearance of a 
BEC in the $T=1$ channel is therefore the unambiguous signature of a 
nonphysical property of a given parametrisation of the nuclear EDF. 
However, it has been repeatedly pointed out that low-density nuclear matter 
can be expected to come close to a BEC-BCS crossover in the $T=1$ channel 
because of the large scattering length in the ${}^{1}S_{0}$ channel in 
the vacuum 
\cite{Brink05a,Broglia13a,Dean03a,Matsuo06a,Lombardo01a,Margueron07a,Sun12a}. 
The nonphysical transition to a BEC can therefore already happen when some 
parameter of the pairing EDF is only mildly off its physical value.

The transition from a ${}^{1}S_{0}$ BCS phase to a BEC has already been found 
in the study of paired INM with contact pairing interactions adjusted to 
reproduce gaps from microscopic models \cite{Margueron07a}. 
Also, a BEC of di-nucleons has been observed in the surface of finite nuclei 
in Refs.~\cite{Rotival09a,RotivalThesis} for specific values of the parameters 
controlling the density dependence of the LO pairing EDF \eqref{eq:EDF:pair:ULB}. 
While there is nothing unusual in the results for \nuc{120}{Sn} displayed in 
Fig.~\ref{fig:sn120:SLy4+ULB-exp}, an instability towards a BEC does clearly
jeopardise the properties of weakly-bound nuclei. This is exemplified
by Fig.~\ref{fig:sn160:SLy4+ULB-exp} for \nuc{160}{Sn}, a weakly-bound nucleus for 
which the chemical potential $\lambda_n$ of neutrons takes a value of about $-630$ keV 
for all pairing parametrisations. For the two parameter sets from Table~\ref{tab:sly4:fit:exp} 
with either $\sigma = 0.4$ or $0.2$ that lead to a BEC in symmetric INM in 
Fig.~\ref{fig:gap:kl:SLy4+ULB-exp}, the normal and pair density of neutrons
develop a long queue that then feeds back to the potentials, thereby 
amplifying the condensation of di-neutrons at large distances. Like in INM at
low density, a dilute gas of di-neutrons is forming outside of the nucleus
that fills the entire size of the numerical box.

The phenomenon of a BEC of neutrons outside of a weakly-bound nucleus
that can emerge in an HFB calculation with an unrealistic 
pairing interaction has to be clearly distinguished from the 
phenomenon of a spurious neutron gas that forms in HF+BCS calculations 
of weakly-bound nuclei even with realistic pairing
interactions~\cite{Dobaczewski84a,Dobaczewski96a}. 
In a full HFB calculation, when all nuclear potentials fall to zero at 
large distance for a system with negative chemical potential $\lambda_q$,
it can be shown that all states in the canonical basis are also
localised, from which then follows that all densities and pair densities 
are localised either \cite{Dobaczewski96a}. In the HF+BCS approximation, 
this formal feature is lost \cite{Dobaczewski96a}. 
By contrast, for a BEC solution of the HFB equations, the condition 
that the potentials fall off to zero is not satisfied anymore for 
the self-consistent solution as evidenced by Fig.~\ref{fig:sn160:SLy4+ULB-exp}.

We have to note, however, that finding or not evidence of such solution 
in calculations of finite nuclei sensitively depends on the 
numerical settings that are used, which in our case are the box size, 
distance between points on the mesh, and the number of single-particle 
states explicitly kept. In too small a basis, the long tail of the 
nonphysical solution might not be fully resolved. And indeed, the HFB 
calculations used to prepare Fig.~\ref{fig:sn160:SLy4+ULB-exp} were run 
in a much larger box with smaller step size and keeping significantly 
more HF states than usually necessary to converge physical solutions 
with respect to the numerical representation when using the two-basis 
method in coordinate space. The onset of signatures of a BEC in finite 
nuclei might also depend on choices made for the pairing cutoff.
The origin of finding a BEC for specific parametrisations of the
nuclear EDF will be investigated elsewhere.

%
%
\section{Adjusting the pairing strength of a density-dependent contact pairing interaction}
\label{sec:fit:fit}


\subsection{General considerations}

Having set up the tool to calculate the pairing gap in INM, we now come back to
the question of adjusting the parameters of the pairing EDF for arbitrary 
parametrisations of the Skyrme ph EDF.
In what follows, we will illustrate under which conditions the pairing
gap in INM $\Delta_q(k_{\lambda,q})$ can be used as a 
proxy to adjust the parameters of the LO pairing EDF \eqref{eq:EDF:pair:ULB} 
in such a way that one finds very similar global trends of pairing correlations 
in HFB calculations of finite nuclei with any (realistic) parametrisation 
of the underlying Skyrme EDF.

For the specific simple LO form of EDF \eqref{eq:EDF:pair:ULB}, all terms 
contribute to the gap in INM. 
To be predictive, the adjustment protocol has to rely on a reference 
calculation for the pairing gap in INM with an effective interaction that well 
describes pairing correlations in finite nuclei. This reference calculation has
to be made with an effective interaction that has no (dominant) terms that 
are only active in finite nuclei. Generalisations to cases where the target
EDF and possibly also the reference EDF contains terms that are only active 
in finite nuclei are possible, but will not be considered here. 
Also, in the absence of an explicit dependence on isospin of the LO 
pairing EDF \eqref{eq:EDF:pair:ULB}, we only consider symmetric INM.
Again, generalisations are possible, but would require a reference 
calculation with a realistic isospin dependence.

We tested several possibilities for such reference calculations, two of 
which will be discussed here. 
The discussion of Figs.~\ref{fig:gap:kl:SLy4+ULB-mix} and 
\ref{fig:gap:kl:SLy4+ULB-exp} has already made it evident that adjusting 
the pairing strength to $\Delta_q(k_{\lambda,q})$ at just one density cannot be 
sufficient. Instead, the full density-dependence, or, equivalently, the full 
$k_{\lambda,q}$ dependence, of the gap has to be used as pseudo-data in the 
parameter adjustment, as each of the three parameters of the LO EDF controls
a different aspect of this curve.

For reasons that will become clear in what follows, we could not identify a 
reference calculation for asymmetric INM of which we can be sure to be
predictive for asymmetric nuclei and which can be expected to be reproducible 
with a LO pairing EDF even when including an isovector dependence, which 
motivated us to stick to the standard form of Eq.~\eqref{eq:EDF:pair:ULB}.
This does, however, not mean that after a parameter adjustment that leads to the
same gaps in symmetric matter all parameter sets will also give the same
predictions for asymmetric matter: different parametrisations of the Skyrme 
ph EDF can have very different isovector effective mass, which then can make
a difference for the evolution of the proton and neutron pairing
gaps in asymmetric matter.

\begin{table}[t!]
\caption{\label{tab:3d:fits:sly4}
Parameter sets \LOPP{}{ULB} of the LO pairing EDF \eqref{eq:EDF:pair:ULB}
obtained adjusting the pairing gaps in symmetric INM obtained for the 
underlying particle-hole Skyrme EDF as indicated to reproduce the pairing gaps 
from SLy4+ULB in symmetric INM. 
The second column lists the isoscalar effective mass of the 
respective Skyrme parametrisation at the saturation point.
The original SLy4+ULB values are given for comparison.
}
\begin{center}
\begin{tabular}{lllll}
\hline\noalign{\smallskip}
  & \multicolumn{1}{c}{$m^*_0/m$} 
  & \multicolumn{1}{c}{$V_0$} 
  & \multicolumn{1}{c}{$\eta$} 
  & \multicolumn{1}{c}{$\sigma$} \\
  & & \multicolumn{1}{c}{(MeV fm$^{3}$)}
  & & \\
\noalign{\smallskip}\hline\noalign{\smallskip}
SLy4        & 0.695 & $-1250.0$ & 1.000 & 1.000  \\
1T2T(0.70)  & 0.700 & $-1250.9$ & 1.000 & 0.993  \\      
1T2T(0.80)  & 0.800 & $-1267.0$ & 1.012 & 0.896  \\   
1T2T(0.85)  & 0.850 & $-1275.3$ & 1.017 & 0.859  \\  
SN2LO1      & 0.709 & $-1251.0$ & 1.009 & 1.013  \\
\noalign{\smallskip}\hline
\end{tabular}
\end{center}
\end{table}


\subsection{Adjusting parameters to infinite nuclear matter pairing gaps from SLy4+ULB}

One successful procedure that we could identify consists in adjusting the 
parameters of the LO pairing EDF \eqref{eq:EDF:pair:ULB} to the gaps obtained 
from HFB calculations of symmetric INM with SLy4+ULB as discussed in 
Sec.~\ref{sec:INM:SLy4+ULB}. 
As pseudo-data, we use the value of $\Delta_q(k_{\lambda,q})$
at 11 points between $k_{\lambda,q} = 0.2 \, \text{fm}^{-1}$
and $k_{\lambda,q} = 1.2 \, \text{fm}^{-1}$ in steps of $0.1 \, \text{fm}^{-1}$.
Depending on the combination of pairing parameters, for 
both larger and also smaller values of $k_{\lambda,q}$ there might 
not be a paired solution of the HFB equations, which would jeopardise
the parameter adjustment. Conversely, using a smaller interval might not 
sufficiently fix the low- and/or high-density behaviour of the gap from the 
reference calculation. All gaps enter the penalty function with the same 
weight.

\begin{figure}[t!]
\centerline{\includegraphics[width=7cm]{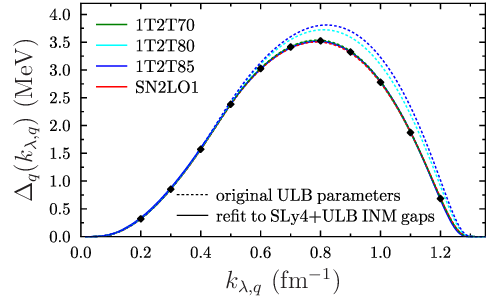}}
\caption{\label{fig:inm:gap:fits2sly4ulb}
Gap at $k_{\lambda,q}$ in symmetric INM for the parameter sets of the 
ph Skyrme EDF as indicated and \LOPP{}{ULB} pairing parameters.
Markers indicate results obtained with SLy4+ULB pairing.
For each Skyrme parametrisation, results obtained with two different parametrisations
of the pairing EDF are shown: one where the original ULB parameters of 
Ref.~\cite{Rigollet99a} are used (dotted lines), and one where the parameters 
were readjusted to reproduce the SLy4+ULB gaps in symmetric INM as indicated 
in Table~\ref{tab:3d:fits:sly4} (solid lines). In the latter case the 
calculated curves almost perfectly fall on top of each other.}
\end{figure}

The pairing parameters resulting from such fit for the 1T2T(0.70), 
1T2T(0.80), 1T2T(0.85), and SN2LO1 parameter sets described in 
Sec.~\ref{sec:edf:ph} are listed in Table~\ref{tab:3d:fits:sly4}.
Whenever possible, we will use the shorthand \LOPP{}{ULB} to refer 
to them in what follows.
Figure~\ref{fig:inm:gap:fits2sly4ulb} compares the gaps in symmetric
INM obtained with these parametrisations with the results from SLy4+ULB 
that defined the penalty function. In all cases the SLy4+ULB gaps are very 
well reproduced. In general, this requires the readjustment of all three 
parameters of the pairing EDF. The reason for this can be easily deduced when 
comparing the results obtained with readjusted pairing parameters with calculations
that use the original ULB pairing parameters that are also displayed with
dotted lines on Fig.~\ref{fig:inm:gap:fits2sly4ulb}: when not readjusting the
pairing parameters to the specific ph interaction, with increasing 
effective mass the curve representing the gap becomes wider, its peak is 
shifted towards larger $k_{\lambda}$, and the peak value increases. As 
illustrated in Sec.~\ref{sec:INM:SLy4+ULB}, the width of the curve and 
position of the peak are controlled by $\eta$ and $\sigma$, whereas $V_0$ mostly 
influences the maximum value. Therefore, the values of all three parameters
have to be readjusted.

It turns out that the parameter $\eta$ controlling the 
volume-to-surface behaviour remains almost constant near 1, whereas the power of
the density dependence $\sigma$ becomes significantly smaller for 
Skyrme parameter sets with larger effective mass. At the same
time, the global amplitude $V_0$ also increases with effective mass. This might 
appear counter-intuitive, because larger effective mass requires weaker 
pairing. $V_0$ indeed has to be reduced when keeping $\sigma = \eta = 1$ 
fixed as done in Ref.~\cite{DaCosta24a}. However, as indicated by 
Table~\ref{tab:sly4:fit:exp}, reducing $\sigma$ at 
fixed effective mass requires a substantial increase
in $V_0$. Combining these two trends, the reduction of $\sigma$ when
increasing $m^*_0/m$ is accompanied by a mild growth of $V_0$.
Note that for different parameter sets of the standard Skyrme NLO EDF
with identical saturation density and isoscalar effective mass $m^*_0/m$ 
(and identical $\hbar^2/2m$), our adjustment procedure leads to exactly 
the same pairing parameters. The reason is that in that case $m^*_0/m$ 
depends on only one coupling constant, which is determined by the 
effective mass at saturation. Indeed, evaluating Eq.~\eqref{eq:ms:q} 
for a NLO EDF and symmetric matter, one finds for the inverse effective 
mass that 
$m/m^*_0 = 1 + \tfrac{2 m}{\hbar^2} \cc{A}^{(2,2)}_{0,\textrm{e}} \, D^{1,1}_{0}$,
such that all standard NLO parameter sets with same 
$2 m \, \cc{A}^{(2,2)}_{0,\textrm{e}} / \hbar^2$
have exactly the same density-dependence of the $k$-independent
$m^*_0/m$ in symmetric INM, and therefore exactly the same constant 
level density at any given $\rho$ in symmetric matter
irrespective of differences in the corresponding equation of state.
Our adjustment protocol therefore reproduces the global scaling of the 
pairing gap with effective mass that is naively expected for the relation 
between the average level density of single-particle states and the 
pairing gap \cite{Jensen86a}. This relation was implied in many publications 
that used the original ULB parameters adjusted for SLy4 also for HFB 
calculations with 
SLy5, SLy6, SLy7 \cite{Chabanat98a},
f$_{0}$, f$_{+}$, f$_{-}$ \cite{Lesinski06a},
the T$IJ$ \cite{Lesinski07a},
SLy5* \cite{Pastore13a}, 
the SLy5s$x$ \cite{Jodon16a}, and
1F2F(0.70), 1T2F(0.70), 1T2T(0.70) \cite{DaCosta24a} 
that all have an effective mass of $m^*_0/m \simeq 0.7$, see for 
example Refs.~\cite{Ryssens21a,Hellemans12a,Jodon16a,Ryssens19b}.

\begin{table}[t!]
\caption{\label{tab:3d:fits:d1s} 
Parameter sets \LOPP{}{D1S} of the LO pairing EDF \eqref{eq:EDF:pair:ULB}
obtained adjusting the pairing gaps in symmetric INM obtained for the 
underlying particle-hole Skyrme EDF as indicated that reproduce the pairing 
gaps from  D1S. The second column lists the isoscalar effective mass of the 
respective Skyrme parametrisation at the saturation point.
}
\begin{center}
\begin{tabular}{lllll}
\hline\noalign{\smallskip}
  & \multicolumn{1}{c}{$m^*_0/m$} 
  & \multicolumn{1}{c}{$V_0$} 
  & \multicolumn{1}{c}{$\eta$} 
  & \multicolumn{1}{c}{$\sigma$} \\
  & & \multicolumn{1}{c}{(MeV fm$^{3}$)}
  & & \\
\noalign{\smallskip}\hline\noalign{\smallskip}
1T2T(0.70) & 0.700 & $-1358.0$ & 0.783 & 0.379  \\  
1T2T(0.80) & 0.800 & $-1397.4$ & 0.823 & 0.361  \\   
1T2T(0.85) & 0.850 & $-1415.8$ & 0.839 & 0.353  \\  
SN2LO1     & 0.709 & $-1334.3$ & 0.783 & 0.400  \\  
\noalign{\smallskip}\hline
\end{tabular}
\end{center}
\end{table}

The connection between effective mass and pairing parameters becomes less 
direct when working with extended forms of the particle-hole EDF.
When complementing the pairing EDF with a dependence on the isovector density
and adjusting the pairing gap also for asymmetric matter, the resulting parameters 
governing the isovector dependence of the pairing EDF would be different as these 
parameter sets of the Skyrme EDF have different isovector effective masses.
Also, the above discussed feature is very specific to the widely-used standard 
form of the NLO Skyrme EDF and neither holds for extended NLO 
Skyrme parameter sets that are complemented with either 
density-dependencies of the gradient terms or three-body gradient 
terms, nor for parametrisations of Skyrme EDFs of higher order in gradients such as
SN2LO1 for which several terms contribute to $m^*_0/m$, see Eq.~\eqref{eq:ms:q}.
Still, the pairing parameters of SN2LO1 listed in Table~\ref{tab:3d:fits:sly4}
remain close to those of SLy4 and 1T2T(0.70) that have an almost identical effective
mass.


\subsection{Adjusting parameters to infinite matter pairing gaps from Gogny D1S}

\begin{figure}[t!]
\centerline{\includegraphics[width=7cm]{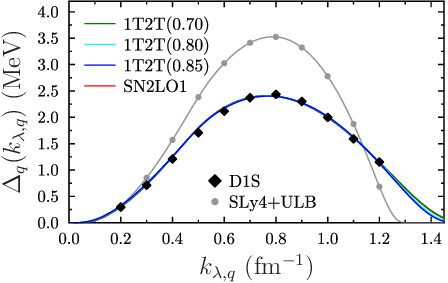}}
\caption{\label{fig:inm:gap:fits2d1s}
Gap $\Delta_q(k_{\lambda,q})$ at the chemical potential of protons and neutrons 
in symmetric INM for the parameter sets of the ph Skyrme EDF as indicated combined
with the respective \LOPP{}{D1S} pairing parameters as listed in Table~\ref{tab:3d:fits:d1s}. 
For most values of $k_{\lambda,q}$ the curves cannot be distinguished on the scale
of the plot.
The $\Delta_q(k_{\lambda,q})$ obtained with the original D1S parametrisation of the 
Gogny force taken from Ref.~\cite{Guillon24m} are represented by filled black diamonds. 
The grey line displays the gap obtained with SLy4+ULB for comparison.
}
\end{figure}

A second possibility for a reference calculation of pairing gaps 
for the adjustment of the parameters of the LO pairing EDF 
that we explored consists in taking results from HFB calculations
with the D1S parametrisation of Gogny's force~\cite{Berger91a}. Unlike 
the vast majority of parametrisations of the Skyrme EDF, parametrisations 
of the Gogny force are constructed to
simultaneously describe the ph and pp parts of the EDF, and D1S is a 
widely-used reference pairing interaction~\cite{Matsuo06a,Garrido99a,Kucharek89a} 
that reproduces well the odd-even staggering of masses~\cite{Dobaczewski15a,Robledo12a} 
and rotational moments of inertia~\cite{Dobaczewski15a}. 

The gaps in symmetric infinite matter obtained from D1S at a few selected 
values of the density have already been used before to constrain the parameters 
of the finite-range regularised pseudopotentials of Ref.~\cite{Bennaceur20a} 
that, like D1S, aim at a simultaneous description of the ph and pp channels 
of the effective interaction.

The parameters of the LO pairing EDF are again adjusted to values 
of $\Delta_q(k_{\lambda,q})$ in symmetric INM calculated at 11 
equidistant values for $k_{\lambda,q}$ with the code of Ref.~\cite{Guillon24m}.
Results of these fits are listed in Table~\ref{tab:3d:fits:d1s}, which 
will be referred to as \LOPP{}{D1S} parameter sets in what follows.

The INM gaps obtained with these parameter sets are compared with the 
original D1S results in Fig.~\ref{fig:inm:gap:fits2d1s}. 
For all \LOPP{}{D1S} parameter sets the $\Delta_q(k_{\lambda,q})$ 
from D1S are closely reproduced at all densities.
Figure~\ref{fig:inm:gap:fits2d1s} also displays the gaps from SLy4+ULB. 
Compared to these, the $k_{\lambda,q}$ dependence of
$\Delta_q(k_{\lambda,q})$ obtained with D1S peaks with a much smaller 
maximum value of about 2.45 instead of 3.55 MeV, but at the same time falls 
off at higher values of $k_{\lambda,q}$ and therefore also higher densities. 

Because of the different shape of the curve representing the D1S gaps in INM, 
it follows from the discussion in Sec.~\ref{sec:role:of:params} that
the parameters of the LO pairing EDF that reproduce the gaps from D1S 
are quite different from those of Table~\ref{tab:3d:fits:sly4} for 
the \LOPP{}{ULB} parameter sets.
First, the values of $\eta$ indicate that reproducing the D1S gaps necessitates
a more ``mixed'' character with about $20\%$ volume contribution, as could 
be expected from Fig.~\ref{fig:sn120:SLy4+ULB-mix} for a distribution 
of $\Delta_q(k_{\lambda,q})$ that reaches up to saturation density. 
Second, with $\sigma \simeq 0.4$, reproducing the D1S gaps requires
also a much more peaked density dependence than it is the case for SLy4+ULB, 
mainly to counterbalance the shift of the maximum of $\Delta_q(k_{\lambda,q})$ 
towards larger $k_{\lambda,q}$ that accompanies an increased volume character,
see Figs.~\ref{fig:sn120:SLy4+ULB-mix} and \ref{fig:gap:kl:SLy4+ULB-exp}.
If the \LOPP{}{D1S} parametrisations were of pure surface character as 
are the parameter sets used to prepare Fig.~\ref{fig:gap:kl:SLy4+ULB-exp}, they 
would actually produce a nonphysical transition to a BEC. As can be seen from 
Fig.~\ref{fig:inm:gap:fits2d1s}, however, this is not the case for any of 
the underlying mean fields.

In the next section, results obtained with the two sets of parametrisations
will be compared with data that are known to be sensitive to pairing 
correlations.

%
%
\section{Pairing correlations in finite nuclei}
\label{sec:paired:nuclei}

In one way or another, pairing correlations affect all properties of nuclei.
We will discuss here three observables that can be modelled with HFB 
calculations whose size is known to scale with the strength of the like-particle 
$T=1$ pairing correlations, all of which have also already been used for 
the parameter adjustment of pairing EDFs
\begin{enumerate}
\item
The odd-even staggering of nuclear masses, or, equivalently, the odd-even 
staggering of nuclear binding energies, which probes the effect of blocking 
a quasiparticle on the total energy. 

\item
The odd-even staggering of nuclear charge radii is sensitive to the change
of the potentials in the single-particle Hamiltonian when blocking a 
quasiparticle~\cite{Fayans94a,Fayans96a,Fayans00a,Fayans01a,Saperstein11a,Borzov22a,Reinhard17a,Reinhard24a}. 

\item
The moment of inertia of collective rotational bands probes 
the effect of superfluidity induced by pairing correlations. In general, 
superfluidity decreases the rotational moment of inertia and thereby increases 
in-band transition energies. 

\end{enumerate}
The odd-even staggering of masses and charge radii is analysed for 
the chains of spherical Sn and Pb isotopes and the chain of deformed
Yb isotopes. For even-even Yb isotopes, we also analyse the yrast
rotational states.

%
\subsection{Odd-even staggering of binding energies}
\label{sec:nuclei:stagger}


\subsubsection{Finite-difference pairing gaps}
\label{sec:oestagger:masses}

Odd-mass nuclei are systematically less bound than what can be inferred
from the average of the binding energies of their even-even neighbours. This
leads to a characteristic odd-even 
staggering of binding energies along isotopic and isotonic chains. 
The energy shift $\Delta_n(Z,N)$ between the energies of odd-mass 
nuclei and the curve that smoothly interpolates the total energy of 
even-even nuclei, denoted as $\mathsf{E}(Z,N)$, is associated with 
the pairing gap.
As we focus here on isotopic chains of elements with even proton number $Z$, 
for the sake of compact notation the proton number is suppressed in what follows. 
Approximations for this gap can be extracted from experimental and 
calculated total energies with finite-difference formulas~\cite{Jensen86a,Duguet01a,Bender00a,Jensen84a,Madland88a}. 
Provided that $\mathsf{E}(N)$ can be Taylor expanded around a 
given nucleus with $N = N_0$, the total energy $E(N)$ 
of the nucleus with $N$ neutrons is given by
\begin{align}
\label{eq:e:taylor}
E(N)
& = \sum_{n=0}^{\infty} \frac{1}{n!} \, 
    \frac{\partial^n \mathsf{E}(N) }{\partial N^n} \bigg|_{N_0} (N-N_0)^n 
    + D (N) 
\end{align}
where $D (N)$ is the contribution to the binding energy that is not 
smooth and therefore not captured by the derivatives, which in most 
cases is $D(N) = 0$ for even $N$ and $D (N) = \Delta_n(Z,N)$ for odd $N$. 
Starting from Eq.~\eqref{eq:e:taylor} one can construct finite difference-formulas 
that provide averages of $D(N)$. The simplest one is a three-point gap 
$\Delta_n^{(3)}(N_0)$ that equals an average of $D(N)$ of three adjacent
isotopes plus a residual mean-field contribution of size
$1/2 \times (-1)^{N_0} \, \partial^2 \mathsf{E}(N) / \partial N^2 \big|_{N_0}$
\cite{Duguet01a,Bender00a}. The latter equals half the derivative of the 
two-neutron separation energy with respect to $N$, which can be as large as a few 
hundreds of keV, and enters $\Delta_n^{(3)}(N)$ with opposite sign for odd and 
even nuclei.
It therefore has been argued in Refs.~\cite{Duguet01a,Bender00a}
that a controlled extraction of $\Delta_n(N)$ is obtained 
through the five-point difference
\begin{align}
\label{eq:Delta:5}
\Delta_n^{(5)}(N_0)
& \equiv - \frac{(-1)^{N_0}}{8} \Big[ E(N_0-2) - 4 E(N_0-1)
    \nn \\
&   \qquad 
     + 6 E(N_0) - 4 E(N_0+1) + E(N_0+2) \Big]
    \nn \\
& = - \frac{(-1)^{N_0}}{8} 
    \Big[ D (N_0-2) - 4 D (N_0-1) 
     \nn \\
&   \qquad 
     + 6 D(N_0) - 4 D(N_0+1) + D(N_0+2) \Big]
   \nn \\
& \quad   
    - \frac{(-1)^{N_0}}{8}
      \frac{\partial^4 \mathsf{E}(N)}{\partial N^4} \bigg|_{N_0}
    + \cdots \, .
\end{align}
For an odd-$N$ isotope, $\Delta_n^{(5)}(N)$ is the weighted average
of the gaps of three odd-$N$ isotopes plus the contribution from
the fourth and higher derivatives of the background, whereas for 
even-$N$ isotopes, $\Delta_n^{(5)}(N)$ is the average of the gaps 
in the two adjacent odd-mass isotopes minus the contribution from
the fourth and higher derivatives of the background. Between major 
shell closures, $\partial^4 \mathsf{E}(N)/\partial N^4$ is in 
general very small. By contrast, any abrupt structural change, such 
as a shell closure or a shape transition, can lead to a 
discontinuity of $\mathsf{E}(N)$. In such case, $D(N)$ has to 
absorb the additional binding from the discontinuity when 
matching $E(N)$ and the Taylor expansion \eqref{eq:e:taylor}.
This can generate large peaks in $\Delta_n^{(3)}$ and $\Delta_n^{(5)}$ 
that have nothing to do with pairing correlations \cite{Duguet01a,Bender00a}.


\subsubsection{Representative examples}
\label{sec:nuclei:stagger:examples}

As representative examples, we choose the chains of Sn, Yb, and Pb isotopes.
For these heavy elements, nu\-cle\-us-dependent beyond-mean-field effects from 
symmetry restoration that cannot be easily absorbed into the parameters of the 
effective interaction can be expected to be less important than for the lighter 
ones~\cite{Bender05a,Bender06a,Bender08a,Delaroche10a}. In addition, 
heavy elements have much longer isotopic chains, which in general helps to 
identify systematic effects.

\begin{figure}[t!]
\centerline{\includegraphics[width=7.5cm]{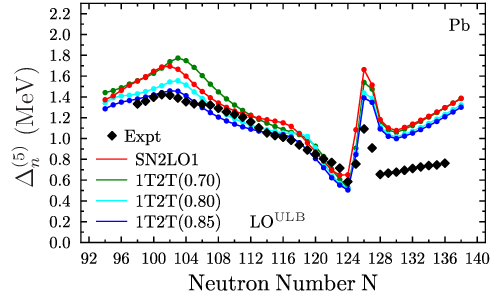}}
\centerline{\includegraphics[width=7.5cm]{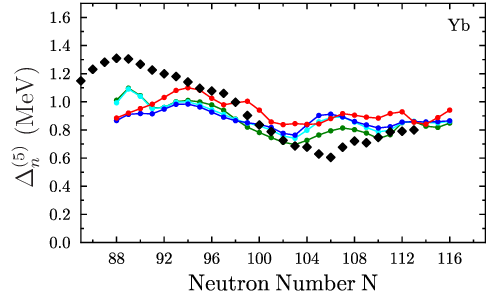}}
\centerline{\includegraphics[width=7.5cm]{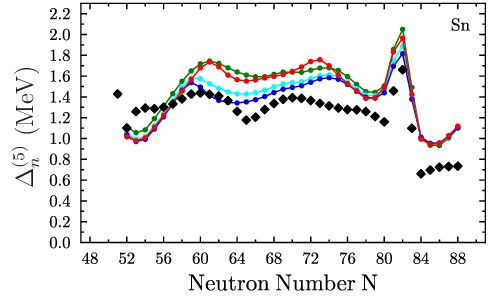}}
\caption{\label{fig:gaps:tosly4+ULB}
Five-point neutron pairing gap $\Delta_n^{(5)}$ for the Sn, Pb, and Yb chains 
calculated with the 1T2T(X) and SN2LO1 parameter sets and \LOPP{}{ULB} pairing parameters. 
}
\end{figure}

Results for the $\Delta_n^{(5)}$ for these three chains
obtained with the four Skyrme parametrisations and the
cor\-re\-spond\-ing \LOPP{}{ULB} pairing parameters  are shown in 
Fig.~\ref{fig:gaps:tosly4+ULB}. With a few exceptions
that will be addressed in what follows, all parameter sets give very similar
results in spite of their different effective mass, and, in case of SN2LO1,
also a different functional form. Indeed, for all nuclei results obtained with
SN2LO1 remain very similar to those from 1T2T(0.70) that has a nearly 
identical effective mass.

Up to the $N=126$ shell closure, the data for $\Delta_n^{(5)}$ of the Pb 
isotopes are quite well described, with some exceptions for the lightest isotopes
for which there are some visible differences for 1T2T(0.70) and SN2LO1. These 
deviations can partially be attributed to the different values of the effective 
mass. Indeed, the spherical $1h_{9/2^-}$ and $1j_{13/2^-}$ shells 
in the single-particle spectrum that are closest to the chemical potential
for these isotopes change their relative distances and their 
distance to the $2f_{7/2^-}$ below and the $3p_{3/2^-}$ above which affects
the amount of pairing correlations found in these nuclei. In particular,
for 1T2T(0.70) and SN2LO1 the $1h_{9/2^-}$ and $1j_{13/2^-}$ shells are
pulled even further down below the $1j_{13/2^-}$ and $3p_{3/2^-}$ 
than is already the case for 1T2T(0.80) in Fig.~\ref{fig:gbandheads:1t2t80:ULB}
with the $1h_{9/2^-}$ getting also much closer to the $2f_{7/2^-}$, whereas
for 1T2T(0.85) the changes are in opposite direction. It seems that with
decreasing effective mass, the single-particle spectrum of the light Pb
isotopes predicted by the four parameter sets used here become more
degraded. In fact, none of the parameter sets reproduces the change of ground state
spin from $9/2^-$ to $3/2^-$ at $N=101$; for 1T2T(0.85) it happens at $N=103$,
whereas SN2LO1, 1T2T(0.70), and 1T2T(0.80) predict that change for $N=105$.
Also, none of the parameter sets predicts the $13/2^+$ state to be 
near-degenerate with the $3/2^-$.

That the calculations overestimate the peak at $N=126$ indicates that 
the discontinuity in total binding energies of even-even nuclei is too
large for this shell closure. This is a consequence of the gap in the 
single-particle spectrum being too large for all parameter sets.

For the heaviest isotopes beyond $N=126$, all parameter sets overestimate
the experimental $\Delta_n^{(5)}$ by about 500~keV. This is a common feature
of many mean-field predictions for pairing gaps in neutron-rich nuclei 
beyond a spherical shell closure that is also observed for the Sn isotopes,
although for these the difference remains about 300~keV.

Interestingly, for the Sn isotopes the experimental data are about 400 keV larger
for the isotopes directly above $N=50$ than for those above $N=82$. The calculations
fall in between, predicting gaps of similar size in both cases that underestimate
data for the lightest isotopes and overestimate the data for the heaviest ones.
Around the neutron shell closures, the different parametrisations yield
very similar gaps, whereas there are now some visible differences
for the mid-shell isotopes. These can again at least partially be attributed 
to a mismatch between calculated and empirical single-particle spectra for
these isotopes.

For Sn isotopes, the relative distance of single-par\-ti\-cle states between
$N=50$ and $N=82$ is not correctly described, see 
Fig.~\ref{fig:gbandheads:1t2t80:ULB}. In particular, the 
$1h_{11/2}$ intruder shell is too high above the normal-parity states,
which is a common problem of most, if not all parametrisations of
nuclear EDF models.

The trough at $N=64$ in the experimental data is probably related to the 
complete filling of the $1g_{7/2^+}$ shell, that should be more isolated 
from the levels around than is the case in the calculations, see 
Fig.~\ref{fig:gbandheads:1t2t80:ULB}. Comparing the 1T2T(x), the
relative positions of the neutron single-particle levels around the chemical
potential are very similar, with the sole exception being the distance between 
the $1g_{7/2^+}$ and the $3s_{1/2^+}$ that grows with effective mass.
This explains why the evolution of the calculated $\Delta_n^{(5)}$ is similar 
for 1T2T(0.70) and SN2LO1, but slightly different for 1T2T(0.80) and 1T2T(0.85). On
the average, it is 1T2T(0.85) that best describes 
the size of $\Delta_n^{(5)}$ around $N=64$.

For the odd Sn isotopes with $64 < N < 82$, all parameter sets give the 
same ground-state spin, which, however, agrees with data only for 
$N = 65$, 67 ($1/2^+$), 71 ($3/2^+$) and 75, 77 ($11/2^-$), mainly 
as a consequence of the $2d_{3/2^+}$ and  $1h_{11/2^-}$
shells being too distant from each other, see Fig.~\ref{fig:gbandheads:1t2t80:ULB}.
For this reason, pairing correlations are probably too much dominated by 
the intruder $1h_{11/2^-}$ shell that becomes the only one with large 
matrix elements of the pairing tensor. Because of its surface-peaked
radial wave function, the pairing matrix elements within the 
spherical $1h_{11/2^-}$ shell are also larger than those of the other 
shells by about 400~keV, and take absolute values that are similar to 
the $\Delta_n^{(5)}$. This effect becomes more pronounced with 
decreasing effective mass of the parameter sets. We therefore
believe that the deviations of the $\Delta_n^{(5)}$ from data
for many of the Sn isotopes below \nuc{132}{Sn} are mainly a direct 
consequence of the incorrectly predicted shell structure, and
not a consequence of a very badly adjusted pairing interaction.
However, as indicated by the analysis of \ref{sec:sly4+ulb}, 
these deviations can be amplified by the so-called time-odd 
terms in the EDF whose size is often correlated with effective mass,
see also \cite{Sun25a}.

That the peak at $N=82$ is too large is again a consequence of the 
gap in the single-particle spectrum being too large at this 
shell closure, but to a lesser extent than for \nuc{208}{Pb}.
For the heavier isotopes for which there are data, all parameter 
sets predict that only the $2f_{7/2^-}$ shell contributes 
significantly to pairing correlations. The reason is that it 
is well separated from the levels below by the major $N=82$ shell
closure, and also from the $3p_{3/2^-}$ level above 
by a $N=90$ sub-shell closure of about  $1.8~\text{MeV}$, see again 
Fig.~\ref{fig:gbandheads:1t2t80:ULB}. The spectrum of excited states 
of \nuc{133}{Sn} suggests that the level density above the $N=82$
shell closure should be larger than what is predicted by the
models, but correcting for this might lead to even stronger
pairing correlations, such that we can only speculate about
the origin of this mismatch.

\begin{figure}[t!]
\centerline{\includegraphics[width=7.50cm]{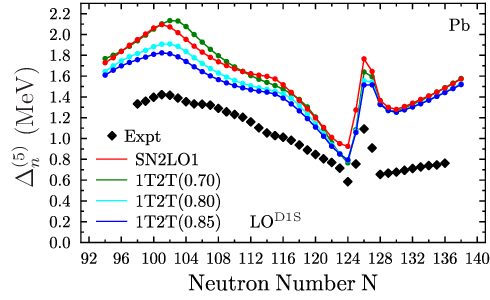}}
\centerline{\includegraphics[width=7.50cm]{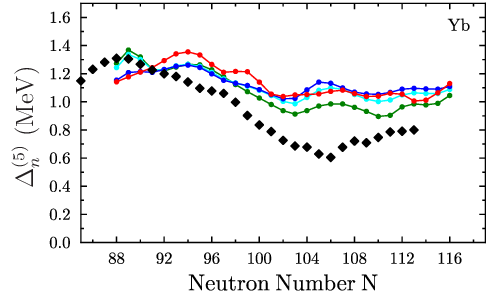}}
\centerline{\includegraphics[width=7.50cm]{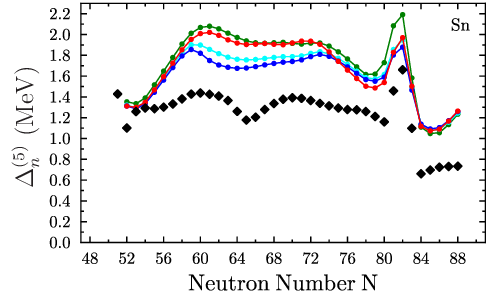}}
\caption{\label{fig:gaps:tod1s}
Same as Fig.~\ref{fig:gaps:tosly4+ULB}, but obtained using \LOPP{}{D1S} pairing parameters.
}
\end{figure}

For the deformed Yb isotopes, the global trend of the data 
for $\Delta_n^{(5)}$ is similarly well described by all
parameter sets, with only small differences between them. 
Similarly to what is found for the Sn isotopes, the experimental 
data are underestimated for neutron-deficient isotopes just above 
the spherical shell closure (which for Yb is $N=82$, though), 
but then well described for the heavier ones, with the exception
of the isotopes around $N=106$ for which the trough in the data 
is not reproduced and a local maximum at $N=104$ is predicted
instead by all parameter sets. The latter can probably 
be attributed to the much larger bunching of calculated levels 
above the $N=104$ gap than what is found in experiment,
see Fig.~\ref{fig:gbandheads:1t2t80:ULB}.
Similarly, the relative order of levels below the $N=104$ gap is not
correctly described, most obviously for the position of the $7/2^+$ 
intruder level that falls in between the $N=100$ and $N=104$ gaps 
predicted by the models. The size of matrix elements $\Delta_{\mu \nu}$
between time-reversed partner levels close to the chemical potential
can differ between the various single-particle states by up to about 
$30 \%$, so blocking the incorrect state can make a visible difference 
for the resulting energy loss.

During our analysis, we noticed that the predictive power of the pairing EDF
for the odd-even staggering of nuclear masses is significantly degraded 
when not keeping all coupling constants of the \textit{particle-hole} Skyrme EDF 
at the values obtained from an underlying Skyrme generator, as is for 
example done for the SLy4 parameter set. An analysis of this finding is given
in \ref{sec:sly4+ulb}.

Figure~\ref{fig:gaps:tod1s} displays the same as Fig.~\ref{fig:gaps:tosly4+ULB}, 
but for calculations with \LOPP{}{D1S} pairing parameters.
These pairing parameters yield much larger $\Delta_n^{(5)}$ than the \LOPP{}{ULB}, 
which in almost all cases largely overestimate the available data. When comparing the 
four ph parametrisations using \LOPP{}{D1S} pairing parameters among each other, 
the differences between them are very similar to what is found with the \LOPP{}{ULB} 
parameters, appearing for the same nuclei and having about the same size and sign.
As both sets of parametrisations yield a quite different spatial form 
factor in finite nuclei, this finding corroborates our earlier conjecture 
that the deviations between the pairing properties found for different
parametrisations of the ph EDF originate
mainly in differences of their single-particle spectra and time-odd terms.


\subsubsection{Why $\Delta_q(k_{\lambda})$ is not representative for the 
               pairing interaction from the Gogny force}
\label{sec:nuclei:problem:D1s}

Pairing matrix elements obtained with the Gogny force D1S are a widely 
used reference in the literature. The lacking performance of the 
\LOPP{}{D1S} pairing parameter sets adjusted to reproduce the 
Gogny D1S pairing gaps $\Delta_q({k_{\lambda,q}})$ in INM when applying 
them to finite nuclei therefore deserves an explanation, in 
particular when recalling that calculations with the original D1S
interaction in the particle-hole and particle-particle channel
describe these data well \cite{Dobaczewski15a,Robledo12a}.
There is apparently a feature of the original Gogny interaction that is 
lost when mapping it onto the LO pairing EDF through pairing gaps 
in infinite matter.

Let us note first that, in spite of the curves for the $\Delta_n^{(5)}$ 
in Fig.~\ref{fig:gaps:tod1s} suggesting that a re-scaling of the global 
pairing strength might bring calculations closer to experiment, we found 
that multiplying $V_0$ by a factor of 0.8 while keeping $\eta$ and 
$\sigma$ constant does not visibly improve the reproduction of the data
for $\Delta_n^{(5)}$.

\begin{figure}[t!]
\centerline{\includegraphics[width=8.0cm]{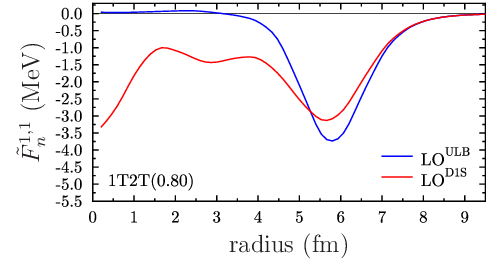}}
\caption{\label{fig:sn120:pairpottod1s}
Pairing potential $\tilde{F}^{1,1}_n(\vec{r})$ found for \nuc{120}{Sn} in calculations with
1T2T(0.80) and either the corresponding \LOPP{}{ULB} or \LOPP{}{D1S} pairing parameters.
}
\end{figure}

Considering that the \LOPP{}{D1S} gaps in INM are at almost all values 
of $k_\lambda$ significantly smaller than those from the \LOPP{}{ULB} 
parameters, see Fig.~\ref{fig:inm:gap:fits2d1s}, it is at first 
surprising why pairing gaps in finite nuclei are well described by the \LOPP{}{ULB} 
parameters, but overestimated by the \LOPP{}{D1S}. The difference is made 
by the small region of momenta with $k_{\lambda,q} \gtrsim 1.15 \, \text{fm}^{-1}$ 
for which the \LOPP{}{D1S} parameter sets still yield finite gaps in INM
while pairing is already almost broken down for the \LOPP{}{ULB}. These
values of $k_{\lambda,q}$ correspond to total densities that range from about 
$0.12~\text{fm}^{-3}$ to saturation density and slightly beyond, 
which are the densities that make most of the bulk volume of finite nuclei. 
When comparing the \LOPP{}{ULB} and \LOPP{}{D1S} pairing parameter sets, 
this then leads to very different pair potentials $\tilde{F}^{1,1}_n(\vec{r})$ 
in  finite nuclei.
This is illustrated by Fig.~\ref{fig:sn120:pairpottod1s} for calculations
of \nuc{120}{Sn} with 1T2T(0.80). \LOPP{}{ULB} pairing parameters lead to 
a surface-peaked $\tilde{F}^{1,1}_n(\vec{r})$ that is even slightly repulsive 
inside the nucleus, whereas \LOPP{}{D1S} pairing parameters yield both a strong 
peak at the surface and a large contribution in the volume, which then results
in too large matrix elements $\Delta_{\mu\nu}$ for \nuc{120}{Sn}. 
It is because of this mix of surface and volume features of the 
\LOPP{}{D1S} parameter sets that, as mentioned above, a simple moderate
rescaling of the overall strength $V_0$ would not significantly improve on the
description of the $\Delta_q^{(5)}$ of finite nuclei in Fig.~\ref{fig:gaps:tod1s}.

\begin{figure}[t!]
\centerline{\includegraphics[width=8.0cm]{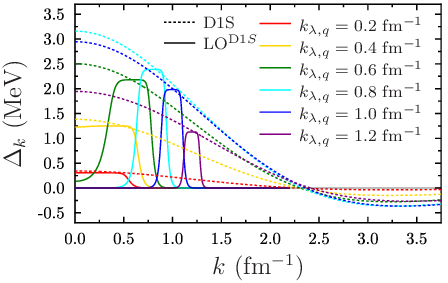}}
\caption{\label{fig:gap:k:tod1s}
Dotted lines: matrix elements $\Delta_k$ as a function of momentum $k$ 
from HFB calculations of symmetric INM with the D1S parametrisation
of the Gogny interaction taken from Ref.~\cite{Guillon24m} 
at values of $k_{\lambda,q}$ as indicated.
Solid lines: matrix elements $\Delta_k$ obtained from INM calculations 
at the same values of $k_{\lambda,q}$ performed with SLy4 and a LO pairing 
EDF whose parameters are adjusted to reproduce the gap $\Delta_q({k_{\lambda,q}})$ 
in symmetric matter predicted by the Gogny D1S interaction.
}
\end{figure}

This explains why the \LOPP{}{D1S} pairing parameters overestimate  
pairing gaps in finite nuclei, but does by itself not explain why our 
adjustment procedure for the \LOPP{}{D1S} pairing parameters yields
unrealistic pairing properties although it starts from a realistic pairing 
interaction. The deeper origin of the problem 
is that for a finite-range pairing interaction like the Gogny 
force the gap at the chemical potential $\Delta_q(k_{\lambda})$ 
is not fully representative for the pairing interaction in infinite 
matter. This can be seen when comparing the 
$k$-dependence of the gap $\Delta_k$ in INM of the original Gogny 
interaction with what is obtained using \LOPP{}{D1S}
parameters for SLy4, see Fig.~\ref{fig:gap:k:tod1s} for six 
different values of $k_{\lambda,q}$ that provide every other of the gaps used 
to adjust the \LOPP{}{D1S} pairing parameters to the gaps 
from D1S as listed in Table~\ref{tab:3d:fits:d1s}.

The gaps $\Delta_k$ obtained with D1S have a strong $k$ dependence and even 
change their sign at high momenta for $k \simeq 2.3~\text{fm}^{-1}$ before falling
off to zero at even higher values of $k$. The sign change has its origin in the
Gogny force being the combination of an attractive long-range part with 
a repulsive short-range part in the $^{1}S_{0}$ channel. 
The momenta at which the $\Delta_k$ become negligibly small depends 
on the shortest range of the interaction, which for D1S
is the one of the repulsive part in the pairing channel.
By contrast, as discussed in Sec.~\ref{sec:INM:SLy4+ULB},
a local LO pairing EDF leads to pairing matrix elements and gaps 
between all plane-wave states that in the absence of a cutoff are
simply constant. The $k$ dependence of the $\Delta_k$ from 
SLy4+\LOPP{}{D1S} visible in Fig.~\ref{fig:gap:k:tod1s} results 
from the energy cutoff \eqref{eq:Delta:inm}.

\begin{figure}[t!]
\centerline{\includegraphics[width=7.5cm]{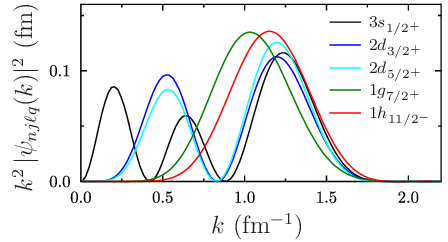}}
\caption{\label{fig:sn120:psi:k}
Square of the reduced radial momentum-space wave function $k \psi_{n j \ell q}(k)$
of the neutron levels around the chemical potential of \nuc{120}{Sn}
calculated with SLy4+ULB pairing. 
}
\end{figure}

There is, however, an important difference between how any energy cutoff acts 
in INM and in finite nuclei. Because of the one-to-one correspondence \eqref{eq:eps} 
between momentum $k$ and single-particle energy $\varepsilon_{\vec{k} \sigma q}$,
it is only in INM that an energy cutoff \eqref{eq:cutoff} also acts like a 
momentum cutoff, although with a density-dependent window size.
This is different for finite nuclei, for which the eigenstates of 
the single-particle Hamiltonian are a wave packet spread over a large 
range of momenta. As an example for the most relevant neutron 
single-particle states in the vicinity of the chemical potential 
$\lambda_n$ of the spherical nucleus \nuc{120}{Sn}, Fig.~\ref{fig:sn120:psi:k}
displays the square of the reduced radial momentum-space 
wave function $k^2 \, | \psi_{n j \ell q}(k)|^2$ obtained from 
the Fourier transformation of the full 3d wave function 
with isospin $q$, total angular momentum $j$ and orbital angular
momentum $\ell$
\begin{align}
\psi_{n j \ell q}(k)
& = \int_0^{L} \! \rmd r \, r^2 \, j_{\ell}(kr) \, \psi_{n j \ell q}(r) \, ,
\end{align}
where $j_\ell(kr)$ is a spherical Bessel function of 
first kind of order $\ell$, $L$ is the size of the radial 
numerical box and the wave function is normalised to
\begin{align}
4 \pi \int_0^{\infty} \! \rmd k \, k^2 \, \big| \psi_{n j \ell q}(k) \big|^2
& = 1 \, .
\end{align}
The quantity $k^2 \, | \psi_{n j \ell q}(k)|^2$ is proportional to the probability 
that a neutron in the given spherical shell takes the momentum $k$.

In finite nuclei the cutoff \eqref{eq:cutoff} of the LO EDFs multiplies the 
contribution from the entire single-particle state, irrespective of its 
decomposition into $k$ components. For \nuc{120}{Sn},
all of the five single-particle states displayed in Fig.~\ref{fig:sn120:psi:k} 
fall inside the pairing window of the LO pairing EDF as can be deduced from 
Fig.~\ref{fig:gbandheads:1t2t80:ULB}. For each of them, 
$k^2 \, | \psi_{n j \ell q}(k)|^2$ has a peak around 
$k \simeq 1.2 \, \text{fm}^{-1}$ 
and is spread up to $k \simeq 1.7 \, \text{fm}^{-1}$.

\begin{figure}[t!]
\centerline{\includegraphics[width=7.5cm]{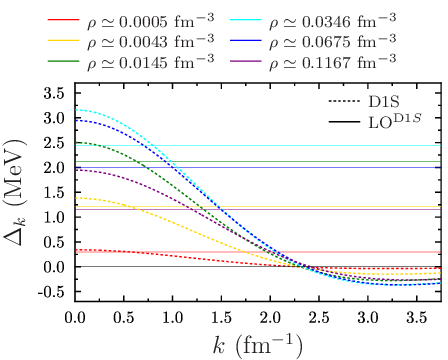}}
\caption{\label{fig:sn120:Delta:k}
Same as Fig.~\ref{fig:gap:k:tod1s}, but the $\Delta_k$ are now plotted as they
contribute for single-particle states inside the pairing window of a finite nucleus. 
For easier connection with finite nuclei, the curves are now labelled by the 
density $\rho$ in INM instead of the corresponding $k_{\lambda}$.
}
\end{figure}

The matrix elements $\Delta_k$ seen by these momentum-space wave
functions\footnote{High-$k$ components of the single-particle states in 
finite nuclei are still implicitly cut by the usual choices for the numerical representation, 
which for the calculations presented here is the highest momentum 
that can be represented on the coordinate-space Lagrange mesh. 
For the spacing of $\Delta x = 0.8~\text{fm}$ used for the 
3d calculations of finite nuclei, this implicit cutoff  
amounts to $k_{\text{max}} = \pi / \Delta x = 3.927~\text{fm}^{-1}$, 
which is outside the range of $k$ displayed in Figs.~\ref{fig:gap:k:tod1s},
\ref{fig:sn120:psi:k},
and~\ref{fig:sn120:Delta:k}.
} 
are therefore what is displayed in Fig.~\ref{fig:sn120:Delta:k}.
Unfortunately, because of the combined mo\-men\-tum- and density-dependence 
of the $\Delta_k$, the matrix elements $\Delta_{n j \ell q}$ of the pair 
potential for states within a given spherical $j$ shell in finite nuclei 
cannot be simply written as the integral over 
$k^2 \, \big| \psi_{n j \ell q}(k) \big|^2$ of Fig.~\ref{fig:sn120:psi:k}
times $\Delta_k$. Still, combining the information from 
Figs.~\ref{fig:sn120:psi:k} and \ref{fig:sn120:Delta:k} qualitatively
explains the very different performance of the Gogny force D1S and
the \LOPP{}{D1S} parameter sets for finite nuclei.

With the exception of the $3s_{1/2^+}$ shell, the dominant $k$ 
contributions to the single-particle states are at momenta at which the 
$\Delta_k$ from the \LOPP{}{D1S} parameters are significantly larger
than the $\Delta_k$ from the original Gogny force at all densities.
As a consequence, for most single-particle states 
in finite nuclei the \LOPP{}{D1S} parameters yield matrix elements 
$\Delta_{n j \ell q}$ of the pair potential that are significantly 
larger than what is obtained with the full Gogny force that the 
LO pairing EDF tries to mimic.
Owing to their structure with a single peak, this effect is most 
pronounced for high-$j$ levels that, because of their high degeneracy, 
provide the dominant contributions to neutron pairing correlations. 

By contrast, at small $k$, the $\Delta_k$ from the full Gogny force
are at most densities larger than the $\Delta_k$ from the \LOPP{}{D1S} 
parameters. For single-particle states that are dominated by 
small-$k$ contributions the opposite could happen, but this mostly 
concerns $s_{1/2^+}$ levels that, because of their small degeneracy, only 
have a minor contribution to pairing correlations in strongly-bound 
heavy nuclei. Still, this might be of relevance when studying halo nuclei
-- for which the very extended spatial wave functions have very large
low-$k$ components -- with pairing EDFs that are adjusted in the spirit 
of what is done for the \LOPP{}{D1S} parameter sets.

In summary, adjusting the parameters of a LO pairing EDF to the INM gaps 
$\Delta_q(k_{\lambda,q})$ obtained from the Gogny force is not sufficient 
to replicate the performance of the Gogny force for pairing gaps in 
finite nuclei. This explains why the \LOPP{}{D1S} parameter sets 
significantly overestimate the $\Delta_n^{(5)}$ displayed in 
Fig.~\ref{fig:gaps:tod1s}.

We expect that a similar mismatch will also be found when adjusting the 
parameters of an LO pairing EDF solely to the $\Delta_q(k_{\lambda,q})$ 
obtained from HFB calculations of INM with other finite-range pairing 
interactions, or when matching the $\Delta_q(k_{\lambda,q})$ from 
more microscopic methods.

%
%
\subsection{Odd-even staggering of charge radii}
\label{sec:nuclei:stagger:r}

Nuclear charge radii exhibit a pronounced odd-even staggering along 
isotopic chains, where in the vast majority of cases the charge radius 
of an odd isotope is smaller than the average of the charge radii of 
its even-even neighbours. There are a few known exceptions, which 
are usually attributed to either shape transitions or to the 
presence of exotic deformation modes. 
It has been pointed out 
\cite{Fayans94a,Fayans96a,Fayans00a,Fayans01a,Saperstein11a,Borzov22a,Reinhard17a,Reinhard24a} 
that in self-consistent mean-field calculations the contribution of a density-dependent 
pairing EDF to the single-particle Hamiltonian \eqref{eq:F11:paircontribution} 
can generate such an odd-even effect.
The mechanism is that for a pairing EDF of form~\eqref{eq:EDF:pair:ULB}, 
its contribution to $F^{1,1}_{q}(\vec{r})$ is proportional to the sum of the absolute 
squares of the proton and neutron pair densities
$\tilde{D}^{1,1}_q(\vec{r})$ and 
$\tilde{C}^{1,1}_q(\vec{r})$ whose overall size scales with the magnitude of pairing 
correlations. As pairing correlations in an odd nucleus  are in general weaker than in the 
adjacent even-even isotopes, the always repulsive pairing contribution to 
$F^{1,1}_{q}(\vec{r})$ is smaller in odd nuclei than in even-even nuclei.
With this, the single-particle energies of all bound neutrons \textit{and} protons 
are slightly lower in odd nuclei than in the adjacent even isotopes, meaning that
the single-particle states are slightly more localised inside odd nuclei.
This ultimately leads to a small systematic shift between the charge radii of the odd and even isotopes. 

The mean-square (ms) charge radii $r_{\text{ch}}^2$ are calculated taking into account
corrections for the internal charge distribution of protons and neutrons, relativistic
corrections from the distribution of magnetisation of protons and neutron 
and the Darwin-Foldy term, as well as a correction for the spurious centre-of-mass
motion of a nuclear mean-field state, leading to the expression 
\cite{Bertozzi72a,Friar75a,Reinhard21a,Bender25c}
\begin{align}
\label{eq:r2:bertozzilike}
r_{\text{ch}}^2
& = \frac{1}{Z} \, \int \! \rmd^3r \, r^2 \, \rho_{\text{ch}}(\vec{r}) \, ,
    \nn \\
& =   \frac{1}{Z} \int \! \rmd^3r \, r^2 \, \rho_p(\vec{r}) 
    + \langle r_{p}^2 \rangle
    + \frac{N}{Z} \, \langle r_{n}^2 \rangle
    \nonumber \\
& \quad
    - \frac{2 \mathcal{D} \, \mu'_p}{Z} 
      \int \! \rmd^3r \, r^2 \, \vnabla \cdot \vec{J}_p(\vec{r})
    \nonumber \\
    &\quad
    - \frac{2 \mathcal{D} \, \mu_n}{Z}  
      \int \! \rmd^3r \, r^2 \, \vnabla \cdot \vec{J}_n(\vec{r})
    + 3 \mathcal{D} 
    \nonumber \\
    &\quad
    - \frac{3 b^2}{2} A^{-1} \, .
\end{align}
where $\rho_p(\vec{r}) = D^{1,1}_p(\vec{r})$ is the point-proton distribution, 
$\vnabla \cdot \vec{J}_q(\vec{r}) = \vnabla \cdot \vec{C}^{1,\nabla \times \sigma}_q(\vec{r})$ 
the divergence of the spin-current vector density of protons ($q=p$) and neutrons ($q=n$),
$\langle r_{p}^2 \rangle = 0.8409 \, \text{fm}^2$ and
$\langle r_{n}^2 \rangle = -0.1155 \, \text{fm}^2$
the ms radii of the internal charge distribution of protons and neutrons,
$\mu'_p = \mu_p - 1/2 = 2.2928473446 \, \mu_N$ and 
$\mu_n = -1.9130427 \, \mu_N$
the anomalous contribution to the magnetic moment of protons and neutrons in units of
the nuclear magneton $\mu_N$, 
$\mathcal{D} = (\hbar^2/2 m_a)^2 = 0.011042234 \, \text{fm}^2 $ a constant 
entering the relativistic corrections, and finally
$b^2 = \hbar^2 / (41 \, \text{MeV} \, m_a \, A^{-1/3})$,
which equals $1.01148879 \, A^{1/3} \, \text{fm}^2$, 
enters the harmonic-os\-cil\-la\-tor estimate of the 
centre-of-mass correction, where $m_a = \tfrac{1}{2} (m_n + m_p)$ is the 
average mass of protons and neutrons (see Ref.~\cite{Bender25c} for details). 

The odd-even staggering of root-mean-square (rms) charge radii along an isotopic
chain can be estimated in analogy to the odd-even staggering of binding energies
discussed in Sec.~\ref{sec:oestagger:masses} through the three-point difference
\begin{align}
\label{eq:Delta:3:r}
& \Delta_{r}^{(3)}(N)
  \nn \\
& = - \frac{(-1)^{N}}{2} 
    \Big[ r_{\text{ch}}(N-1) - 2 r_{\text{ch}}(N) + r_{\text{ch}}(N+1) \Big]
\end{align}
that estimates the ``gap'' between the surface that interpolates
between the charge radii of odd-$N$ nuclei and the surface that 
interpolates between the charge radii of even-even nuclei. We use 
a convention for the sign of $\Delta_{r}^{(3)}$ such that positive 
values indicate normal staggering when the radii of even-even isotopes
are larger than the average of the radii of the two adjacent odd-$A$ nuclei.

\begin{figure}[t!]
\centerline{\includegraphics[width=7.5cm]{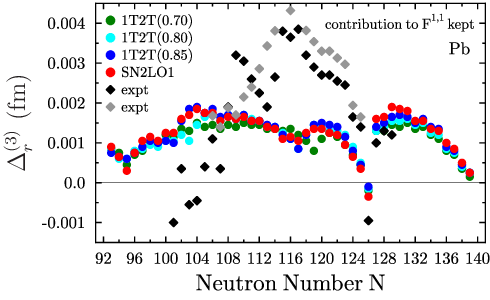}}
\centerline{\includegraphics[width=7.5cm]{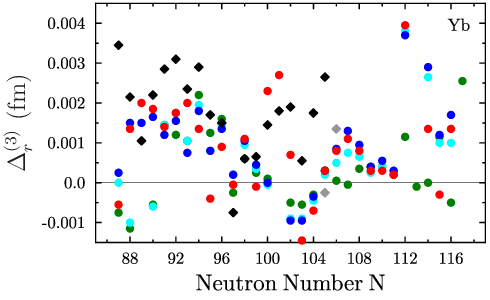}}
\centerline{\includegraphics[width=7.5cm]{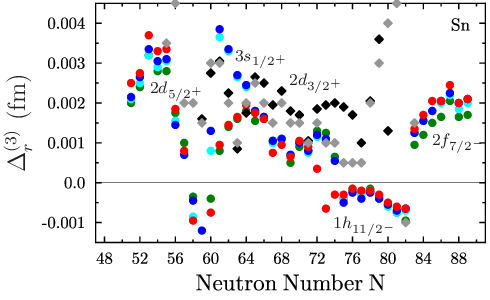}}
\caption{\label{fig:delta3r:tosly4+ULB}
Three-point gap $\Delta_{r}^{(3)}$ of the rms charge radii
of the Pb, Yb, and Sn isotopic chains calculated with the 1T2T(X) and SN2LO1 parameter 
sets and \LOPP{}{ULB} pairing parameters,
taking into account the contribution of the pairing EDF to $F^{1,1}_{q}(\vec{r})$.
Experimental data taken from different sources are plotted with either black or grey
diamonds (see text). Quantum numbers indicate the blocked ground-state configuration 
of Sn isotopes obtained with 1T2T(0.80). For the Yb isotopes with $N=89$ and 101, 
some calculated values are below the range of values shown, whereas for $N=113$ and 117 
some are above.
}
\end{figure}

\begin{figure}[t!]
\centerline{\includegraphics[width=7.5cm]{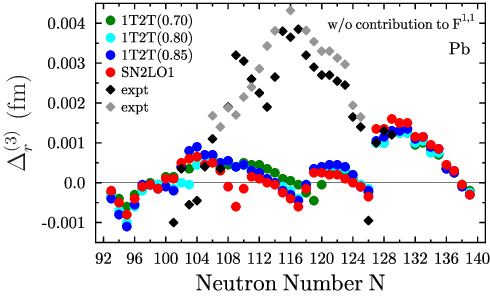}}
\centerline{\includegraphics[width=7.5cm]{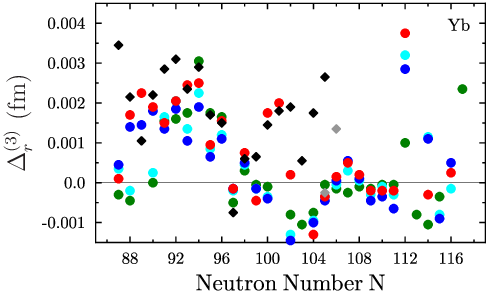}}
\centerline{\includegraphics[width=7.5cm]{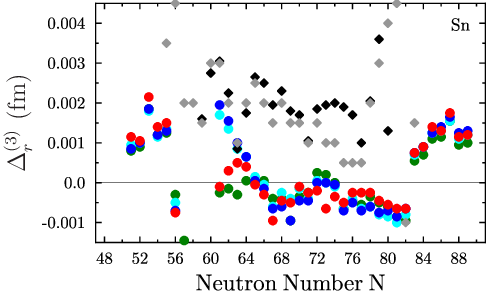}}
\caption{\label{fig:delta3r:tosly4+ULB:nopot}
Same as Fig.~\ref{fig:delta3r:tosly4+ULB}, but omitting the contribution of the 
pairing EDF to $F^{1,1}_{q}(\vec{r})$ in the calculations. For the Yb isotopes 
with $N=89$, 101, 103, 115 and 116, some calculated values are below the range of values 
shown, whereas for $N=113$ and 117 some are above.
}
\end{figure}

For radii, using the three-point expression \eqref{eq:Delta:3:r} instead of a 
higher-order formula is sufficient because -- contrary to the systematics of 
binding energies -- there is no significant contribution to the radii that 
scales with the square of the nucleon numbers. Some authors use an alternative
definition of $\Delta_{r}^{(3)}(N)$ \cite{Gustafsson25x,Takacs25x} without 
the alternating sign $(-1)^{N}$ that, for nuclei with normal staggering 
of radii, leads to positive values 
for even isotopes and negative values for odd ones. In our opinion, such 
convention makes the identification of systematic trends in the deviation 
from experiment unnecessarily difficult.

Figure~\ref{fig:delta3r:tosly4+ULB} displays results for $\Delta_{r}^{(3)}$ 
of the rms charge radii obtained from the same calculations used to prepare
Fig.~\ref{fig:gaps:tosly4+ULB}. For most isotopes in each of the three chains,
theory predicts normal odd-even staggering of charge radii, although with
some striking differences between the isotopic chains. Results obtained 
with the four parameter sets are overall very similar; in particular, 
there is no clearly identifiable impact of the effective mass, and results 
obtained with SN2LO1 are very similar to those obtained with the 
three NLO parameter sets.

Before comparing the model predictions with experiment, we analyse the 
effect of keeping or not the contribution \eqref{eq:F11:paircontribution} 
from the pairing EDF to the potential $F^{1,1}_q(\vec{r})$ that enters 
the single-particle Hamiltonian. To this aim, Fig.~\ref{fig:delta3r:tosly4+ULB:nopot}
displays results for $\Delta_{r}^{(3)}$ calculated in the same manner
as in Fig.~\ref{fig:delta3r:tosly4+ULB}, except that the contribution
\eqref{eq:F11:paircontribution}  
from the pairing EDF to the potential $F^{1,1}_q(\vec{r})$ is omitted when
solving the HFB equations.
In particular for the chains of spherical Sn and Pb isotopes, this makes 
a large systematic difference: $\Delta_{r}^{(3)}$ values calculated
without the pairing contribution to the potential $F^{1,1}_q(\vec{r})$ 
are systematically smaller than those obtained with the pairing 
contribution, in some cases even exhibiting inverted
odd-even staggering. For the deformed Yb isotopes, however, the differences
between both sets of calculations are more subtle. This is possibly a
degeneracy effect: for spherical nuclei, blocking a sub-state in a 
highly-degenerate $j$-shell yields a spatial change to $F^{1,1}_q(\vec{r})$ 
that has a large overlap with all other non-blocked orbits from the same
$j$-shell that have practically the same radial wave function, whereas in 
deformed nuclei there is only the conjugate state that has a similar orbit.

The odd-even staggering of rms charge radii is a small effect on the order
of a few per mille of the total size of $r_{\text{ch}}$.
The odd-even staggering of ms charge radii is clearly and systematically
visible in the measurements of the isotopic shift of ms charge radii, see for 
example Refs.~\cite{Anselment86a,deWitte07a,Renth24a} for Pb isotopes. 
After conversion of isotopic shifts to rms charge radii, however, the propagated
error bars of the rms charge radii usually take values of a few times 
$10^{-3}\, \text{fm}$, which is already comparable in size to the typical values
of $\Delta_{r}^{(3)}$, whose error bar would be even larger. As it is 
customary in most of the existing literature on the odd-even staggering of 
rms charge radii, we omit the error bars of experimental data in 
Figs.~\ref{fig:delta3r:tosly4+ULB} and \ref{fig:delta3r:tosly4+ULB:nopot}
that otherwise would be illegible.

A related issue is that experimental data taken from different sources 
also often disagree on the order of a few times $10^{-3} \, \text{fm}$. In 
Figs.~\ref{fig:delta3r:tosly4+ULB} and \ref{fig:delta3r:tosly4+ULB:nopot}, 
black diamonds indicate experimental values constructed from the rms charge 
radii tabulated in Ref.~\cite{Angeli13a}, whereas grey diamonds 
indicate data from more recent measurements of isotopic shifts. For Pb isotopes
with $N \simeq 112$, the $\Delta_{r}^{(3)}$ computed from the isotopic shifts 
reported in Ref.~\cite{Renth24a} have the opposite slope compared to those obtained 
from Ref.~\cite{Angeli13a}. Besides covering a wider range in $N$ than the data 
tabulated in Ref.~\cite{Angeli13a}, the recent data for Sn isotopes reported
in Ref.~\cite{Gustafsson25x} significantly differ from those of Ref.~\cite{Angeli13a}
for the isotopes just below the $N=82$ shell closure. Note, however, 
that data for rms charge radii reported in Ref.~\cite{Gustafsson25x} are limited to 
three digits after the decimal point, which consequently results in a 
``binning'' of the $\Delta_{r}^{(3)}$ values in discrete steps of $0.0005 \, \text{fm}$. 
Because of this, one cannot attribute a significance to any disagreement of theory 
-- or other experiments -- with the $\Delta_{r}^{(3)}$ from Ref.~\cite{Gustafsson25x} 
that remains smaller than that.
There also are new experimental data for the isotopic shifts of \nuc{175}{Yb}
and \nuc{177}{Yb} \cite{Flanagan12a,Li21a}. Isotope
\nuc{177}{Yb} was not included in 
Ref.~\cite{Angeli13a}, but the rms radius of \nuc{175}{Yb} resulting from 
this new measurement is significantly larger than the value tabulated 
there. This changes the experimental $\Delta_{r}^{(3)}$ 
value of \nuc{175}{Yb} from $+0.00265 \, \text{fm}$ when computed from 
data listed in Ref.~\cite{Angeli13a} to $-0.00025 \, \text{fm}$ when 
using the data from Ref.~\cite{Flanagan12a}. While the large difference
between experiments found for \nuc{175}{Yb} is most probably exceptional, all
of this nevertheless indicates that some of the experimental data for $\Delta_{r}^{(3)}$ 
might have larger uncertainties than usually acknowledged, such
that the comparison between theory and data has to be made with great caution.

It also has to be recalled that the charge radii of different 
configurations of a given odd-$A$ nucleus are different,
as is for example indicated by available data for low-lying spin-isomeric 
states in odd-mass Pb \cite{Anselment86a,deWitte07a,Renth24a} and 
Sn \cite{Yordanov20a} isotopes.
Such configuration dependence is also found in calculations, independent 
of including or not the contribution of the pairing EDF to the potential 
$F^{1,1}_q(\vec{r})$, Eq.~\eqref{eq:F11:paircontribution}. The absolute 
size of the charge radius of a given
configuration, and the differences between them, are of course 
different when including or not the contribution from the pairing 
EDF to the potential. An example where
this is clearly visible on Figs.~\ref{fig:delta3r:tosly4+ULB} and 
\ref{fig:delta3r:tosly4+ULB:nopot} are the $\Delta_{r}^{(3)}$ values
of Sn isotopes around $N = 61$, for which different parameter 
sets predict different ground-state configurations that have 
significantly different charge radii. 
Because of this sensitivity, it cannot
be expected that theory correctly predicts $\Delta_{r}^{(3)}$ values
unless it also predicts the correct ground-state spin. 
And as already illustrated in Sec.~\ref{sec:nuclei:stagger:examples}, the
theoretical predictions for the ground-state spin of odd nuclei
often do disagree with experiment. One could of course
block the same configuration as found in experiment, which for calculating 
total charge radii and their isotopic shifts is a valid strategy. But if 
the explanation of the origin of the odd-even staggering of charge radii 
sketched above is correct, in view of the smallness of this effect it is 
not obvious that blocking a single-particle level that has an incorrect
distance to the chemical potential will induce the correct rearrangement 
effects with changing potential depth when comparing odd and even-even
isotopes.

The mismatch in ground-state configuration between theory and experiment 
found for some nuclei in Fig.~\ref{fig:gbandheads:1t2t80:ULB}
therefore explains some of the large deviations found between calculated
and observed $\Delta_{r}^{(3)}$ values in Figs.~\ref{fig:delta3r:tosly4+ULB}
and \ref{fig:delta3r:tosly4+ULB:nopot}. For the odd-mass Pb isotopes between 
$105 \leq N \leq 117$, theory and experiment agree on the ground-state
spin $3/2^{-}$, but above and below they often disagree. For the 
heavier isotopes up to the $N=126$ shell closure, experiment
assigns $5/2^{-}$ for $119 \leq N \leq 123$, and $1/2^{-}$ only for $N=125$, 
whereas theory predicts 
$1/2^{-}$ ground states for all of these. Going to lighter isotopes, theory
predicts the transition to $9/2^{-}$ ground states at $N=103$, whereas
in experiment this is observed at $N=99$. The latter could be the reason 
why the inverted staggering at around $N \simeq 102$ is not reproduced
by theory, and why theory underestimates the staggering around 
$N \simeq 124$. But for the Pb isotopes around $N \simeq 110$ there
is a large systematic mismatch between theory and experiment for 
$\Delta_{r}^{(3)}$ despite the fact that the ground-state spin is 
correctly predicted.

For the Sn isotopes, the evolution of the calculated $\Delta_{r}^{(3)}$ 
with $N$ is much more discontinuous than what is found for the Pb isotopic chain.
Interestingly, almost all of these discontinuities occur where the 
ground-state configuration changes in the model description. 
With 1T2T(0.80), one finds $5/2^+$ up to $N=59$, then $1/2^+$ for $61 \leq N \leq 67$, 
$3/2^+$ for $69 \leq N \leq 73$, $11/2^-$ for $75 \leq N \leq 81$, 
and finally $7/2^-$ for $N$ above. With the exception of the $5/2^+$ 
configurations that yield an arch-like dependence of $\Delta_{r}^{(3)}$,
each of the other ground-state configurations leads to a distinctly different 
size of $\Delta^{(3)}_{r}$. Sometimes, the other three parameter sets 
predict the change of ground-state configuration to happen for different 
isotopes, in which case there are large differences between the calculated 
$\Delta_{r}^{(3)}$ from different parameter sets. In particular, 
the other parameter sets predict that either \nuc{109}{Sn} (with SN2LO1) 
or \nuc{111}{Sn} (with 1T2T(0.70) and 1T2T(0.85)) has a $7/2^+$ ground state, 
which explains the large deviations between models around $N = 60$.

For the Yb isotopes, the calculations roughly follow the 
global trend of the experimental data for $\Delta_{r}^{(3)}$, 
but again with many differences in detail. 
Because of the ground-state configuration being deformed,
all sin\-gle-particle levels in the fully paired states are just 
twofold degenerate. The observed ground-state spin 
of the blocked configurations changes each time when going 
to the next odd-mass isotope, whereas in the model calculations
there are a few cases where two consecutive odd-mass isotopes 
are predicted to have the same blocked configuration. 
As already discussed in Sec.~\ref{sec:nuclei:stagger:examples}, 
neither the ground-state spin nor the spectrum of
low-lying 1qp states of the Yb isotopes is always correctly predicted. 
This can be expected to have an impact on the predictions for 
$\Delta_{r}^{(3)}$, but in not entirely foreseeable ways.
The largest deviations between theory and experiment 
are found for the isotopes around $N \simeq 88$ and $N \simeq 100$.

In spite of the uncertainties of the data and the many differences found
between data and experiment, the \LOPP{}{ULB} parameter sets
yield an odd-even staggering of charge radii of reasonable size, 
in particular for Sn and Yb isotopes, if, and only if, the contribution 
of the pairing EDF to $F^{1,1}_q(\vec{r})$ is included.
However, the large discrepancy found for Pb isotopes with $112 \leq N \leq 122$
-- which is the region for which the odd-even staggering of masses is very
well described -- either points to the need for additional terms in the pairing 
EDF (such as the gradient terms used in 
Refs.~\cite{Fayans96a,Fayans00a,Fayans01a,Saperstein11a,Borzov22a,Reinhard17a,Reinhard24a})
or to missing microscopic physics in the model, or to an ill-controlled aspect of the 
particle-hole part of the EDF. A candidate for the latter are the  
time-odd terms of the ph EDF. As explained in \ref{sec:sly4+ulb}, some of them 
have a sizeable contribution to the odd-even staggering of binding energies,
and thereby can also introduce an odd-even effect to the eigenvalues
of the single-particle Hamiltonian that then feeds back onto the localisation 
of the spatial wave functions.

We note in passing that mean-field models that employ an extended 
pairing EDF with an additional dependence on the square of the gradient 
of the normal density and that are fine-tuned to describe the isotopic shifts 
and their odd-even staggering of Ca isotopes~\cite{Reinhard17a,Reinhard24a} 
largely overestimate the available data for the odd-even staggering of the 
charge radii of many spherical Sn~\cite{Gustafsson25x} and Pb 
isotopes \cite{Renth24a}, and even more so those for the deformed 
Yb isotopes \cite{Takacs25x}. This possibly indicates that the mechanism 
that drives the evolution of charge radii in the 
$pf$-shell region is different from the mechanism behind the smaller 
and more regular odd-even staggering found for most heavy nuclei.

We will not detail results for $\Delta_{r}^{(3)}$ obtained with 
the \LOPP{}{D1S} parametrisations. From the discussion in 
Sec.~\ref{sec:nuclei:stagger} it cannot be expected that they well
describe the pair densities that are at the origin of the variation
of $F^{1,1}_q(\vec{r})$ between odd and even isotopes.
Overall, the \LOPP{}{D1S} parameters also predict mostly positive 
values of $\Delta_{r}^{(3)}$ of Sb, Pb, and Yb isotopes
when including the contribution of the pairing
EDF to the single-particle potential $F^{1,1}(\vec{r})$. The actual
values for $\Delta_{r}^{(3)}$, however, tend to be somewhat smaller 
than what is found with the \LOPP{}{ULB}.

%
%
\subsection{Rotational moment of inertia}
\label{sec:nuclei:rot}

Shortly after the successful explanation of superconductivity in 
solids with the BCS model, it was pointed out that the presence of 
pair correlations could strongly affect the rotational moments 
of inertia of atomic nuclei, which to a large extent resolved the
discrepancy between earlier models and available data 
\cite{Belyaev59a,Nilsson61a}.

A widely-used approach to model nuclear collective rotation 
microscopically is to determine the energy $E(J)$ of rotating 
states with angular momentum $J$ in a 
self-consistent cranked HFB calculation\footnote{The fully 
variational cranked HFB method is not to be confused 
with the so-called ``cranking approximation'' \cite{Ring80a,Brink05a,Belyaev59a,Nilsson61a} 
for the calculation of moments of inertia. The latter is perturbative and 
neglects the nuclear response to rotational motion, in particular 
from the time-odd terms in the EDF. The moment of inertia from cranked HFB 
(often in the limit of small $\omega_c$) is also known as the Thouless-Valatin 
\cite{Ring80a} moment of inertia. We are, however, interested in the 
evolution of the moment of inertia along the rotational band.
} 
that minimises the Routhian  $E^{\omega}$~\cite{Ring80a,deVoigt83a,Ragnarsson95a}
\begin{equation}
\label{eq:routian}
E^{\omega}
= E(J) - \omega_c \langle \hat{J}_z \rangle
\end{equation}
that represents the energy of the nucleus in a frame rotating with the 
angular frequency $\omega_c$. Like in the majority of applications of 
this model, the symmetries imposed in the \textsf{MOCCa} code used 
here imply principal-axis cranking, where the direction of the angular 
momentum vector is aligned with a major axis of a triaxial nuclear shape. 
For sake of compact notation, throughout this paper it is assumed that 
angular momenta are in units of $\hbar$.

In many implementations of cranked HFB, $E^{\omega}$ and $\langle \hat{J}_z \rangle$ 
are obtained from minimising $E^{\omega}$ for a set of systematically varied values 
of $\omega_c$. In our calculations, however, the Lagrange parameter $\omega_c$ is 
adjusted to reproduce prescribed values of 
$\langle \hat{J}_z \rangle = \sqrt{J(J+1)}$ that can be associated with the 
energy levels at angular momentum $J$ in the rotational band. With this, 
$\gamma$-ray energies and moments of inertia can be calculated in exactly the 
same way as in the evaluation of experimental data.

For a schematic axially-symmetric rigid rotor with 
constant moment of inertia $\Theta$ and energy spectrum 
$E(J) = J(J+1)/2\Theta$, the $\gamma$ ray transition 
energies  between the levels at $J$ and $J-2$ 
\begin{align}
\label{eq:Egamma}
E_\gamma(J)
& = E(J) - E(J-2)
\end{align}
would depend linearly on $J$, $E_\gamma(J) = (4J - 2 )/2\Theta$. 
In the cranked HFB description of nuclear rotation, however,
the moment of inertia changes with increasing $J$. The reason is that the 
expectation value of the
many-body angular momentum $\hat{J}_z$ entering the Routhian~\eqref{eq:routian}
is generated microscopically from the one-body current 
$\vec{j}_0(\vec{r})$ and spin density $\vec{s}_0(\vec{r})$
distributions
\begin{align}
\langle \hat{J}_z \rangle
& = \int \! \rmd^3r \, \vec{e}_z \cdot
    \big[ \vec{r} \times \vec{j}_0(\vec{r}) 
          + \tfrac{1}{2} \vec{s}_0(\vec{r}) \big] \, .
\end{align}
To obtain a given finite value of $\langle \hat{J}_z \rangle$, the angular
momenta of the single-particle states progressively align with the axis of 
collective rotation. Because the angular momenta of the two nucleons in a conjugate 
pair of the $J=0$ ground state point into opposite directions, one of them tends
to align more with the rotational axis than the other. The resulting mismatch 
between the spatial orbits of the conjugate states generates the time-odd 
densities  $\vec{j}_0(\vec{r})$ and $\vec{s}_0(\vec{r})$, and in general also 
weakens pairing correlations \cite{deVoigt83a,Banerjee73a,Bengtsson79a}. 
In specific cases, including some of the Yb isotopes discussed here, 
the single-particle energies of the two nucleons in some of the pairs 
changes so quickly in opposite direction with $J$ such that a multi-quasiparticle 
configuration is found as the energetically lowest (``yrast'') state when solving
the cranked HFB equations. At such transition, the moment of inertia often 
dramatically changes, which produces a so-called ``backbending'' when plotting 
the moment of inertia as a function of $\omega$. The same is found when plotting 
the moments of inertia deduced from the observed transitions between yrast 
states~\cite{deVoigt83a,Banerjee73a,Bengtsson79a}.

\begin{figure}[t!]
\centerline{\includegraphics[width=7.0cm]{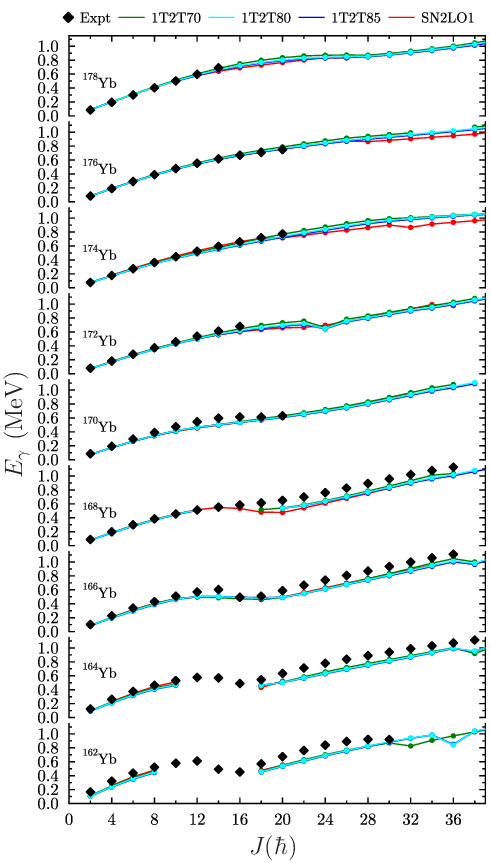}}
\caption{\label{fig:Egamma:tosly4+ulb}
$\gamma$-ray transition energies $E_\gamma$ of yrast states of the even-even 
Yb isotopes as indicated calculated with the 1T2T(X) and SN2LO1 parameter sets 
and \LOPP{}{ULB} pairing parameters (see text). Note that for many states and nuclei, 
results obtained with the four parametrisations cannot be distinguished on the scale 
of the figure.
}
\end{figure}

In cranked HFB, the nucleus' response to rotation is also
affected by the size of the often not very well controlled time-odd terms 
in the EDF, which can lead to significant differences between the
moments of inertia predicted by different parameter sets 
\cite{Hellemans12a,Sun25a,Dudek95a}.

For the purpose of probing the predictive power of the pairing 
interaction, we found it however more straightforward to analyse 
$E_\gamma$ as a function of angular momentum $J$ instead of the 
more frequently discussed kinematical 
or dynamical moments of inertia that highlight effects from 
the alignment of single-particle orbits
with the rotational axis. When plotting $E_\gamma$ as a 
function of $J$, as done in Fig.~\ref{fig:Egamma:tosly4+ulb}, 
a backbend is signalled by a decrease of $E_\gamma$ with $J$, i.e.\ a
substantial increase of the collective moment of inertia.
Because of the coexistence of two different solutions of the cranked HFB 
equations \eqref{eq:routian} with same $\omega_c$ at a backbend, fully 
self-consistent HFB calculations do not always converge for angular 
momenta for which $E_\gamma$ decreases even when fixing the 
expectation value of $\hat{J}_z$, which explains the absence 
of some calculated points.

Note that in the analysis of rotational bands, the value of $E_\gamma$ can
be associated with the empirical rotational frequency $\omega(J-1) = E_\gamma(J)/2$
\cite{deVoigt83a}, which in general takes a similar (albeit not identical) 
value as the Lagrange parameter $\omega_c$ in Eq.~\eqref{eq:routian}.

Results from cranked HFB calculations for the yrast band of even-even Yb isotopes 
obtained with the \LOPP{}{ULB} pairing parameters are displayed in 
Fig.~\ref{fig:Egamma:tosly4+ulb}. We limited the range of isotopes to those 
for which data are available and whose calculated deformation energy surface has 
a pronounced minimum at large deformation.

The angular momenta at which the backbendings are found in experiment
are in general fairly well reproduced by the calculations, which implies 
that, in spite of the many problems with the relative distances between 
single-particle levels pointed out above, the position of the 
quickly aligning levels relative to the chemical potentials is 
quite well described by all four parameter sets.

For most Yb isotopes, calculations closely follow data for $E_\gamma$ 
up to the backbend, but then tend to underestimate the values of $E_\gamma$ 
above, an exception being the lightest isotopes \nuc{162-164}{Yb}, 
for which the $E_\gamma$ are somewhat underestimated at all $J$. 
Also, in almost all cases, the results from all four parameter sets nearly
fall on top of each other, notable exceptions being high-spin states of 
\nuc{174}{Yb} and \nuc{176}{Yb} for which SN2LO1 predicts slightly smaller $E_\gamma$
for angular momenta for which there are no data available yet. Still, compared to the 
many small differences found between the parameter sets for the $\Delta_n^{(5)}$ and 
$\Delta_{r}^{(3)}$, the similarity of results for $E_\gamma$ is quite striking.
The \LOPP{}{ULB} parameter sets quite satisfactorily describe pairing correlations 
for these nuclei, irrespective of the effective mass of the parametrisation, or 
the detailed form of the EDF.

We will not discuss results of $E_\gamma$ obtained with the \LOPP{}{D1S} 
parametrisations in detail. What is found is that the $E_\gamma$ values 
are systematically slightly larger than those obtained with the \LOPP{}{ULB},
as is expected because of the overall stronger pairing interactions.
For the \LOPP{}{D1S} parameter sets all calculated 
curves also fall almost perfectly on top of
each other when plotting these results on the same scale as in
Fig.~\ref{fig:Egamma:tosly4+ulb}.

%
\section{Summary, conclusions, and outlook}
\label{sec:summary}

Motivated by the well-known observation that the adjustment of the parameters 
of the pairing interaction is compromised by insufficiencies of the single-particle
spectra predicted by even the best available parametrisations of the particle-hole
EDF, we explored the possibility to use pairing properties of the 
idealised model system of infinite matter for a more controlled parameter 
adjustment of effective pairing interactions that accompany a given 
paramerisation of the mean fields.

To that aim, we implemented a solver for the HFB equations
of infinite homogeneous non-polarised nuclear matter at arbitrary 
isospin asymmetry that can handle very general local Skyrme EDFs in the
ph and pp channels.

Limiting ourselves for the present study to the widely-used density-dependent 
LO local pairing EDF, we studied the correlation between the pairing gap in INM 
and pairing properties of finite nuclei. As reference calculations for the 
gaps in INM we utilised two time-tested pairing interactions. One is the
``ULB'' parametrisation of the LO local pairing EDF, 
originally adjusted for the SLy4 parametrisation of the
Skyrme NLO EDF and also widely used since with other parametrisations 
with the same isoscalar effective mass $m^*_0/m = 0.7$. The other is the 
D1S parametrisation of the finite-range Gogny interaction. 

To cover a wide range of underlying mean-fields, we adjusted pairing parameters
for three parametrisations of the NLO Skyrme EDF that differ in their
effective mass and for one parametrisation of the N2LO Skyrme EDF that
yields a more intricate momentum dependence of the mean fields than 
the standard NLO parameter sets.
For each of these, the parameters of the LO local pairing EDF were
adjusted to a set of gaps $\Delta_q(k_{\lambda,q})$ at the chemical
potential that covers a wide range of densities of symmetric infinite 
nuclear matter and were obtained with either of the two reference 
calculations. The main observations of our study are 
\begin{enumerate}
\item
The pairing gap in INM can be used as a proxy to transfer the
global performance of a given combination of Skyrme and LO pairing parameter 
sets for finite nuclei to a Skyrme parameter set with different 
effective mass. This is most useful when generating families of
Skyrme parametrisations that have systematically varied effective
mass, or that are of different order in gradients,
in particular in view of the robustness and numerical 
simplicity of HFB calculations of INM.

\item
Using our adjustment strategy, global trends of the odd-even 
staggering of masses and rotational moments of inertia are described 
with very similar quality across parameter sets. Nonetheless,
there remain local differences between the performance of the 
such adjusted combined Skyrme and pairing parametrisations for 
specific nuclei. These can be attributed to deficiencies
in the description of single-particle spectra that are differently 
pronounced for each of the four parameter sets used here, and also to 
differences in the size of time-odd terms in the ph part of the EDF.
Because of the interdependencies between the parameters of the
Skyrme ph EDF, for standard NLO EDFs some of the latter 
are correlated to the effective mass. For an improved modelling of the 
odd-even staggering of nuclear masses and rotational moments of inertia 
it is therefore not entirely sufficient to fine-tune the effective 
pairing interacting, but  also to better control shell structure and 
time-odd terms as predicted by the ph part of the EDF.

\item
By contrast, when attempting to replicate the pairing properties of a 
finite-range interaction with a local pairing EDF, it is not 
sufficient to reproduce the density dependence of the pairing gap
$\Delta_q(k_{\lambda,q})$ at the chemical potential of INM. The main reason 
is that the matrix elements $\Delta_k$ from finite-range interactions 
have a strong momentum dependence that is not captured by the gap 
at $k_{\lambda,q}$. For the specific case of pairing matrix elements
from Gogny D1S, the mapping onto a LO pairing EDF yields much stronger
pairing correlations than the original Gogny force, which then 
overestimates pairing gaps in finite nuclei.

\item
A similar mismatch for the pairing properties of finite nuclei 
can be expected when trying to map the gaps at the chemical potential 
obtained from more microscopic calculations. 
In either case, a local pairing EDF with higher-order terms in 
gradients will be needed to simultaneously map the gap in INM and 
describe finite nuclei. Going to NLO in derivatives would add two new degrees 
of freedom that will help in this respect: pair-kinetic terms that lead 
to a $k$-dependence of the gap in INM, and terms with gradients of 
densities that are only active in inhomogeneous systems.

\item
To reproduce a given infinite-matter reference calculation for $\Delta_q(k_{\lambda,q})$ 
for ph interactions with different effective mass, one has in general to 
readjust all of the three parameters of the LO pairing EDF that control 
strength, volume-to-surface ratio, and power of the density dependence.

\item
For parametrisations of the pairing EDFs with very small values for the power 
of the density dependence we find a nonphysical transition to a 
Bose-Einstein condensate of di-nucleons at low density in INM. This signals a 
spurious instability of the effective interaction that leads to
unrealistic results for finite nuclei. The same instability can also
be found for more complex pairing interactions \cite{Guillon24m}. The 
conditions for its appearance will be analysed elsewhere.
\end{enumerate}
Besides the practical points already mentioned above, there are also 
some more conceptual issues that need to be clarified by future work.
These concern
\begin{enumerate}
\item
The transfer of the predictive power for pairing properties of
finite nuclei between different \textit{types} of pairing EDFs. 
Until there is a satisfying answer, our procedure for the parameter 
adjustment relies on having a predictive parametrisation for each 
given type of pairing EDF to start with. 

\item 
The generalisation of the procedure proposed here to 
more general local pairing EDFs with kinetic 
and gradient terms. Such EDFs generate a momentum
dependence of the $\Delta_k$, but with reasonable parametrisations 
their variations are much smaller than what 
is found for finite-range interactions.

\item
The strategy to set up reference calculations for the pairing gap in INM.
There is no reason to believe that the gaps from the phenomenological 
SLy4+ULB parametrisation of the pairing EDF that we successfully 
employ as a reference is an optimal one. Our observation that
the momentum dependence of the gaps $\Delta_k$ in INM has
a significant impact on the pairing properties of nuclei can be 
expected to also  apply to calculations of the pairing gap in 
nuclear matter from first principles. In fact, when solely aiming at
the best effective modelling of pairing correlations in finite nuclei
at the HFB level, in view of the different correlation modes of INM and
finite nuclei, it is even not evident that a reference HFB calculation
of INM has to give a realistic microscopic description of pairing gaps 
in INM itself. 

\end{enumerate}
In spite of these remaining questions, the strategy to transfer
the pairing properties of SLy4+ULB to other 
Skyrme parametrisations proposed here has already been successfully used 
for the description of nuclear phenomena across the chart of
nuclei, including also Skyrme parameter sets not considered
in the analysis presented here.


\section*{Acknowledgements}

Illuminating discussions on various aspects of nuclear pairing that we had over 
the years with Jacek Do\-ba\-czew\-ski,
Paul-Henri Heenen, Jacques Meyer, and Wouter Ryssens
that inspired part of this work are gratefully acknowledged. 
The computations were performed using HPC resources from the CC-IN2P3 of the CNRS.


\begin{appendix}

%
\section{Solving the HFB equations for homogeneous infinite nuclear matter}
\label{sec:HFB:INM}

In this appendix, we sketch how the usual representation of general HFB theory 
as discussed in \cite{Ring80a,Bender19a} simplifies for the case of $^{1}S_{0}$ 
pairing correlations in homogeneous infinite nuclear matter. 

In INM, the spatial part of a nucleons' single-particle wave function 
is an eigenstate of momentum, i.e.\ a plane wave.
For $^{1}S_{0}$ pairing correlations, the Bogoliubov 
transformation then only mixes single-particle states with $\vec{k}$ and 
spin $\sigma$ and their time-reversed states with $-\vec{k}$ and spin 
$-\sigma$ (as usual ``spin'' will be used synonymously 
for spin projection).
With this, the single-particle states remain eigenstates of 
the isospin operator $\hat{\tau}_{3}$ and can be chosen to be eigenstates
of $\hat{\sigma}_z$. As a consequence, the full spinors representing single-particle
states in INM can be chosen as
\begin{align}
\label{eq:plane:wave:0}
\Psi_{\vec{k}, \sigma, q}(\vec{r})
& = \langle \vec{r} | a^\dagger_{\vec{k} \sigma q} | - \rangle
  = \psi_{\vec{k} \sigma q} (\vec{r}) \, \chi_\sigma \, \xi_q
    \nn \\
& = \frac{1}{(2\pi)^{3/2}} \, e^{+\iunit \vec{k} \cdot \vec{r}} \, 
    \chi_\sigma \, \xi_q
    \, ,
\end{align}
where we distinguish between the full spinor $\Psi_{\vec{k} \sigma q}(\vec{r})$
in spin-isospin space and the spatial wave function of its spin-isospin 
components $\psi_{\vec{k} \sigma q} (\vec{r})$. 
The $\chi_\sigma$ and $\xi_q$ are unit spinors in spin- and isospin-$1/2$
spaces, respectively
\begin{alignat}{3}
\chi_+
& = \twospinor{1}{0}_{\sigma} \, \, \quad 
\chi_-
& = & \twospinor{0}{1}_{\sigma} \, , \quad
    \\   
\xi_n
& = \twospinor{1}{0}_{\tau} \, \, \quad 
\xi_p
& = & \twospinor{0}{1}_{\tau} \, , 
\end{alignat}
with 
\begin{align}
\hat{\sigma}_z \, \chi_\sigma
& = \sigma \, \chi_\sigma \, , \quad 
\hat{\vsigma}^2 \, \chi_\sigma
  = 3 \, \chi_\sigma \, , \\
\hat{\tau}_3 \, \xi_n
& = + \xi_n \, , \quad 
\hat{\tau}_3 \, \xi_p
  = - \xi_p \, , \quad 
\hat{\tau}^2 \, \xi_q
  = 3 \, \xi_q \, . 
\end{align}
The basis states are pairwise connected by the time-reversal operator
$\hat{T} = -\iunit \sigma_y \hat{K}$ \cite{MessiahII}
\begin{align}
\label{eq:t-reversal}
\hat{T} \, \Psi_{\vec{k}, \sigma, q}(\vec{r})
& = \hat{T} \, \psi_{\vec{k}, \sigma, q} (\vec{r}) \, \chi_\sigma \, \xi_q
    \nn \\
& = \sigma \, \psi^*_{\vec{k}, -\sigma, q}(\vec{r}) \, \chi_{-\sigma} \, \xi_q
    \nn \\
& = \sigma \, \Psi_{-\vec{k}, -\sigma, q}(\vec{r}) \, ,
\end{align}
which fixes the relative phase between states with $\pm \vec{k}$. We recall 
that for Fermionic single-particle states one has
$\hat{T}^2 \, \Psi_{\vec{k}, \sigma, q}(\vec{r}) 
= - \Psi_{\vec{k}, \sigma, q}(\vec{r})$ \cite{MessiahII}.
To better follow the phase factors, we continue to use a notation 
with separate momentum and spin indices. In specific matrix elements, 
$\sigma = +1$ is indicated by $\uparrow$, and $\sigma = -1$ by $\downarrow$.
In the canonical basis of the HFB problem, the Bogoliubov quasiparticle 
vacuum takes the form of a BCS state \cite{Bender19a}, which for INM 
can be written as
\begin{align}
\label{eq:BCS}
| \text{BCS}_q \rangle
& = \prod_{\vec{k}} 
    \big(  u_{k,q} 
         + v_{k,q} a^\dagger_{\vec{k} \uparrow q} a^\dagger_{-\vec{k} \downarrow q} 
    \big) \, 
    | - \rangle \, .
\end{align}
This expression is schematic in so far as in INM the product over $\vec{k}$ is 
over a continuous spectrum of momenta $\vec{k}$ that have all possible 
orientations in space, with states having the same absolute value of 
$k = |\vec{k}|$ being degenerate and having the same occupation amplitudes 
$u_{k,q}$ and $v_{k,q}$.

With the symmetry choices made here, the canonical basis can be chosen 
to be equal to the basis of eigenstates of the single-particle Hamiltonian, 
such that for homogeneous INM the HFB approach becomes numerically 
equivalent to a HF+BCS calculation. 
Some papers on the mean-field description of pairing correlations in INM
even go one step further and neglect the feedback of pairing correlations
on the mean-fields and only solve the gap equation of BCS theory for fixed
single-particle spectrum \cite{Chamel08a}.
This, however, is not what is done here. The rest of this appendix outlines 
how symmetry considerations can be used to construct closed expressions 
for the matrix elements $u_{q,k}$ and $v_{q,k}$ of the Bogoliubov 
transformation of full HFB theory in the canonical basis.

Note that, in contrast to the usual representation of BCS states for finite nuclei, 
the product in Eq.~\eqref{eq:BCS} runs over all $\vec{k}$ and not just half of them. 
By contrast, the product is limited to half of the combinations of $\vec{k}$ with 
spins $\sigma$, i.e.\ the product does not explicitly contain 
$a^\dagger_{\vec{k} \downarrow q} a^\dagger_{-\vec{k} \uparrow q}$: 
this operator is already implicitly contained in the many-body wave function 
through $a^\dagger_{-\vec{k} \uparrow q} a^\dagger_{\vec{k} \downarrow q} = 
-a^\dagger_{\vec{k} \downarrow q} a^\dagger_{-\vec{k} \uparrow q}$.
Numerically, the continuous variable $\vec{k}$ will be discretised with
$n_k$ points for its modu\-lus, see \ref{sec:discrete}.

As we only consider $T=1$ pairing between like nucleons, for sake of compact 
notation the rest of this subsection will be for a generic type 
of nucleons without keeping track of an isospin index.
When treating asymmetric matter, the occupation amplitudes, 
the elements of the normal and anomalous density matrices, the pairing 
gaps, single-particle energies and chemical potentials will have of course
have different values for protons and neutrons.

The usual sign change in $v_k$ when going to the conjugate state
concerns the occupation amplitude of the state with opposite 
$\vec{k}$ \textit{and} opposite spin (for which there is no 
explicit symbol with our choice for the representation of 
the formalism). The normalisation of the BCS state~\eqref{eq:BCS}
\begin{align}
1
& = \langle \text{BCS}| \text{BCS} \rangle 
  =  \prod_{\vec{k}} \big( u_{k}^* u_{k} + v_{k}^* v_{k} \big)
\end{align}
leads to the condition $|u_{k}|^2 + |v_{k}|^2 = 1$ for all $\vec{k}$. 
The $u_k$ and $v_k$ can in general be complex. Up to a further 
common phase of both $u_k$ and $v_k$, a possible choice is \cite{Gross91a,Cohen19a}
\begin{align}
\label{eq:can:uk}
u_k
& = \sin (\vartheta_k)  \, , 
    \\
\label{eq:can:vk}
v_k
& = e^{\iunit \varphi_k} \cos (\vartheta_k) \, .
\end{align}
with $0 \leq \vartheta_k \leq \pi$ and $0 \leq \varphi_k \leq 2 \pi$.
As we are interested only in systems with a single gap in the 
$S=0$ channel for a given nucleon species, we can 
choose $\varphi_k = 0$ for all $\vec{k}$, i.e.\ $u_k$ and $v_k$ being 
real without loss of generality~\cite{Gross91a,Cohen19a}. 
When $\bar{\mu}$ labels the time-reversed state of $\mu$, the normal and 
anomalous density matrices are given by \cite{Ring80a,Bender19a}
\begin{align}
\rho_{\mu \nu}
& \equiv \langle \text{BCS} | a^\dagger_{\nu} a_{\mu}| \text{BCS} \rangle 
  = \big( V^* V^T \big)_{\mu \nu}
  = v_{\mu}^2 \, \delta_{\mu \nu} \, ,
          \nn \\
\kappa_{\mu \nu} 
& \equiv \langle \text{BCS} | a_{\nu} a_{\mu}| \text{BCS} \rangle 
   = \big( V^* U^T \big)_{\mu \nu}
   = v_{\mu} u_{\nu} \, \delta_{\nu \bar{\mu}} \, ,
           \nn \\
\kappa^{*}_{\mu \nu} 
& \equiv \langle \text{BCS} | a^\dagger_{\mu} a^\dagger_{\nu} | \text{BCS} \rangle 
   =  v_{\mu} u_{\nu} \, \delta_{\nu \bar{\mu}} \, .
\end{align}
%

For INM the non-vanishing elements of the $\rho$ and $\kappa$ matrices become
\begin{align}
\rho_{\pm \vec{k} \sigma, \pm \vec{k} \sigma}
& \equiv \langle \text{BCS} | a^\dagger_{\pm \vec{k} \sigma} a_{\pm \vec{k} \sigma}| \text{BCS} \rangle
  \nn \\
& = V_{+\vec{k} +\sigma  , -\vec{k} -\sigma} \, V_{+\vec{k} +\sigma, -\vec{k} -\sigma}
  \nn \\
& = v_{k}^2 \, ,
          \\
\kappa_{+\vec{k} \uparrow, -\vec{k} \downarrow} 
& \equiv \langle \text{BCS} | a_{-\vec{k} \downarrow} a_{+\vec{k} \uparrow} | \text{BCS} \rangle 
      \nn \\
& = V_{+\vec{k} \uparrow  , -\vec{k} \downarrow} \, U_{-\vec{k} \downarrow, -\vec{k} \downarrow}
  \nn \\
& = + u_{k} v_{k} \, ,
    \\
\kappa_{-\vec{k} \downarrow, +\vec{k} \uparrow} 
& \equiv \langle \text{BCS} | a_{+\vec{k} \uparrow} a_{-\vec{k} \downarrow} | \text{BCS} \rangle 
      \nn \\
& =  V_{-\vec{k} \downarrow, +\vec{k} \uparrow} \, U_{+\vec{k} \uparrow  , +\vec{k} \uparrow}
  \nn \\
& = - u_{k} v_{k} \, ,
    \\
\kappa_{-\vec{k} \uparrow, +\vec{k} \downarrow} 
& \equiv \langle \text{BCS} | a_{+\vec{k} \downarrow} a_{-\vec{k} \uparrow} | \text{BCS} \rangle 
      \nn \\
& = V_{-\vec{k} \uparrow  , +\vec{k} \downarrow} \, U_{+\vec{k} \downarrow  ,+\vec{k} \downarrow }
  \nn \\
& = + u_{k} v_{k} \, ,
    \\
\kappa_{+k \downarrow, -k \uparrow} 
& \equiv \langle \text{BCS} | a_{-\vec{k} \uparrow}  a_{+\vec{k} \downarrow} | \text{BCS} \rangle 
     \nn \\
& = V_{+\vec{k} \downarrow, -\vec{k} \uparrow} \, U_{-\vec{k} \uparrow  , -\vec{k} \uparrow} 
  \nn \\
& = - u_{k} v_{k} \, ,
\end{align}
and zero otherwise. The $U$ and $V$ matrices are the usual building blocks of the
Bogoliubov transformation \cite{Ring80a,Bender19a,Blaizot86a}.
In addition, with the choice $\varphi_k = 0$ for the phases of $v_k$,
the matrix elements of $\rho$ and $\kappa$ are real with
\begin{align}
\kappa^{*}_{\vec{k} \sigma, \vec{k}' \sigma'} 
& \equiv \langle \text{BCS} | a^\dagger_{\vec{k} \sigma} a^\dagger_{\vec{k}' \sigma'} | \text{BCS} \rangle 
= \kappa_{\vec{k} \sigma, \vec{k}' \sigma'} \, .
\end{align}
From these relations, the $U$ and $V$ matrices of the Bogoliubov transformation 
in the canonical basis can be constructed. In the canonical basis, one of these 
has to be symmetric and the other skew-symmetric \cite{Ring80a,Bender19a}. 
Equations~\eqref{eq:can:uk} and~\eqref{eq:can:vk} follow the usual convention 
where $U$ is the symmetric matrix and $V$ the skew-symmetric one. In addition,
these relations follow the most common phase convention where in the canonical 
basis the symmetric matrix $U$ is chosen to be positive semi-definite.
The matrix elements of the skew-symmetric matrix $V$ then will not have a
universal sign. The full $U$ and $V$ matrices are block diagonal 
in $2 \times 2$ matrices that connect single-particle states linked by 
time-reversal. For given $\pm \vec{k}$ there are two such sub-matrices which 
each consist of the two single-particle states that are coupled in the two-body 
wave functions. These two-by-two blocks become
\begin{alignat}{3}
\label{eq:U:+u-d}
\left( 
    \begin{array}{cc}
    U_{+\vec{k} \uparrow  , +\vec{k} \uparrow} & U_{+\vec{k} \uparrow  , -\vec{k} \downarrow} \\
    U_{-\vec{k} \downarrow, +\vec{k} \uparrow} & U_{-\vec{k} \downarrow, -\vec{k} \downarrow}
    \end{array}
    \right) 
& = \left( 
    \begin{array}{cc}
      u_{k} &  0     \\
        0    & u_{k}
    \end{array}
    \right) \, , 
    \\
\label{eq:V:+u-d}
\left( 
    \begin{array}{cc}
    V_{+\vec{k} \uparrow  , +\vec{k} \uparrow} & V_{+\vec{k} \uparrow  , -\vec{k} \downarrow} \\
    V_{-\vec{k} \downarrow, +\vec{k} \uparrow} & V_{-\vec{k} \downarrow, -\vec{k} \downarrow}
    \end{array}
    \right) 
& = \left( 
    \begin{array}{cc}
      0  & +v_k \\
     -v_k &  0
    \end{array}
    \right)  \, ,
    \\
\intertext{and}
\label{eq:U:+d-u}
\left( 
    \begin{array}{cc}
    U_{-\vec{k} \uparrow  , -\vec{k} \uparrow} & U_{-\vec{k} \uparrow  , +\vec{k} \downarrow} \\
    U_{+\vec{k} \downarrow, -\vec{k} \uparrow} & U_{+\vec{k} \downarrow, +\vec{k} \downarrow}
    \end{array}
    \right) 
& = \left( 
    \begin{array}{cc}
      u_{k} & 0     \\
      0     & u_{k}
    \end{array}
    \right)  \, ,
    \\
\label{eq:V:+d-u}
\left( 
    \begin{array}{cc}
    V_{-\vec{k} \uparrow  , -\vec{k} \uparrow} & V_{-\vec{k} \uparrow  , +\vec{k} \downarrow} \\
    V_{+\vec{k} \downarrow, -\vec{k} \uparrow} & V_{+\vec{k} \downarrow, +\vec{k} \downarrow}
    \end{array}
    \right) 
& = \left( 
    \begin{array}{cc}
      0  & +v_k \\
    - v_k & 0
    \end{array}
    \right) 
  \, ,
\end{alignat}
Note that the signs represent the relative phases between the non-zero elements 
of the $V$ matrix as they enter the definition of the BCS state \eqref{eq:BCS}, but 
they do not indicate the actual sign of the \textit{numerical} value of $\pm v_k$. The
numerical value of $v_k$ is determined by the sign of the gap, see Eq.~\eqref{eq:v},
such that $-v_k$ can actually be positive and $+v_k$ be negative.

The $u_k$ and $v_k$ are determined by the HFB equations for the quasiparticle
wave functions~\cite{Ring80a,Bender19a,Blaizot86a}
\begin{equation}
\label{eq:HFB:abstract:1}
\fourmat{h - \lambda}{\Delta}{-\Delta^*}{-h^* + \lambda}
\twospinor{U}{V} 
= E \twospinor{U}{V} \, ,
\end{equation}
for each nucleon species $q = n$, $p$, where $\lambda$ is a Lagrange multiplier
for the adjustment of the average particle number per volume (i.e.\ of the 
density) and where the matrices $h$ and $\Delta$ are given by
\cite{Ring80a,Bender19a,Blaizot86a}
\begin{align}
\label{eq:HFB:matel}
\hat{h}_{\nu \mu}
& \equiv \frac{\delta E}{\delta \rho_{\mu \nu}} \, ,
\quad
\Delta_{\nu \mu} 
  \equiv \frac{\delta E}{\delta \kappa^{*}_{\nu \mu}} \, ,
\quad
\Delta^{*}_{\nu \mu} 
  \equiv \frac{\delta E}{\delta \kappa_{\nu \mu}} \, .
\end{align}
Even when the numerical values of the $\kappa_{\nu \mu}$ are real and 
therefore equal to those of $\kappa^{*}_{\nu \mu}$, for the derivation of the 
HFB equations one has to treat $\kappa_{\nu \mu} = -\kappa_{\mu \nu}$ 
and $\kappa^{*}_{\nu \mu} = - \kappa^{*}_{\mu \nu}$ as independent degrees of 
freedom that are varied separately, otherwise one does not obtain the 
usual matrix structure of the HFB equation.

Because of the definitions~\eqref{eq:HFB:matel}, the matrices 
$\hat{h}_{\vec{k} \sigma, \vec{k} \sigma}$ and 
$\Delta_{-\vec{k} -\sigma, +\vec{k} +\sigma}$ have the same 
$2 \times 2$ block structure between single-particle states 
linked by time-reversal like the $\rho$ and $\kappa$ matrices, 
respectively
\begin{alignat}{3}
\label{eq:hmat}
\varepsilon_{k}
& \equiv \hat{h}_{\vec{k} \sigma, \vec{k} \sigma} 
& = & \frac{\delta E}{\delta \rho_{\vec{k} \sigma, \vec{k} \sigma}}
    \\
\label{eq:Deltamat}
\Delta_{k} 
& \equiv  \Delta_{-\vec{k} \downarrow, +\vec{k} \uparrow} 
& = & \frac{\delta E_{\text{pair}}}{\delta \kappa^{*}_{-\vec{k} \downarrow, + \vec{k} \uparrow}} \, ,
           \\
\label{eq:DeltaSmat}
\Delta^{*}_{k} 
& \equiv  \Delta^{*}_{-\vec{k} \downarrow, +\vec{k} \uparrow} 
& = & \frac{\delta E_{\text{pair}}}{\delta \kappa_{-\vec{k} \downarrow, + \vec{k} \uparrow}} \, .
\end{alignat}
where we use directly that the matrix representing the single-particle Hamiltonian 
is diagonal by construction and replace its non-vanishing matrix elements by its 
eigenvalues $\varepsilon_k$ that will be called single-particle energies. 
That the gaps $\Delta_{k}$ are elements of skew-symmetric matrices 
is again completely hidden by the compact notation
\begin{align}
\label{eq:Delta:k:symmetry}
\Delta_{k}
& = + \Delta_{-\vec{k} \downarrow, +\vec{k} \uparrow} 
  = - \Delta_{+\vec{k} \uparrow  , -\vec{k} \downarrow}
    \nn \\
& = + \Delta_{+\vec{k} \downarrow, -\vec{k} \uparrow}
  = - \Delta_{-\vec{k} \uparrow  , +\vec{k} \downarrow} \, ,
\end{align}
and therefore has to be explicitly kept track of with suitable phase factors.
For a pairing interaction that at a given $k$ is attractive, 
and our choice of phases, $\Delta_{k}$ takes a positive value.

The matrix representation  of the HFB Hamiltonian in the canonical
basis of homogeneous non-polarised INM therefore decomposes 
into $2 \times 2$ blocks that only mix single-particle states with
$\pm \vec{k}$ and opposite spin 

\begin{strip}
\rule[-1ex]{\columnwidth}{1pt}\rule[-1ex]{1pt}{1.5ex}
\begin{align}
\label{eq:HFB:+u-d}
\left( 
    \begin{array}{cccc}
     h_{+\vec{k} \uparrow  , +\vec{k} \uparrow} - \lambda & 0 & 0                        & \Delta_{+\vec{k} \uparrow  , -\vec{k} \downarrow} \\
     0                             & h_{-\vec{k} \downarrow  , -\vec{k} \downarrow} - \lambda  & \Delta_{-\vec{k} \downarrow, +\vec{k} \uparrow} & 0 \\
     0                             & -\Delta_{+\vec{k} \uparrow  , -\vec{k} \downarrow} &  -h_{+\vec{k} \uparrow  , +\vec{k} \uparrow} + \lambda & 0  \\
    -\Delta_{-\vec{k} \downarrow, +\vec{k} \uparrow} & 0                                & 0 & - h_{-\vec{k} \downarrow  , -\vec{k} \downarrow} + \lambda  \\
    \end{array}
    \right) 
&
    \left( 
    \begin{array}{cc}
    U_{+\vec{k} \uparrow  , +\vec{k} \uparrow} & U_{+\vec{k} \uparrow  , -\vec{k} \downarrow} \\
    U_{-\vec{k} \downarrow, +\vec{k} \uparrow} & U_{-\vec{k} \downarrow, -\vec{k} \downarrow} \\
    V_{+\vec{k} \uparrow  , +\vec{k} \uparrow} & V_{+\vec{k} \uparrow  , -\vec{k} \downarrow} \\
    V_{-\vec{k} \downarrow, +\vec{k} \uparrow} & V_{-\vec{k} \downarrow, -\vec{k} \downarrow}
    \end{array}
    \right)
    \nn \\
&
  = E_{k}
    \left( 
    \begin{array}{cc}
    U_{+\vec{k} \uparrow  , +\vec{k} \uparrow} & U_{+\vec{k} \uparrow  , -\vec{k} \downarrow} \\
    U_{-\vec{k} \downarrow, +\vec{k} \uparrow} & U_{-\vec{k} \downarrow, -\vec{k} \downarrow} \\
    V_{+\vec{k} \uparrow  , +\vec{k} \uparrow} & V_{+\vec{k} \uparrow  , -\vec{k} \downarrow} \\
    V_{-\vec{k} \downarrow, +\vec{k} \uparrow} & V_{-\vec{k} \downarrow, -\vec{k} \downarrow}
    \end{array}
    \right) \, .
\end{align}
\hfill\rule[1ex]{1pt}{1.5ex}\rule[2.3ex]{\columnwidth}{1pt}
\end{strip}%
\noindent
Unlike the case of finite nuclei for which one has a full $2 n_k \times 2 n_k$ 
matrix problem that in general has to be solved with numerical diagonalisation 
methods, the solutions of the HFB problem \eqref{eq:HFB:+u-d} 
can be given in a closed form.

Substituting the matrix elements of the quasiparticle Hamiltonian by
\eqref{eq:hmat}, \eqref{eq:Deltamat}, and \eqref{eq:DeltaSmat}, 
as well as the elements of $U$ and $V$ by
\eqref{eq:U:+u-d} and \eqref{eq:V:+u-d}, the HFB equation 
\eqref{eq:HFB:+u-d} is equivalent to the two coupled equations 
\begin{align}
\label{eq:HFB:INM:cc:1}
( \varepsilon_{k} - \lambda ) u_k + \Delta_k v_k 
& = E_k u_k \, ,
    \\
\label{eq:HFB:INM:cc:2}    
\Delta_k u_k - ( \varepsilon_{k} -\lambda ) v_k
& = E_k v_k \, .
\end{align}
The eigenvalues $E_k$ of Eq.~\eqref{eq:HFB:+u-d} can be deduced 
from the squares of Eqs.~\eqref{eq:HFB:INM:cc:1} and~\eqref{eq:HFB:INM:cc:2}
\begin{align}
( \varepsilon_{k} - \lambda )^2 u_k^2 + 2 ( \varepsilon_{k} - \lambda ) \Delta_k u_k v_k + \Delta_k^2 v_k^2 
& = E_k^2 u_k^2 \, ,
    \\
\Delta_k^2 u_k^2 - 2 ( \varepsilon_{k} -\lambda ) \Delta_k u_k v_k + ( \varepsilon_{k} -\lambda )^2 v_k^2
& = E_k^2 v_k^2 \, ,
\end{align}
Taking the sum of these equations and substituting the resulting
$u_k^2 + v_k^2$ factor in front of each non-vanishing term by $1$ this yields
\begin{align}
\label{eq:Equasi2}
( \varepsilon_{k} -\lambda )^2 + \Delta_k^2 
& = E_k^2 \, .
\end{align}
Recalling that the elementary excitations of the HFB ground state are given 
by the quasiparticle energies \cite{Ring80a,Bender19a} that each bring an 
excitation energy of $E_k$, the paired HFB quasiparticle vacuum with lowest 
energy is necessarily the state from which one can only create 
excitations with positive quasiparticle energy, see \cite{Bender19a} for 
a detailed discussion. Consequently, the quasiparticle energies of the solutions 
of the HFB equations that give the ground state of paired INM are given by
\begin{align}
\label{eq:Equasi}
E_k
& = + \sqrt{( \varepsilon_{k} -\lambda )^2 + \Delta_k^2} \, .
\end{align}
The values of the amplitudes $u_k$ and $v_k$ can also 
be deduced from Eqs.~\eqref{eq:HFB:INM:cc:1} and \eqref{eq:HFB:INM:cc:2}.
Multiplying \eqref{eq:HFB:INM:cc:1} with 
$v_k$ and \eqref{eq:HFB:INM:cc:2} with $u_k$ 
\begin{align}
( \varepsilon_{k} -\lambda ) u_k v_k + \Delta_k v_k^2 
& = E_k u_k v_k \, ,
    \\   
\Delta_k u_k^2 - ( \varepsilon_{k} -\lambda ) u_k v_k
& = E_k u_k v_k \, ,
\end{align}
and taking the sum of these two equations yields
\begin{align}
\label{eq:uv}
u_k v_k 
& = \frac{\Delta_k}{2 E_k} \, ,
\end{align}
where we used once again that $u_k^2 + v_k^2 = 1$. For the phase convention 
$u_k \geq 0$, the numerical value of $v_k$ has the same sign as $\Delta_k$
\begin{align}
\label{eq:u}
u_k
& = + \sqrt{u_k^2} \, ,
    \\
\label{eq:v}
v_k
& = \text{sign}(\Delta_k) \sqrt{v_k^2} \, .
\end{align}
In the most general case,
the sign of the numerical value of $u_k^2 - v_k^2$ is also fixed by the sign 
of $E_k$. Assuming $\Delta_k \neq 0$ and inserting~\eqref{eq:Equasi2} 
for $E_k^2$ back into either Eq.~\eqref{eq:HFB:INM:cc:1} or \eqref{eq:HFB:INM:cc:2}, 
yields in both cases the relation
\begin{align}
\label{eq:BCS:relation}
2 ( \varepsilon_{k} - \lambda ) u_k v_k 
& = \Delta_k \big( u_k^2 - v_k^2 \big) \, .
\end{align}
Inserting Eq.~\eqref{eq:uv} into Eq.~\eqref{eq:BCS:relation}, one finds
\begin{align}
\label{eq:u2-v2}
\big( u_k^2 - v_k^2 \big)
& = \frac{ \varepsilon_{k} - \lambda }{E_k}
\, .
\end{align}
For positive $E_k$, the sign of $\big( u_k^2 - v_k^2 \big)$, and thereby 
the relative size of $u_k^2$ and $v_k^2$, is determined by the sign of 
$( \varepsilon_{k} - \lambda )$: for $\varepsilon_{k} > \lambda$ one has 
$u_k^2 > v_k^2$, whereas for $\varepsilon_{k} < \lambda$ one finds 
$u_k^2 < v_k^2$. 

The occupation probabilities can be directly calculated from
Eq.~\eqref{eq:u2-v2} without passing through quadratic equations
for $v_k^2$ and $u_k^2$ that would require an analysis of the 
solution to be kept. Using once more that $u_k^2 + v_k^2 = 1$ to 
substitute either $v_k^2$ or $u_k^2$ in Eq.~\eqref{eq:u2-v2}, one 
finds for the occupation numbers that
\begin{align}
1 - 2 v_k^2
& = \frac{\varepsilon_{k} - \lambda }{E_k} \, ,
    \\
2 u_k^2 - 1
& = \frac{\varepsilon_{k} - \lambda }{E_k} \, ,
\end{align}
which, inserting Eq.~\eqref{eq:Equasi}, leads to the well-known expressions
for the BCS ground state~\cite{Ring80a}
\begin{align}
\label{eq:v2}
v_k^2
& = \frac{1}{2} \bigg[ 1 - \frac{\varepsilon_{k} -\lambda}
                          {\sqrt{( \varepsilon_{k} -\lambda )^2 + \Delta_k^2}} \bigg] \, ,
    \\
\label{eq:u2}
u_k^2
& = \frac{1}{2} \bigg[ 1 + \frac{\varepsilon_{k} -\lambda}
                        {\sqrt{( \varepsilon_{k} -\lambda )^2 + \Delta_k^2}} \bigg] \, .
\end{align}
Positive quasiparticle energy $E_k > 0$ therefore implies that 
$v_k^2 < 1/2$ for $\varepsilon_{k} > \lambda$ and 
$v_k^2 > 1/2$ for $\varepsilon_{k} < \lambda$ without the need for 
additional considerations.

With this, for given mean-field potentials and chemical potential $\lambda$, 
the $n_k$ solutions of HFB equations for INM with positive eigenvalues 
can be directly determined from the matrix elements 
$h_{\vec{k} \sigma, \vec{k} \sigma}$ and 
$\Delta_{\vec{k} \sigma, -\vec{k} -\sigma}$
without the need to numerically solve a system of coupled integro-differential 
equations for the quasiparticle wave functions. 
The eigenvalues are given by Eq.~\eqref{eq:Equasi}, 
whereas Eqs.~\eqref{eq:U:+u-d}, \eqref{eq:V:+u-d}, \eqref{eq:v}, and \eqref{eq:u} 
define the eigenvectors in the plane-wave basis~\eqref{eq:plane:wave:0}. Note 
that the spatial wave functions of the upper and lower components of the 
quasiparticle wave functions are plane waves with opposite $\vec{k}$.

We note in passing that there is a second set of $n_k$ solutions of the 
HFB equations with negative quasiparticle energies \cite{Bender19a,Blaizot86a} for
which the roles of $U$ and $V$ are exchanged.
These are only relevant when either solving the HFB equations for 
explicit quasiparticle excitations or when solving the HFB equations at 
finite temperature, neither of which is needed here.

%
\section{Implementing the HFB equations for the Skyrme energy density functional}
\label{sec:INM:Skyrme}

To numerically solve the HFB equations in INM for the Skyrme EDF, the single-particle
energies $\varepsilon_{k}$ and the gaps $\Delta_k$ have to be calculated for this
specific EDF. The relevant expressions are given in this appendix, using the  
notation of Ref.~\cite{Ryssens21a}. Unlike~\ref{sec:HFB:INM}, the
isospin indices will again be explicitely given.


\subsection{Local densities}
\label{sec:EDF:densities}

With the symmetries of homogeneous isotropic non-polarised INM, all normal and 
spin currents are zero, $C^{A,B}_q = C^{A,B \sigma}_q = 0$, as are all 
spin densities $D^{A,B \sigma}_q = 0$. As a consequence, for standard NLO Skyrme 
EDFs, one just has to calculate the local density $D^{1,1}_q$ and the local 
scalar kinetic density $D^{(\nabla, \nabla)}_q$ of protons and neutrons $q=p$, $n$
\begin{align}
\label{eq:D11}
D^{1,1}_q
& = \sum_{\sigma} \int \! \rmd^3 k \, 
    \rho_{\vec{k} \sigma q, \vec{k} \sigma q} \, 
    \big| \psi_{k \sigma q} (\vec{r}) \big|^2 \, ,
    \\
\label{eq:Dnn}
D^{(\nabla, \nabla)}_q
& = \sum_{\sigma} \int \! \rmd^3 k \,  
    \rho_{\vec{k} \sigma q, \vec{k} \sigma q} \, \vec{k}^2 \, 
    \big| \psi_{k \sigma q} (\vec{r}) \big|^2 \, .
\end{align}
The numerical evaluation of the momentum-space integrals is sketched 
in~\ref{sec:discrete}. For N2LO EDFs, one has in addition to construct 
the kinetic tensor density $D^{\nabla, \nabla}_{\mu\nu q}$ that in isotropic 
INM only has diagonal components that are all equal
$D^{\nabla, \nabla}_{q,xx} = D^{\nabla, \nabla}_{q,yy} = D^{\nabla, \nabla}_{q,zz}$
\begin{align}
D^{\nabla, \nabla}_{q,\mu\nu}
& = \sum_{\sigma} \int \! \rmd^3 k \, 
    \rho_{\vec{k} \sigma q, \vec{k} \sigma q} \, k_{\mu} \, k_{\nu} \,  
    \big| \psi_{k \sigma q} (\vec{r}) \big|^2
    \nn \\
& = \tfrac{1}{3} \, \delta_{\mu \nu} \, D^{(\nabla, \nabla)}_q \, ,
\end{align}
and the higher-order scalar kinetic density
\begin{align}
D^{\Delta, \Delta}_q
& = \sum_{\sigma} \int \! \rmd^3 k \, 
    \rho_{\vec{k} \sigma q, \vec{k} \sigma q} \, \vec{k}^4 \, 
    \big| \psi_{k \sigma q} (\vec{r}) \big|^2 \, .
\end{align}
For all of these, the matrix elements of $\rho_{\vec{k} \sigma q, \vec{k} \sigma q}$
are simply given by the occupation numbers $v_k^2$ of the single-particle states 
of the nucleon species $q$ calculated from Eq.~\eqref{eq:v2}.

From the proton and neutron densities, the usual isoscalar and isovector 
densities that enter the compact formulation of the Skyrme EDF are obtained as
\begin{align}
D^{A,B}_0
& = D^{A,B}_n + D^{A,B}_p \, ,
    \\
D^{A,B}_1
& = D^{A,B}_n - D^{A,B}_p \, .
\end{align}
%

\subsection{Local pair densities}
\label{sec:EDF:pairdensities}

From the symmetries assumed here, it also follows that many 
of the local pair densities are zero, i.e.
$\tilde{C}^{A,B}_q = \tilde{C}^{A,B \sigma}_q = \tilde{D}^{A,B \sigma}_q  = 0$.
In addition, with the choice of real occupation amplitudes~\eqref{eq:can:uk}
and~\eqref{eq:can:vk}, all non-vanishing local pair densities are 
real~\cite{Ryssens21a}. 

For the pair EDF~\eqref{eq:EDF:pair:ULB} used here, only the local pair 
density $\tilde{D}^{1,1}_q$ of protons and neutrons needs to be evaluated
\begin{align}
\tilde{D}^{1,1}_q
& = \sum_{\sigma} (-\sigma) 
    \langle \text{BCS} | a_{\vec{r}-\sigma} a_{\vec{r}\sigma} | \text{BCS} \rangle 
    \nn \\ 
& = \sum_{\sigma} \int \! \rmd^3 k \, 
    \kappa_{\vec{k} \sigma, -\vec{k} -\sigma} \, 
    (-\sigma) \,
    \psi_{\vec{k} \sigma q} (\vec{r}) \, \psi_{-\vec{k} -\sigma q} (\vec{r})
    \nn \\
& = 2 \int \! \rmd^3 k \, u_k v_k \, 
    \big| \psi_{\vec{k} \sigma q} (\vec{r}) \big|^2 \, ,
\end{align}
where we use that for a time-reversal invariant system it follows 
from Eq.~\eqref{eq:t-reversal} that 
$- \sigma \, \psi_{-\vec{k} -\sigma q} (\vec{r})
= \psi^*_{\vec{k} \sigma q} (\vec{r})$.

For convenience, the cutoff factors $f^2_k$ of Eq.~\eqref{eq:cutoff}
are absorbed into effective pair densities
\begin{align}
\label{eq:Dp11cut}
\tilde{D}^{1,1}_q
& \to 2 \int \! \rmd^3 k \, u_k v_k \, f^2_k \,
    \big| \psi_{\vec{k} \sigma q} (\vec{r}) \big|^2
    \, ,
\end{align}
as done in Ref.~\cite{Ryssens21a}.
Note that the sums and differences of the pair densities $\tilde{D}^{A,B}_q$
of protons and neutrons do not represent isoscalar and isovector objects. 
Isoscalar and isovector pair densities have to be constructed differently
\cite{Perlinska04a}, reason being that pair densities do not transform like 
hermitian matrices, but like two-body wave functions \cite{Ryssens21a}.
In fact, the pairing EDF of Eq.~\eqref{eq:EDF:pair:ULB}
actually represents the pairing energy in the $T=1$ channel.


\subsection{The energy density functional}
\label{sec:EDF:ph}

The total energy of a piece of nuclear matter of volume $V$ is obtained 
as the sum of kinetic energy, and the contributions of the time-even part 
of the Skyrme EDF at LO, NLO, and N2LO, as well as the pairing EDF
\begin{align}
\label{eq:Etot:INM}
E
& = \int_{V} \! \rmd^3 r \, \mathcal{E}
    \nn \\
&  = \int_{V} \! \rmd^3 r 
    \Big[   \mathcal{E}_{\mathrm{kin}} 
          + \mathcal{E}^{(0)}_{\mathrm{Sk}, e}
          + \mathcal{E}^{(2)}_{\mathrm{Sk}, e}
          + \mathcal{E}^{(4)}_{\mathrm{Sk}, e}
          + \mathcal{E}^{(0)}_{\mathrm{pair}, e}
    \Big] 
    \, .
\end{align}
Compared to the calculations of finite nuclei, the centre-of-mass correction
vanishes, and the Coulomb interaction between protons is omitted. 

As the total energy scales with $V$, it is more useful to look at the energy 
per particle, which equals the energy density divided by the total density 
$E/A = \mathcal{E} / \rho$, which for homogeneous INM are both 
numbers instead of spatial functions.
The kinetic energy density is given by
\begin{align}
\label{E:kin}
\mathcal{E}_{\mathrm{kin}}
& = \frac{\hbar^2}{2m} D_0^{(\nabla,\nabla)} \, .
\end{align}
For the sake of keeping the calculation of INM simple, one is obliged to use
the same mass $m$ for protons and neutrons, even when using a parametrisation
of the Skyrme EDF that was adjusted using their different physical masses. 
Otherwise the saturation point of INM is not in symmetric matter, which 
introduces numerous practical and conceptual difficulties.

For homogeneous INM all derivatives of local
densities are zero. The few remaining LO, NLO, 
and N2LO contributions from the full energy density of the ph EDF as defined
in Ref.~\cite{Ryssens21a} are then
\begin{align}
\mathcal{E}^{(0)}_{\mathrm{Sk}, \mathrm{e}}
& = \sum_{t=0,1}
      \Big[ 
      \cc{A}_{t,\textrm{e}}^{(0,1)}    \big( D^{1,1}_t \big)^2
    + \cc{A}^{(0,2)}_{t,\textrm{e}}    \big( D^{1,1}_0 \big)^\alpha 
                                       \big( D^{1,1}_t \big)^2
      \Big] \, , 
\\
\label{eq:SkTeven:2}
\mathcal{E}^{(2)}_{\mathrm{Sk},\mathrm{e}}
& = \sum_{t=0,1}
    \cc{A}^{(2,2)}_{t,\textrm{e}} D_t^{1,1} D_t^{(\nabla,\nabla)} 
      \, , 
      \\
\label{eq:SkTeven:4}
  \mathcal{E}^{(4)}_{\mathrm{Sk},\mathrm{e}}
  = & \sum_{t=0,1}
     \Big[
     \cc{A}^{(4,2)}_{t,\textrm{e}} D^{1,1}_t \, D^{\Delta, \Delta}_t
   + \cc{A}^{(4,3)}_{t,\textrm{e}} D^{(\nabla, \nabla)}_t D^{(\nabla, \nabla)}_t
\nonumber \\
   & \quad
   + \cc{A}^{(4,4)}_{t,\textrm{e}} \sum_{\mu\nu} D^{\nabla, \nabla}_{t, \mu\nu}
                                                 D^{\nabla, \nabla}_{t, \mu\nu}  
    \Big] \, .
\end{align}
The definition of the coupling constants
$\cc{A}^{(n,j)}_{t,\textrm{e}}$ in terms of the usual 
coupling constants $t_i$ and $x_i$ of the EDF generator 
is given in Ref.~\cite{Ryssens21a}.


\subsection{The single-particle Hamiltonian}
\label{sec:sph}

In a plane wave basis, the matrix $h_{\vec{k} \sigma q, \vec{k}' \sigma' q'}$ that represents 
the single-particle Hamiltonian in INM is diagonal, such that the non-vanishing matrix
elements are automatically its eigenvalues $\varepsilon_{\vec{k} \sigma q}$, 
usually called single-particle energies.
For the Skyrme EDF used here, these are given by
\begin{align}
\label{eq:h}
h_{\vec{k} \sigma q, \vec{k} \sigma q}
& = \frac{\delta \mathcal{E}}{\delta \rho_{\vec{k} \sigma q, \vec{k} \sigma q}}
    \nn \\
& = \bigg( \frac{\hbar^2}{2m} + F^{(\nabla,\nabla)}_q \bigg) \vec{k}^2
    + F^{1,1}_q 
     \nn \\
& \quad 
    + \sum_{\mu} F^{\nabla,\nabla}_{q,\mu\mu} \, k_{\mu}^2
    + F^{\Delta,\Delta}_q \, \vec{k}^2 \, \vec{k}^2 \, ,
\end{align}
where the potentials are defined as the functional derivatives of the 
interaction part of the EDF with respect to the local densities, 
$F^{A,B}_q \equiv \delta \mathcal{E} / \delta D^{A,B}_q$.

Because of the isotropy of non-polarised INM, all three remaining
components to the N2LO kinetic tensor potential are actually equal
$F^{\nabla,\nabla}_{q,x x} = F^{\nabla,\nabla}_{q,y y} = 
F^{\nabla,\nabla}_{q,z z}$, such 
that one can substitute the kinetic tensor term through
$F^{\nabla,\nabla}_{q,xx} \, k_{x}^2 
+ F^{\nabla,\nabla}_{q,yy} \, k_{y}^2
+ F^{\nabla,\nabla}_{q,zz} \, k_{z}^2
= F^{\nabla,\nabla}_{q,\mu\mu} \, \vec{k}^2$, where $\mu$ is the index of any component 
$x$, $y$, or $z$, and over which summation is not needed anymore.

Using the shorthand introduced in Ref.~\cite{Ryssens21a} for the combinations of 
coupling constants that appear in contributions from the same ($q' = q$) or 
the other ($q' \neq q$) nucleon species to the single-particle potentials
\begin{align}
\cc{A}^{(i,j)}_{qq',x} 
& =  \big( \cc{A}^{(i,j)}_{0,x} + \cc{A}^{(i,j)}_{1,x} \big) \, \delta_{qq'}
    +\big( \cc{A}^{(i,j)}_{0,x} - \cc{A}^{(i,j)}_{1,x} \big) \big( 1 - \delta_{qq'} \big) 
    \nn \\
& = \cc{A}^{(i,j)}_{0,x} 
   + \big( 2 \delta_{qq'} - 1 \big)  \cc{A}^{(i,j)}_{1,x} \, , 
\end{align}
the contribution from the Skyrme EDF to the potentials 
$F^{A,B}_q$ for the nucleon  species $q = p$, $n$ that are associated 
with time-even densities through
$F^{A,B}_q \equiv \delta \mathcal{E}_{\text{Sk}} / \delta D^{A,B}_q$
are given by
\begin{align}
\label{eq:F11:Sk}
F^{1,1}_q
& = \sum_{q' = p,n} \bigg[
      2 \, \cc{A}^{(0,1)}_{qq',\text{e}} D^{1,1}_{q'}
    + 2 \, \cc{A}^{(0,2)}_{qq',\text{e}} \big( D^{1,1}_{0} \big)^\alpha \, D^{1,1}_{q'} 
     \nn \\
& \quad 
     + \cc{A}^{(2,2)}_{qq',\textrm{e}} D^{(\nabla,\nabla)}_{q'}
     + \cc{A}^{(4,2)}_{qq',\textrm{e}} D^{\Delta, \Delta}_{q'}
      \bigg]
     \nn \\
& \quad
     + \alpha \, \cc{A}^{(0,2)}_{0,\text{e}} \, 
       \big( D^{1,1}_{0} \big)^{\alpha-1} \, \big( D^{1,1}_{0} \big)^2
     \nn \\
& \quad
     + \alpha \, \cc{A}^{(0,2)}_{1,\text{e}} \, 
       \big( D^{1,1}_{0} \big)^{\alpha-1} \, \big( D^{1,1}_{1} \big)^2
     \\
\label{eq:Fnknk:Sk}
F^{(\nabla,\nabla)}_q
& = \sum_{q' = p,n} \Big[ 
        \cc{A}^{(2,2)}_{qq',\textrm{e}} \, D^{1,1}_{q'} 
    + 2 \cc{A}^{(4,3)}_{qq',\textrm{e}} \, D^{(\nabla, \nabla)}_{q'}
    \Big] \, ,
    \\
\label{eq:Fnmnk:Sk}
F_{q, \mu\nu}^{\nabla,\nabla}  
& = \sum_{q' = p,n} 2 \cc{A}^{(4,4)}_{qq',\textrm{e}} \, D^{\nabla,\nabla}_{q', \mu\nu} \, ,
    \\
\label{eq:Fnmnmnknk:Sk}
F_{q}^{\Delta,\Delta}
& = \sum_{q' = p,n} \cc{A}^{(4,2)}_{qq',\textrm{e}} \, D^{1,1}_{q'} \, . 
\end{align}
%

\subsection{The effective mass of N2LO EDFs}
\label{sec:mstar:N2LO}

For Skyrme EDFs at NLO, at given density and asymmetry the effective masses of 
neutrons and protons in infinite matter might be different, but they are identical
for all nucleons of the same species, irrespective of their momentum.
This is different for N2LO EDFs with their higher-order gradient terms.
The effective $k$-mass is defined through a derivative of the single-particle 
Hamiltonian \cite{Jaminon89a}, that for non-relativistic EDFs can be expressed as
\begin{align}
\label{eq:ms:q}
\frac{m}{m^*_q(k_0)}
& = \frac{m}{\hbar^2 k_0}
    \frac{\partial}{\partial k} 
    h_{\vec{k} \sigma q, \vec{k} \sigma q}
    \Big|_{k=k_0}
    \nn \\
& = 1 + \frac{2m}{\hbar^2}\bigg( F^{(\nabla,\nabla)}_q 
    + F^{\nabla,\nabla}_{q,\mu\mu} 
    + 2 F^{\Delta,\Delta}_q \, \vec{k}^2 \bigg) \, .
\end{align}
where the index $\mu$ is again either $x$, $y$, or $z$.
With this, the single-particle Hamiltonian \eqref{eq:h} 
can be rewritten as 
\begin{align}
\label{eq:effmassapprox}
h_{\vec{k} \sigma q, \vec{k} \sigma q}
& = \frac{\hbar^2}{2m^*_q(k)} \vec{k}^2
    + F^{1,1}_q 
    - F^{\Delta,\Delta}_q \, \vec{k}^4
    \nn \\
& = \frac{\hbar^2}{2m^*_q(k)} \vec{k}^2 + U_q(\vec{k}^2)
    \, .
\end{align}
For NLO EDFs the higher-order potential $F^{\Delta,\Delta}_q = 0$
vanishes, such that the single-particle energy is the kinetic energy evaluated
of nucleons with constant effective mass $m^*_q/m$ plus a constant. For N2LO EDFs,
however, the $\vec{k}^4$ contribution to the single-particle Hamiltonian
is double-counted in the now $k$-dependent effective mass \eqref{eq:ms:q}. 
It therefore has to be subtracted when rewriting 
$h_{\vec{k} \sigma q, \vec{k} \sigma q}$ in the form~\eqref{eq:effmassapprox}, 
meaning that both the effective mass and the effective potential 
$U_q(\vec{k}^2) = F^{1,1}_q - F^{\Delta,\Delta}_q \, \vec{k}^4$ 
become $k$ dependent.

We note in passing that for finite-range interactions for which the 
effective mass emerges from the non-locality of the exchange term,
one also naturally finds that both the effective mass and 
the potential depend on $k$ in a manner that is much more involved than 
Eqs.~\eqref{eq:ms:q} and~\eqref{eq:effmassapprox}.


\subsection{The pairing EDF}
\label{sec:EDF:pp}

In the notation of Ref.~\cite{Ryssens21a}, for non-polarised homogeneous INM 
the LO pairing EDF of Eq.~\eqref{eq:EDF:pair:ULB} becomes
\begin{align}
\label{eq:EDF:pair:ULB:new}
\mathcal{E}^{(0)}_{\mathrm{pair}, e}
& = \sum_{q=p,n} 
    \Big[  \tilde{\cc{A}}^{(0,1)}_{qq}           
           + \tilde{\cc{A}}^{(0,2)}_{qq} \big( D^{1,1}_0 \big)^\sigma 
    \Big] \, 
    \tilde{D}^{1,1*}_{q} \, \tilde{D}^{1,1}_{q} \, ,
\end{align}
where we keep the distinction between $\tilde{D}^{1,1}_{q}$ and $\tilde{D}^{1,1*}_{q}$
for the sake of a more transparent derivation of the pair potentials and gaps 
along the lines of the standard textbook discussion of the HFB equations
\cite{Ring80a,Bender19a,Blaizot86a} in spite of the possibility to choose 
all quantities to be real in the INM calculations reported here.
The pair density $\tilde{D}^{1,1}_{q}$ entering Eq.~\eqref{eq:EDF:pair:ULB:new} 
and all other relations in this section is the one with cutoffs defined in
Eq.~\eqref{eq:Dp11cut}.

The compact coupling constants of Eq.~\eqref{eq:EDF:pair:ULB:new} are 
related to the traditional ones of Eq.~\eqref{eq:EDF:pair:ULB} through
\begin{align}
\tilde{\cc{A}}^{(0,1)}_{qq}
& = \tfrac{1}{4} \, V_0 \, ,
    \\
\tilde{\cc{A}}^{(0,2)}_{qq} 
& = - \frac{\eta}{4 \rho_{\text{ref}}^\sigma} \, V_0 \, .
\end{align}
For the pair potential $\tilde{F}^{1,1}_q$ this yields
\begin{align}
\label{eq:Fp11}
\tilde{F}^{1,1}_q
& =   \frac{\delta \mathcal{E}^{(0)}_{\mathrm{pair}, e}}
           {\delta \tilde{D}^{1,1*}_q}
  =   \tilde{\cc{A}}^{(0,1)}_{qq} \tilde{D}^{1,1}_{q}
    + \tilde{\cc{A}}^{(0,2)}_{qq}
      \big( D^{1,1}_0 \big)^\sigma \, \tilde{D}^{1,1}_{q} \, ,
\end{align}
and the complex conjugate of this expression for $\tilde{F}^{1,1*}_q$.
With this, the gaps $\Delta_k$ defined through Eq.~\eqref{eq:Deltamat} 
are given by
\begin{align}
\label{eq:Delta:inm}
\Delta_k
& = \Delta_{-\vec{k} \downarrow, +\vec{k} \uparrow} 
  = \frac{\delta E_{\mathrm{pair}}}{\delta \kappa^{*}_{-\vec{k} -\downarrow, + \vec{k}}}
    \nn \\
& = \frac{\delta \mathcal{E}^{(0)}_{\mathrm{pair}, e}}{\delta \tilde{D}^{1,1*}_q}
    \frac{\delta \tilde{D}^{1,1*}_q}{\delta \kappa^{*}_{-\vec{k} -\downarrow, + \vec{k}}}
  = - f_k^2 \, \tilde{F}^{1,1}_q \, .
\end{align}
The pairing EDF also adds a further contribution to $F^{1,1}_{q}$ 
of Eq.~\eqref{eq:F11:Sk}
\begin{align}
\label{eq:F11:full}
F^{1,1}_{q}
& = \frac{\delta \mathcal{E}_{\mathrm{Sk}, \mathrm{e}}}{\delta D^{1,1}_q}
    + \sigma \, \tilde{\cc{A}}^{(0,2)}_{qq} \,
    \big( D^{1,1}_0 \big)^{\sigma - 1} \!
    \sum_{q' = p,n} \big| \tilde{D}^{1,1}_{q'} \big|^2
\end{align}
that is included in our calculations of INM. We note in passing that for
homogeneous INM this contribution to the potential leads to a constant positive 
shift of $F^{1,1}_{q}$ that, for realistic parametrisations, does not exceed 
a few hundred keV and shifts all single-particle energies $\varepsilon_{k}$
by the same amount without affecting the single-particle wave functions. As
a consequence, including this term or not in INM calculations neither affects any of the normal 
or pair densities, nor any other normal or pair potential than $F^{1,1}_{q}$,
nor the pairing gaps $\Delta_k$, or the total binding energy. The only other
quantities that change are the chemical potentials $\lambda_q$, but without
affecting the $k_{q,\lambda}$.


\section{Discretisation of momentum-space integrals}
\label{sec:discrete}

For isotropic INM as considered here, the 3d integral over a function 
$f(k)$ of the momentum can be reduced to a 1d-integral that is evaluated 
with a numerical quadrature
\begin{equation}
\int \! \rmd^3 k \, f(k)
= 4 \pi \int \! \rmd k \, k^2 \, f(k)
\to 4 \pi \sum_{i} w_i \, k_i^2 \, f(k_i)
\end{equation}
with integration weights $w_i$ and abscissas $k_i$.
The following considerations entered the set-up of 
the discretisation:
\begin{enumerate}
\item
Because of the pairing cutoff \eqref{eq:cutoff}, the integration can 
be stopped at some $k_{\text{max}}$ that is sufficiently larger than 
the momentum at which the cutoff $f_{k,q}$ makes the gaps 
$\Delta_k$ fall to negligibly small values.

\item
The occupation numbers $v_k^2$ are known to vary quick\-ly around the 
chemical potential, whereas their values much more slowly approach 1
below or fall off to 0 above. The $u_k v_k$ factors multiplying 
pair densities behave similarly. To achieve better convergence of 
integrals weighted with either $v_k^2$ or $u_k v_k$ with respect 
to their numerical discretisation, it is therefore
advantageous to have a denser mesh for momenta close to the 
chemical potentials $\lambda_q$, in particular in the weak-pairing 
limit when the $\Delta_k$, and with that the interval over which 
$v_k^2$ falls from 1 to 0, become very small. To achieve this, 
for each nucleon species the numerical integration interval 
$[0,k_{\text{max},q}]$ is divided into four sub-intervals 
\begin{align}
\int \! \rmd k 
& \to \bigg[   \int_{0}^{k_1} + \int_{k_1}^{k_2} 
             + \int_{k_3}^{k_4} + \int_{k_4}^{k_5}    
      \bigg] \, \rmd k
\end{align}
where $k_2 = k_{\lambda,q}$ is the momentum corresponding to the 
chemical potential and $k_3 = k_{\lambda,q} + \epsilon$. The values of
$k_1$ and $k_4$ are the momenta corresponding to the single-particle 
energies $\varepsilon_{\lambda,q} \mp 0.2 \, \text{MeV}$.

\item
In asymmetric matter, the values of $k_1$, $k_2$, $k_3$,
$k_4$, and $k_5$ are different for protons and neutrons. 

\item
Setting the upper limit $k_2$ of the second interval to either 
$k_{\text{F},q}$ or $k_{\lambda,q}$, and to have the lower limit $k_3$ 
of the third interval at an only infinitesimally larger value permits
the precise evaluation of the momentum-space integrals in the HF limit
when pairing breaks down, such that $k_2 = k_{\lambda,q} = k_{\text{F},q}$ 
with $v_{k_2}^2 = 1$ and $v_{k_3}^2 = 0$. 

\item
If $k_{\lambda,q}$ does not fall on a boundary of a sub-interval of the
numerical integration, one artificially enforces $u_k v_k \neq 0$ 
around the chemical potentials as the HF solution with $u_k v_k = 0$
that yields the targeted density cannot be numerically represented.

\item
The numerical integration is performed with an composite Newton-Cotes 
quadrature with eight equidistant points in the sub-intervals. At given 
total number of integration points, using such higher-order quadrature 
formula yields a more precise description of local higher-order kinetic 
densities $D^{\nabla,\nabla}_q$ and $D^{\Delta,\Delta}_q$ than using a 
simple trapezoidal formula \textit{at exactly the same numerical cost}.

\item
For INM in the BCS phase and $k_{\lambda,q}^2 > 0$, these integrals
are discretised with
\begin{alignat}{3}
n_{k_1}
& = \tfrac{n_k}{3} + 1  & \quad & \text{points in $[0,k_1]$,}
    \\
n_{k_2}
& = \tfrac{n_k}{6}      & \quad & \text{points in $[k_1,k_2]$,}
     \\
n_{k_3}
& = \tfrac{n_k}{6} + 1  & \quad & \text{points in $[k_3,k_4]$,}
    \\
n_{k_4}
& = \tfrac{n_k}{3}     & \quad & \text{points in $[k_4,k_5]$,}
\end{alignat}
such that the two discretisation points at $k_1$ and $k_4$ contribute
to two of the sub-integrals.

\item
In case of a BEC solution for the nucleon species $q$ for which $k_{\lambda,q}$
is outside of the spectrum of physical plane-wave states, the integrals cannot be 
set up as described above. In this case, all four sub-integrals are discretised 
in equidistant steps from $k = 0$ up to a suitably chosen $k_{\text{max},q}$ 
above the pairing cutoff.

\item
Because of setting up the integral with four regions 
that all are discretised with a composite eight-point quadrature and shared
points on some of the boundaries, $n_k - 2$ has to be a multiple of 42. 
Calculations reported here were performed with in total 3782 discretisation 
points, which in the majority of cases are many more than needed to achieve 
numerical convergence, but this ensures convergence also in extreme cases.

\end{enumerate}


\section{Solution of the self-consistent problem}
\label{sec:self-consistency}

The HFB equations for INM can be solved with a straightforward simplification
of the the two-basis method described in Ref.~\cite{Ryssens19a} that is 
used to solve the HFB equations for finite nuclei.

Solving the HFB equations can be broken down to nested ``subproblems''
as defined in Ref.~\cite{Ryssens19a}, where inside each 
iteration that approaches the solution of the
HFB problem one has to solve (i) a linear sub-problem that determines 
the eigenvalues and eigenvectors of the single-particle Hamiltonian 
$h$ for given potentials, (ii) a pairing sub-problem that solves the HFB 
matrix equation in the basis of eigenstates of $h$, and (iii) a 
non-linear sub-problem for the updates 
of the self-consistent-fields (SCFs) that enter
the calculation of the matrices that represent $h$ and $\Delta$.

For INM, the linear sub-problem of diagonalising the single-particle
Hamiltonian $h$ for given potentials is analytically solved as the 
eigenvectors of $h$ are known a priori to be the plane waves of 
Eq.~\eqref{eq:plane:wave:0}, while the eigenvalues of $h$ can be
calculated from the analytical expression~\eqref{eq:h}.

Concerning the pairing sub-problem, for given chemical potential 
$\lambda_q$ and set of $h_k$ and $\Delta_k$, the numerical values 
of the eigenvalues and eigenvectors of the HFB matrix equation 
\eqref{eq:HFB:equation} are calculated from the analytical expressions 
constructed in \ref{sec:HFB:INM}. To solve the
pairing sub-problem, however, the chemical potentials $\lambda_q$ of 
protons and neutrons have to be numerically adjusted 
such that $\rho_p$ and $\rho_n$ calculated through Eq.~\eqref{eq:D11}
combine to the targeted values of $\rho$ and $I$ defined through 
Eqs.~\eqref{eq:rho} and \eqref{eq:I}. This problem can be mapped on finding 
the roots of two coupled non-linear equations, one for protons and another one
for neutrons. In case of either symmetric or pure neutron matter, 
there is even just 
one independent chemical potential, which then can be efficiently determined 
with a combination of bisection and Brent's method. For the more general case 
of arbitrarily asymmetric INM, the two independent chemical potentials of 
protons and neutrons are adjusted with Newton's method. 

In either case, $k_{\lambda,q}$ is then determined by finding the
momentum that yields $h_{k_{\lambda,q }\sigma, k_{\lambda,q} \sigma} = \lambda_q$
from Eq.~\eqref{eq:h}. Even for Skyrme EDFs at N2LO, this is a simple bi-quadratic
equation that can be analytically solved.

As the integration mesh as defined in \ref{sec:discrete} is centred
at the momentum $k_{\lambda,q}$ corresponding to the chemical potential
$\lambda_q$, the integration mesh however has to be re-centred to $k_{\lambda,q}$ 
for each update of $\lambda_q$.

As a numerically simpler and more robust alternative, the HFB equations 
can also be solved at fixed $k_{\lambda,q}$~\cite{Guillon24m} (which
is equivalent to fixed $\lambda_q$), in which case 
the integration mesh does not have to be re-centred. In such scheme,
however, one has no control over the resulting values of $\rho_n$ and $\rho_p$, 
which makes it impossible to trace the density-dependence of asymmetric matter 
at a fixed arbitrary asymmetry, or to scan BECs when the chemical potential 
is below the potential depth in INM.
Still, a scheme with fixed $k_{\lambda,q}$ offers a very robust procedure for 
the analysis of the BCS phase of both symmetric and neutron matter.

Compared to the treatment of finite nuclei within the two-basis method as
described in Ref.~\cite{Ryssens19a}, the canonical basis of the HFB problem 
for INM does not have to be explicitly constructed, as the plane-wave states
simultaneously diagonalise the single-particle Hamiltonian 
$h_{\vec{k} \sigma q, \vec{k}' \sigma' q}$ and the density matrix
$\rho_{\vec{k} \sigma q, \vec{k}' \sigma' q}$. Instead, the normal 
and pair densities can be directly calculated in the plane-wave basis 
as described in \ref{sec:EDF:densities} and 
\ref{sec:EDF:pairdensities} without a basis change. With this, the 
HFB calculation of INM has the numerical simplicity of what for finite nuclei 
is usually called the HF+BCS calculation \cite{Ring80a}, but without loss 
of generality. 

For INM, the non-linear SCF updates also become quite simple: the 
potentials and densities are numbers instead of spatial functions, such 
that there are no spatial oscillations to be attenuated when going 
from one iteration to the next. Still, to stabilise
the iterations, the normal potentials $F^{A,B}$ and pair potentials 
$\tilde{F}^{A,B}$ are linearly mixed going from the iteration $i$
to the next
\begin{align}
F^{A,B(i+1)}
= \big( 1 - \beta \big) \, F^{A,B(i)}
  + \beta \, F^{A,B(i+1)}_{|\Psi\rangle}  \, ,
    \\
\tilde{F}^{A,B(i+1)}
=  \big( 1 - \beta \big) \, \tilde{F}^{A,B(i)}
  + \beta \, \tilde{F}^{A,B(i+1)}_{|\Psi\rangle} \, , 
\end{align}
where the $F^{A,B(i)}$ are the ``mixed'' potentials constructed at 
the end of iteration $i$ that enter the HFB Hamiltonian at 
iteration $i+1$, and the $F^{A,B(i+1)}_{|\Psi\rangle}$ are the potentials
obtained directly from the densities constructed from the solution of 
the HFB equations at iteration $i+1$.
The optimal value for the mixing parameter $\beta$ depends on the 
properties of the parametrisations of the Skyrme and pair EDFs used,
and also the density and asymmetry at which the calculation is 
performed. In some cases the HFB equations can actually 
be converged without such mixing, $\beta = 1$. A robust 
choice that worked for almost all calculations we made so far is 
$\beta = 0.1$.

In the limit of disappearing pairing, the code does in general not 
find a HF solution by itself when starting from a weakly paired solution,
reason being that $k_{\lambda,q}$ continues to change by a tiny amount 
at each iteration. When the code detects that the $u_k v_k$ factors are 
significantly different from zero only for a very few integration points 
around $k_{\lambda,q}$, it sets the occupation numbers to those of 
a HF calculation for which $k_{\text{F},q}$ and $k_{\lambda,q}$ are 
equal and correspond to the targeted density $\rho_q$ through 
Eq.~\eqref{eq:kF}. 

At convergence, the internal consistency between the calculation of the 
EDF \eqref{eq:Etot:INM}, the calculation of the potentials $F^{A,B}_q$ of
Eqs.~\eqref{eq:F11:Sk}, \eqref{eq:Fnknk:Sk}, \eqref{eq:Fnmnk:Sk}, 
\eqref{eq:Fnmnmnknk:Sk}, \eqref{eq:Fp11}, and \eqref{eq:F11:full},
and the elements of the matrices $h_{\vec{k} \sigma q; \vec{k}' \sigma' q}$
\eqref{eq:h} and $\Delta_{\vec{k} \sigma q; \vec{k}' \sigma' q}$ 
\eqref{eq:Delta:inm} can be verified through an alternative 
calculation of the total energy through the expression \cite{Ryssens15a}
\begin{align}
\label{eq:magic}
\mathcal{E}_{\mathrm{tot},sp}
& =  \tfrac{1}{2} \mathcal{E}_{\mathrm{kin}}
   + \tfrac{1}{2} \mathcal{E}_{\mathrm{sp}}
   + \tfrac{1}{2} \mathcal{E}_{\Delta}
   + \mathcal{E}_{\mathrm{re}}
\end{align}
that, besides the kinetic energy $\mathcal{E}_{\mathrm{kin}}$ 
of Eq.~\eqref{E:kin}, is composed of the weighted sum over single-particle 
energies
\begin{align}
\mathcal{E}_{\mathrm{sp}}
& = \frac{8 \pi}{(2\pi)^3} \sum_q
    \int \! \rmd k \, k^2 \, v_{k,q}^2 \, \varepsilon_{k,q} \, ,
\end{align}
the pairing energy obtained summing up gaps
\begin{align}
\mathcal{E}_{\Delta}
& = - \frac{1}{2} \frac{8 \pi}{(2\pi)^3} \sum_q
    \int \! \rmd k \, k^2 \ u_{k,q} \, v_{k,q} \, f_{k,q}^2 \, \Delta_{k,q} \, ,
\end{align}
and the rearrangement energy from the density-de\-pen\-dent terms 
that for the EDF used here reads
\begin{align}
\mathcal{E}_{\text{re}}
& = \frac{\alpha}{2} \sum_{t=0,1} 
    \cc{A}^{(0,2)}_{t,\textrm{e}} \big( D^{1,1}_0 \big)^\alpha \big( D^{1,1}_t \big)^2
    \nn \\
& \quad  
   +\frac{\sigma}{2}
    \sum_{q = p,n}  \tilde{\cc{A}}^{(0,2)}_{qq} \,
    \big( D^{1,1}_0 \big)^{\sigma1} \! \big| \tilde{D}^{1,1}_{q} \big|^2
    \, .
\end{align}
In our implementation, for NLO and N2LO EDFs, the energy per 
particle calculated either directly from the EDF of Eq.~\eqref{eq:Etot:INM} 
expressed through the local densities as outlined in~\ref{sec:EDF:ph}
and~\ref{sec:EDF:pp}, or as $\mathcal{E}_{\mathrm{tot},sp}/\rho$ from
Eq.~\eqref{eq:magic} typically differs by less than $10^{-2}$ keV.

\begin{figure*}[t!]
\centerline{\includegraphics[width=5.33cm]{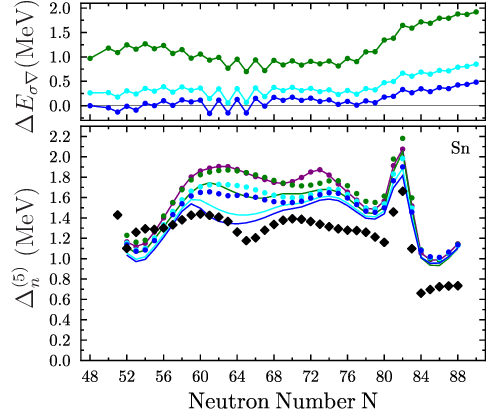}
 \includegraphics[width=5.33cm]{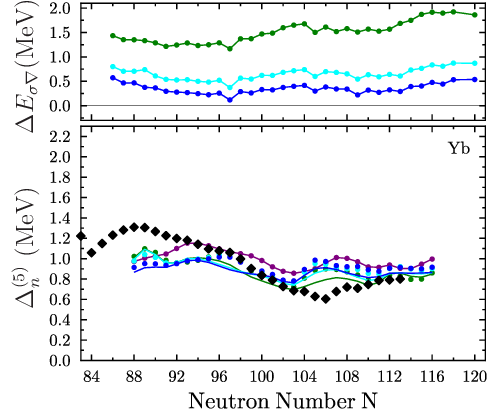}
  \includegraphics[width=5.33cm]{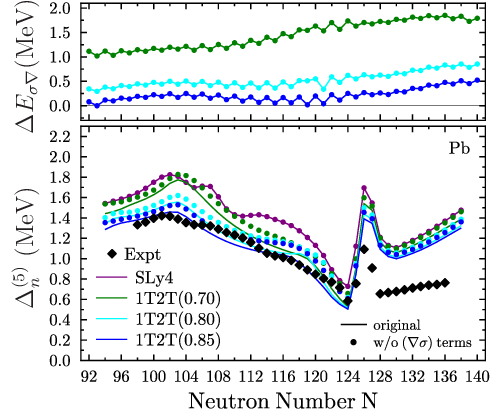}
 }
\caption{\label{fig:gaps:sly4:ULB}
The lower panels show the pairing gap $\Delta^{(5)}_n$ 
for the chains of Sn, Yb, and Pb,  isotopes calculated with the original 
SLy4+ULB parameter set (purple) compared to calculations with the 1T2T(X) 
parameter sets combined with \LOPP{}{ULB} pairing parameters. 
For the 1T2T(X) results from two different sets of calculations 
are shown that either do (solid lines) or do not (dots) take into 
account the terms in the EDF that contain two gradients and two Pauli 
matrices (see text).
The upper panel displays the difference in (negative) total ground-state energy
$\Delta E = E_{\text{w/o} (\sigma \nabla)} - E_{\text{org}}$
between these two choices. Positive values indicate that the 
calculation without spin-gradient terms yields more binding energy.
}
\end{figure*}


\section{The role of spin-gradient terms}
\label{sec:sly4+ulb}

A point not addressed in the main text concerns the predictive
power of the original SLy4+ULB parametrisation for the pairing 
properties of finite nuclei or, more specifically, if SLy4+ULB
shares the predictive power of the 
1T2T(X) parametrisations plus a pairing EDF adjusted
to reproduce the pairing gap of SLy4+ULB in symmetric INM. 

Our unexpected observation is that the data for  
$\Delta_{n}^{(5)}(Z,N)$ are significantly less well reproduced
by the original SLy4+ULB parametrisation than by the
the 1T2T(X) + \LOPP{}{ULB} parameter sets. 
This is illustrated by the bottom panels of Fig.~\ref{fig:gaps:sly4:ULB}, 
which compare results for the Sn, Yb, and Pb isotopic chains
obtained with SLy4+ULB (purple line) with results 
obtained with the 1T2T(X) + \LOPP{}{ULB} (data taken from 
Fig.~\ref{fig:gaps:tosly4+ULB} and represented by solid lines in the 
same colour as there).
For SLy4, the difference between calculated and experimental gaps 
is particularly large for the spherical Sn and Pb isotopes just below 
\nuc{132}{Sn} and \nuc{208}{Pb}, respectively.

It turns out that this difference in performance 
is mainly caused by the terms in the NLO EDF that in one way or another 
couple two gradients with two Pauli spin matrices 
\begin{align}
\label{eq:sigmasigmanablanabla}
\mathcal{E}_{\sigma \nabla}(\vec{r})
& = A^{(2,2)}_{t,o} 
    \bigg[ \vec{D}^{1,\sigma}_{t}(\vec{r}) 
           \cdot \vec{D}^{(\nabla,\nabla)\sigma}_{t} (\vec{r})
           \nn \\
& \qquad \qquad - \sum_{\mu \nu} 
           C^{1,\nabla \sigma}_{t,\mu\nu}(\vec{r}) \, 
           C^{1,\nabla \sigma}_{t,\mu\nu}(\vec{r})
    \bigg]
     \nn \\
& \quad + A^{(2,1)}_{t,o}  
          \vec{D}^{1,\sigma}_{t} (\vec{r}) \cdot \Delta \vec{D}^{1,\sigma}_{t} (\vec{r}) \, ,
\end{align}
represented again in the notation of Ref.~\cite{Ryssens21a}.
Galilean invariance of the nuclear EDF \cite{Dudek95a,Engel75a} requires that
two out of these three terms have the same coupling constant $A^{(2,2)}_{t,o}$ 
with opposite sign (but none of these terms contributes to the EDF 
of isotropic homogeneous non-polarised INM).

For reasons of numerical and phenomenological convenience, 
the coupling constants of these spin-gradient 
terms have been explicitly set to zero 
during the parameter adjustment of SLy4, see \cite{Chabanat98a},
which is common practice for the majority of Skyrme parameter
sets found in the literature \cite{Ryssens15a}. 
By contrast, all of these terms are kept for the 1T2T(X) and also
SN2LO1.
The influence of these terms on binding energies and the 
$\Delta^{(5)}_n(Z,N)$ values of Sn, Pb, and Yb isotopes can also
be deduced from Fig.~\ref{fig:gaps:sly4:ULB}. The dots on the 
lower panel represent $\Delta^{(5)}_n(Z,N)$
obtained from self-consistent
calculations with modified 1T2T(X) + \LOPP{}{ULB} parametrisations
for which the coupling constants of the spin-gradient terms
of Eq.~\eqref{eq:sigmasigmanablanabla} are artificially set to zero as 
always done for SLy4.\footnote{Note that a 
modified SLy4 parametrisation to which spin-gradient terms 
are added with coupling constants as dictated by the two-body 
Skyrme generator has a finite-size spin instability, which prohibits
us from doing the analysis also the other way around.
} 
The such modified 1T2T(X) parameter 
sets do of course not minimise the penalty function of their
parameter adjustment anymore and should therefore not be used
for production calculations. With few exceptions, the 
modified 1T2T(X) parameter sets yield significantly larger 
$\Delta^{(5)}_n(Z,N)$, with the effect being much more 
dramatic for the Sn and Pb chains than for the deformed Yb 
isotopes.

Interestingly, the effect of the spin-gradient terms seems to scale 
with effective mass. In particular, the $\Delta^{(5)}_n(Z,N)$ obtained
from the modified 1T2T(0.70) are very similar to those of SLy4 that 
has a near-identical effective mass. The effective-mass dependence 
of the spin-gradient terms can be better isolated when plotting the 
difference in (negative) total energy between a calculation without 
spin-gradient terms and a full calculation
\begin{align}
\Delta E_{\sigma \nabla}(Z,N) 
& \equiv E_{\text{w/o} (\sigma \nabla)}(Z,N) - E_{\text{org}}(Z,N) \, ,
\end{align}
which is done in the upper panels on Fig.~\ref{fig:gaps:sly4:ULB}. 
The spin-gradient terms are mostly repulsive. For the 1T2T(X),
the energy loss from the spin-gradient terms indeed becomes much larger 
with decreasing $m^*_0/m$. This seems to be an indirect consequence of the 
coupling constants of the spin-gradient terms of Eq.~\eqref{eq:sigmasigmanablanabla} 
depending on different combinations of the same coupling constants 
$t_1$, $x_1$, $t_2$, and $x_2$ that determine the difference between the
bare nucleon mass and the isoscalar and isovector effective masses 
of a NLO Skyrme EDF.

A global shift of all binding energies does not affect
$\Delta^{(5)}_n(Z,N)$, though. More important for the purpose 
of our study is the clearly visible odd-even staggering of 
$\Delta E_{\sigma \nabla}(Z,N)$ that is much more pronounced for the 
chains of spherical Sn and Pb isotopes than for the chain of deformed Yb isotopes.
The size of the odd-even staggering of $\Delta E_{\sigma \nabla}(Z,N)$
also tends to grow with decreasing effective mass, while exhibiting also
a dependence on the quantum numbers of the blocked neutron.

The comparison of results from calculations with the modified and the original
1T2T(X) clearly indicates that the spin-gradient  terms can bring a sizeable 
contribution to the odd-even staggering of total energies. For odd nuclei, the 
time-even $C^{1,\nabla \sigma}_{t,\mu\nu} C^{1,\nabla \sigma}_{t,\mu\nu}$
spin-current tensor terms, which are mainly determined by the overall 
filling of orbits \cite{Lesinski07a,Bender09t}, usually take a similar 
size as in adjacent even-even nuclei. What is different for odd nuclei 
is that there is an additional contribution from the time-odd
$\vec{D}^{1,\sigma}_{t} \cdot \vec{D}^{(\nabla,\nabla)\sigma}_{t}$
terms that leads to spin-dependent effective masses.
These two types of terms usually contribute with opposite sign to 
the total energy, and in some cases the time-odd contribution 
reaches half the size of its time-even partner terms.
Combined, these terms generate a contribution to the odd-even staggering 
of masses that has nothing to do with pairing correlations.
For the 1T2T(X) the combined contribution of these two terms is
repulsive -- as is the case for the majority of Skyrme parametrisations
that take them into account \cite{Lesinski07a,Bender09t} -- 
leading to a smaller energy loss for odd than for even isotopes.

In addition, for most Skyrme NLO EDFs the time-odd 
$\vec{D}^{1,\sigma}_{t} \cdot \Delta \vec{D}^{\sigma}_{t}$ terms
that also only contribute to odd-mass nuclei are in general attractive and, 
depending on quantum numbers of the blocked quasiparticle, can be 
as large as a few 100 keV, which then further increases the binding 
energy of odd isotopes.

The presence of the spin-gradient terms of course affects the total size
of the other time-odd spin terms though self-consistent polarisation effects
when minimising the total energy, as it does more generally affect the size 
of all other terms, but the overall net effect is that the contribution of 
the spin-gradient terms of Eq.~\eqref{eq:sigmasigmanablanabla} in most cases
reduces the difference in binding energy between odd nuclei and the 
adjacent even-even ones.

Through Eq.~\eqref{eq:Delta:5}, the spin-gradient terms then ultimately 
lead to smaller $\Delta^{(5)}_n(Z,N)$ 
for \textit{all} isotopes. Because of the interdependence of the coupling 
constants in the Skyrme EDF, for the 1T2T(X) the overall size of the 
odd-even effect from spin-gradient terms depends on effective mass.
For the SLy4 parametrisation that by construction is to be used without 
spin-gradient terms, one consequently finds larger odd-even staggering 
of total energies and systematically larger $\Delta^{(5)}_n(Z,N)$ than 
what is found when using the (original) 1T2T(X) parameter sets. 

The absence of spin-gradient terms actually introduces a violation of 
the Pauli principle by the Skyrme EDF (in addition to the one already 
present because of density-dependent terms and the use of different 
EDFs for the particle-hole and particle-particle channels)
which is accompanied by spurious self-interactions
\cite{Stringari78a,Lacroix09a,Bender09p}
whose net effect on binding energies might be on an energy scale 
that is relevant for the phenomena discussed here \cite{Tarpanov14b}. 
Because of the difficulty to isolate the net contribution from 
self-interactions to the total energy, however, it cannot be easily 
judged if the difference in performance of SLy4 and the 1T2T(X) for 
$\Delta^{(5)}_n(Z,N)$ is mainly related to relevant physics brought by 
spin-gradient terms to the 1T2T(X) or by spurious self-interactions 
contained in SLy4.

The quantum-number dependence of the contribution of the spin-gradient terms
to $\Delta^{(5)}_n(Z,N)$ could also be partially responsible for  
deviations between theory and experiment when the ground-state configuration 
is incorrectly described, and the effective-mass dependence of this 
contribution found for the 1T2T(X) could be partially responsible for 
the systematic spread between the predictions from the 1T2T(X) for 
$\Delta^{(5)}_n(Z,N)$ as found for certain nuclei. 

The difference in performance between SLy4 and the 1T2T(X) also raises the 
question to which extent the very satisfying performance of the \LOPP{}{ULB} 
pairing parameters that are adjusted to reproduce the gaps from SLy4+ULB 
pairing in INM might be fortuitous.

It is to be noted that, compared to the NLO EDF, the N2LO EDF of SN2LO1 
contains several additional spin-gradient terms \cite{Ryssens21a} with 
independent coupling constants, and there are also additional 
time-even terms contributing to $m^*_0/m$, which complicates 
the situation and might lead to different correlations between
the terms in the EDF than what is found for the 1T2T(X).

\end{appendix}




\newpage{\pagestyle{empty}\cleardoublepage}

\end{document}